\documentclass{iopart}

\usepackage{graphicx,epsfig}
\usepackage{longtable}
\usepackage{color}
\usepackage[sort&compress,numbers]{natbib}
\usepackage{doi}
\usepackage{hyperref}

\usepackage{bm}
\usepackage{capt-of}
\usepackage{relsize}
\usepackage{xcolor}
\usepackage{amssymb}
\usepackage[normalem]{ulem}
\usepackage{mathptmx}
\usepackage{longtable}

\newcommand\nomenclature[2]{#1 & #2 \\}

\newcommand{\nuc}[2] {$^{#1}$#2}

\newcommand{\be}{\begin{equation}}
\newcommand{\ee}{\end{equation}}
\newcommand{\bea}{\begin{eqnarray}}
\newcommand{\eea}{\end{eqnarray}}

\begin{document}

\topical{Electromagnetic reactions on light nuclei}

\author{Sonia Bacca$^{1,2}$ and Saori Pastore$^3$}

\address{$^1$TRIUMF, 4004 Wesbrook Mall, Vancouver, BC, V6T 2A3, Canada}
\address{$^2$Department of Physics and Astronomy, University of Manitoba, Winnipeg, MB, R3T 2N2, Canada }
\address{$^3$Department of Physics and Astronomy, University of South Carolina, Columbia, SC 29208, USA}

\ead{bacca@triumf.ca, pastores@mailbox.sc.edu}

\begin{abstract}
 Electromagnetic reactions on light nuclei are fundamental to advance our understanding of
 nuclear structure and dynamics. The perturbative nature of the electromagnetic probes allows to clearly
 connect measured cross sections with the calculated structure properties of nuclear targets. 
 We present an overview on recent theoretical {\it ab-initio} calculations of electron-scattering and
 photonuclear reactions involving light nuclei.  We encompass both the conventional approach
 and the novel theoretical framework provided by chiral
 effective field theories. Because both strong and electromagnetic interactions are involved
 in the processes under study, comparison with available experimental data provides stringent
 constraints on both many-body nuclear Hamiltonians and electromagnetic currents.
 We discuss what we have learned from studies on electromagnetic observables of light nuclei,
 starting from the deuteron and reaching up to nuclear systems with mass number $A=16$.
\end{abstract}
%\noindent{\it Keywords\/}:ab-initio,
%\pacs{put PACS here}
\submitto{\JPG}
%\maketitle

\tableofcontents

\section{Introduction}

In this review we report on recent theoretical {\it ab-initio} calculations of 
electromagnetic (e.m.) reactions on light nuclei. E.m.~reactions are particularly
suited to test the extent and predictive power of nuclear theories.
In fact, the small e.m.~coupling strength, characterized by the fine-structure constant
$\alpha \sim 1/137$, allows for a perturbative treatment of the e.m.~interaction.
Contributions beyond the leading order 
term in the $Z\alpha$-expansion, where $Z$ is the number of protons,
that is beyond the Born approximation, are sufficiently small to be safely
disregarded in light nuclei. This leaves us with relatively simple reaction mechanisms
and with formal expressions for the cross sections in which the nuclear
structure content can be easily isolated from the well know one associated with
the external structureless probes~\cite{DeForest66,Donnelly75,Donnelly84}.
In many cases, experimental data of e.m.~observables can be accessed with great accuracy,
providing us with stringent constraints on nuclear models. A theoretical understanding
and control of nuclear e.m.~structure and dynamics is a necessary prerequisite
for studies on weak induced reactions.  The experimental
data acquisition for this kind of processes is comparatively
more involved owing to the tinier cross sections and to the fact that
neutrinos are chargeless particles and, thus, they are hard to collimate
and detect. Therefore, in order to address important issues, such as the recently
observed anomaly in the measured cross section of
quasi-elastic neutrino scattering off $^{12}$C~\cite{Aguilar08}, it is imperative to first
validate our theoretical understanding of e.m.~reactions on light nuclei.

In this review we discuss mainly electron scattering reactions
and photonuclear reactions as possible doorways to access nuclear properties. 
Alternative e.m.~reactions, very useful especially in the case of exotic nuclei,
are, {\it e.g.}, Coulomb scattering processes. We refer to them in a few instances, 
and redirect the reader to dedicated reviews (see, {\it e.g.}, Ref.~\cite{Gade2008} 
for more details).

In the Born approximation for the electron and photonuclear cross sections,
the single photon exchanged transfers a four-momentum
$q^\mu=\left(\omega,{\bf q}\right)$ to the target nucleus (with
$\omega=\left |{\bf q}\right|$, for real photons)\footnotemark[1]\footnotetext{Here and in what follows
we use the convention $\hbar=c=1.$}.
Photons, then, probe the e.m.~charge and current distributions of nuclei with spatial
resolution $\propto 1/\left |{\bf q}\right|$.
In this review, we are concerned with processes occurring below
the pion production threshold, that is we consider transferred energies
$\omega < m_\pi$, where $m_\pi \sim 140$ MeV is the pion mass. At these
energies, pions only appear as virtual particles being exchanged among
nucleons. Ground-states properties, such as nuclear elastic form factors,
are accessed via elastic electron-scattering reactions (where $\omega=0$).
Observables associated with inelastic processes (where $\omega\neq0$)
are, for example, e.m.~transition strengths 
and inelastic response functions. Other interesting e.m.~processes,
not covered in the present review, are, for example, Compton scattering
reactions on light nuclei, for which we refer the reader to the
review article of Ref.~\cite{compton_scattering}. 

In the theoretical framework we discuss,
commonly referred to as {\it ab-initio}, nuclei are described in terms
of point-like nucleons interacting among themselves via many-body
forces. Nuclear forces are in practice phenomenological, in that they
are constructed so as to reproduce available experimental data. For
example, nucleon-nucleon (NN) potentials invoke parameters which are
fitted, with a $\chi^2$/datum close to one, to a wide number of NN scattering data
and to the  deuteron binding energy.
Potentials fulfilling the characteristics outlined above are called
realistic. Most of the potentials utilized in the calculations
we present explicitly include one-pion exchange (OPE) mechanisms to
describe the long-range part of the nuclear interaction, and, while being
phase-shift equivalent, they implement different schemes
to parametrize the intermediate- and short-range parts, including
multiple-pion or heavy-meson exchanges~\cite{Machleidt89}.
In view of their crucial role in reproducing
the energy spectrum of light nuclei, three-nucleon (3N) forces are accounted
for in most of the results we present. Models for 3N potentials 
involve, for example, multiple-meson exchanges
and virtual excitations of nucleonic degrees of freedom, {\it e.g.}, $\Delta$-resonances~\cite{Machleidt89}. 

We discuss non-relativistic theoretical frameworks (for recent reviews on
relativistic calculations of e.m.~properties of $A=2$ and $3$ nuclei see,
{\it e.g.}, Refs.~\cite{Gilman,Pinto09,Marcucci14} and references therein).
Nuclear wave functions are then solutions of the Schr\"odinger
equation with a Hamiltonian consisting of the sum of the nucleons' kinetic
energies along with NN and 3N potentials. The average nucleon velocity,
for example, in $A=8$ nuclei is of the order of $\sim 0.2$. 
This motivates the non-relativistic description of nuclear systems.
Relativistic effects are taken into account as kinematic corrections to
the non-relativistic nuclear operators that act on the nuclear
wave functions, and they are given by higher order terms in the $\left| {\bf p} \right|/m$
expansion of the corresponding covariant operators, where ${\bf p}$ and
$m$ are the nucleon's momentum and mass, respectively.

Implicit in the definition of {\it ab-initio} calculations is the requirement 
that the computational methods utilized to solve the many-body Schr\"odinger equation
provide solutions that are numerically exact or obtained within controlled approximation schemes. 
 
Within the microscopic description outlined above,
nuclear e.m.~currents are also expressed as an expansion in many-body
operators. The major contribution to this expansion is provided by the
non-relativistic single-nucleon e.m.~current operator, referred to as
the non-relativistic Impulse Approximation (IA) operator. Thus,
in the limit of $|{\bf q}|\rightarrow0$, the time-like component of
the nuclear e.m.~current reduces to the protons' charges inside the nucleus
to give the total nuclear charge, while the space-like component consists of
the single-nucleon vector current generated by moving protons
(convection current) and that associated with the nucleons' spins
(spin-magnetization current). 

Nuclear e.m.~currents and nuclear potentials are linked by the 
continuity equation, resulting from the gauge invariance of the theory,
and expressing that the charge is a conserved quantity. 
For example, to the long-range OPE part of the NN interaction
correspond two-body OPE e.m.~currents, which involve a photon interacting
with virtual pions being exchanged among the nucleons in a pair. Currents
relying on meson-exchange mechanisms are called
meson-exchange currents (MEC). 
The seminal derivations of MEC corrections date back to the late '40s, and were
carried out by Villars in Ref.~\cite{Villars47} and Miyazawa in Ref.~\cite{Miyazawa51} for 
nuclear static magnetic moments. The 1972 work by Riska and Brown~\cite{Riska72}
provided the first strong evidence for the need to incorporate two-body OPE e.m.~currents,
in addition to the IA terms. That study was focused
on evaluating the cross section of thermal neutron radiative capture on proton, and MEC contributions
were found to provide the missing $\sim 10\%$ correction required to reach agreement with the experimental datum. 
MEC have been been widely studied  to high levels of sophistication and accuracy (see, for example,
Refs.~\cite{Chemtob71,Fabian79,rho79,Towner84,Buchmann85,Riska84,Riska89} for early developments on the topic). 
In their most recent formulation, described, for example, in Refs.~\cite{Marcucci05,Marcucci08}
and references therein, MEC include, in addition to the standard convection
and spin-magnetization single-nucleon operators, two- and three-body components
constructed from NN and 3N nuclear potentials so as to satisfy the
continuity equation. This ensures that e.m.~currents and nuclear interactions
are consistent in describing the short and intermediate range dynamics.
The continuity equation does not uniquely constrain components
of e.m.~currents that are orthogonal to the momentum carried by the external photon
field. This introduces a degree of model dependence in that
transverse MEC are not uniquely defined.

The theoretical method described above, which is here referred to as the
`conventional' approach\footnotemark[1],\footnotetext{The conventional approach
is in the literature often referred to as the `Standard Nuclear
Physics Approach' or SNPA.}
has been successfully applied to study a wide variety of nuclear e.m.~observables.
The  most comprehensive review on its application to light nuclei was released
in 1998 by Carlson and Schiavilla in Ref.~\cite{Carlson98}. Since then,
a number of additional calculations have appeared in the literature. Among these,
we highlight the very recent and computationally demanding {\it ab-initio} 
Green's function Monte Carlo calculations of the $^{12}$C elastic e.m.~form
factors and sum rules of longitudinal and transverse response functions,
that include MEC corrections and 3N forces, carried out by Lovato and
collaborators in Ref.~\cite{Lovato13}. In their study, the calculated MEC
contribution to the transverse sum rule is found to significantly increase
(by up to $\sim 50\%$) the IA results, corroborating the importance of many-body
effects in nuclear systems. 

The recent history of few-body nuclear physics has witnessed
the tremendous development of chiral effective field theories ($\chi$EFTs) that 
systematically describe the interactions of nucleons among themselves and
with external electroweak probes.  $\chi$EFTs present the advantage of providing a direct connection
between the theory of nuclei, expressed in terms of non-relativistic nucleons interacting
via many-body potentials, and quantum chromodynamics (QCD), describing
the dynamics of the underlying constituents of matter, {\it i.e.}, quarks and gluons.
The effective nuclear Lagrangians are expressed in terms of degrees of freedom
which are bound states of QCD, such as nucleons, pions, and $\Delta$-isobars,
and are constructed so as to preserve all the symmetries, in particular chiral symmetry,
exhibited by the underlying theory of QCD in the low-energy regime relevant to
nuclear physics. In this energy regime, QCD does not have a simple solution
because the strong coupling constant becomes too large and 
perturbative techniques cannot be applied to solve it.
However, chiral symmetry dictates that pions couple among themselves
and to other composite degrees of freedom by powers of momenta generically denoted by $Q$.
The effective Lagrangians describing these interactions can be expanded
in powers of $Q/\Lambda_\chi$, where $\Lambda_\chi\sim1$ GeV
represents the chiral-symmetry breaking scale and characterizes
the convergence of the expansion.
Therefore, $\chi$EFTs provide an expansion of the Lagrangian in powers of a small
momentum as opposed to an expansion in the strong coupling constant,
restoring, in practice, the possibility of applying perturbative techniques also in the
low-energy regime of interest. The coefficients of the chiral expansion are called
Low Energy Constants (LECs). They are unknown and need to be fixed by comparison
with the experimental data. The transition amplitudes obtained from the effective
Lagrangians are also expanded in terms of $(Q/\Lambda_\chi)^\nu$. This, in principle,
allows to evaluate nuclear observables to any degree $\nu$ of desired accuracy, with an
associated theoretical error roughly given by $(Q/\Lambda_\chi)^{(\nu+1)}$. 

Since the pioneering work of Weinberg~\cite{Weinberg90,Weinberg91,Weinberg92}
released in the early nineties, this calculational scheme has been widely utilized in nuclear physics and
nuclear $\chi$EFT has developed into an intense and prolific field of research.
Here, we focus on calculations of nuclear e.m.~observables carried out within $\chi$EFT
formulations in which pions and nucleons are retained as relevant degrees of freedom,
and limit the discussion to reactions occurring within the low-energy regime of
applicability of the theory. Nuclear many-body operators constructed from pion and nucleon
 interactions, involve multiple pion exchange contributions as well as
contact-like interaction terms. Heavier degrees of freedom, such as nucleons'
excited states and/or heavier mesons, are `integrated out' and their interactions
are implicitly accounted for through the LECs of the theory. 
Nuclear two-- and three--body interactions were first investigated in the late `90s
by Ord\`o\~nez, Ray, and van Kolck within the standard
time ordered perturbation theory framework~\cite{vanKolck94,Ordonez92,Ordonez96}.
Currently, chiral NN (3N) potentials commonly used in {\it ab-initio} calculations include
up to next-to-next-to-next-to leading order or N3LO  (next-to-next-to leading order or N2LO) 
corrections
in the chiral expansion~\cite{Epelbaum09,Epelbaum12,Machleidt11}.
Work is in progress to explicitly incorporate the $\Delta$-isobar
among the relevant degrees of freedom of nuclear $\chi$EFTs to develop both nuclear potentials
(see, {\it e.g.}, Refs.~\cite{Krebs11,Krebs13,Piarulli14}) and e.m.~currents (see, {\it e.g.},
Ref.~\cite{Pastore08} for preliminary work on this topic).
For comprehensive reviews on EFTs applied to few-body nuclear systems we remind, {\it e.g.},
to Refs.~\cite{vanKolck93,Bernard95,Beane00,Bedaque02,Scherer05,Epelbaum09,Platter09,Machleidt11,Epelbaum12}.

Electroweak currents have also been described in  $\chi$EFT
formulations with pions and nucleons. Interactions of nuclei with external electroweak probes
have been first studied in covariant perturbation theory by Park, Min, and Rho in Refs.~\cite{Park93,Park96},
where two-body electroweak current operators have been constructed up to N3LO 
accuracy, that is up one--loop corrections.
More recently, $\chi$EFT e.m.~currents up to N3LO have been derived within two different implementations of
time ordered perturbation theory: one is by the JLab-Pisa group (see Refs.~\cite{Pastore08,Pastore09,Pastore11,Piarulli12})
and the other one is by the Bochum-Bonn group (see Refs.~\cite{Kolling09,Kolling11}).
Two-body $\chi$EFT e.m.~currents have been recently used in a number of {\it ab-initio} calculations
of e.m.~observables in $A=2$--$9$ nuclei. These calculations constitute one of the topics
discussed in the present review. 

With this review we hope to provide a general presentation of {\it ab-initio}
calculations of e.m.~observables of nuclear systems with mass number up to $A=16$.
Because the scope of this review is rather wide, topics
are not covered exhaustively. We focus on investigations developed in
the years after the release of the 1998 review by Carlson and Schiavilla~\cite{Carlson98}.
We highlight results (where available) that are comprehensive of many-body components in the 
e.m.~current operators, and results (where available) that use chiral potentials and/or chiral
e.m.~currents. The emphasis is on comparing theoretical calculations with experimental data. 
We hope that this review will prove to be a useful and compact report
on the present status of theoretical {\it ab-initio} studies on nuclear e.m.~reactions
at low-energies. 

This review is structured as follows: in Section~\ref{sec:Hamiltonians-cnt} we
briefly present nuclear potentials and e.m.~currents commonly utilized in 
{\it ab-initio} calculations; Sections~\ref{sec:electron} and~\ref{sec:photon}
are devoted to electron scattering reactions and e.m.~reactions involving real photons,
respectively; we summarize and present an outlook in Section~\ref{sec:conclusions}. 
In the very last section of this review a list of acronyms and abbreviations is provided. 
%%%%

\section{Nuclear Hamiltonians and electromagnetic currents}
\label{sec:Hamiltonians-cnt}

This section is devoted to briefly describe the theoretical framework of {\it ab-initio}
calculations, and we divide it into three subsections.
The first one discusses the conventional approach developed to construct
nuclear potentials and consistent e.m.~currents. The second one presents
nuclear potentials and e.m.~currents from a $\chi$EFT perspective, in which nuclear
dynamics are given in terms of pion and nucleon degrees of freedom. 
In the last subsection, we define conventions and notations utilized throughout
the course of this review. 

\subsection{The conventional approach}
\label{subsec:SNPA}

Within the {\it ab-initio} framework, nuclei are conceived as a collection
of non-relativistic nucleons interacting via many-body potentials. Thus, 
nuclear wave functions are solution of the Schr\"odinger equation
\begin{equation}
\label{eq:Lipp}
 H | \Psi_n \rangle = E_n | \Psi_n \rangle \ ,
\end{equation}
with a nuclear Hamiltonian $H$ given by
\begin{equation}
\label{eq:hamiltonian}
 H = \sum_i t_i + \sum_{i<j} v_{ij} + \sum_{i<j<k} V_{ijk} + \dots \ .
\end{equation}
Here, $t_i$ is the non-relativistic single-nucleon kinetic energy, while 
$v_{ij}$ and $V_{ijk}$ are NN and 3N potentials, respectively.
Higher order terms in the many--body expansion are represented by
the dots. 

Many computational techniques (for a review see, {\it e.g.}, Ref.~\cite{orlandini2012}) have been developed to 
obtain accurate solutions of Eq.~(\ref{eq:Lipp}).
Methods commonly utilized in $A\leq4$ systems
rely on the Faddeev decomposition of the Schr\"odinger equation. Among these are,
for example, the Faddeev-Yakubovsky scheme~\cite{Yakubovsky66}
and the  hyperspherical harmonics (HH) method~\cite{Kievsky92,Kievsky94,Kievsky97,Viviani05PRC, Viviani05, Kievsky08}. 
A method that can also tackle nuclei with mass number $A>4$ is
the Lorentz integral transform (LIT)~\cite{efros1994, Efl07}  used in combination with
the effective interaction hyperspherical harmonics (EIHH) expansion~\cite{EIHH,barnea2001}.
Other powerful and accurate {\it ab-initio} methods employed to solve
the few-- and many--body nuclear problem include,
for example, Quantum Monte Carlo (QMC) methods~\cite{Carlson87,Wiringa91,pudliner1997,Wiringa00,pieper2001,Pervin07},
such as the variational (VMC) and the Green's function Monte Carlo (GFMC) approaches
(currently available for $A\leq12$ systems), no-core shell model (NCSM)
methods~\cite{navratil2009,Barrett13},
coupled-cluster (CC) methods~\cite{dean2004, hagen2010b, CC_review}, and the 
Fermionic Molecular Dynamics (FMD) approach~\cite{Neff2008}.

Calculations of the energy spectrum of light nuclei show that, while 3N forces are essential to reach
agreement with the experimental data, their net contribution is much smaller than that associated with
NN forces. For example, in GFMC calculations of $A=3-8$ nuclei ground-state energies~\cite{pieper2001},
the $V_{ijk}$ contribution is found to be up to $\sim 8\%$ of that due
to $v_{ij}$. 
The convergence pattern exhibited by the NN and 3N nuclear operators emerges naturally
in the chiral expansion of nuclear forces (see next subsection), and 4N forces
are predicted to be suppressed with respect to leading 3N forces. 4N forces are 
neglected in all the results shown  here. 
However, one has to keep in mind that, due to a significant cancellation between the kinetic and
the $v_{ij}$ terms of the nuclear Hamiltonian, 3N forces can provide up to $\sim 30\%$ of
the total calculated energy~\cite{pieper2001}.

The calculations we present use several modern  NN and 3N potentials. In particular, at large inter-nucleon
distances, they all assume that the NN interaction results from the exchange of one pion among the nucleons, as
it is schematically represented by the  diagram illustrated in panel $(b)$ of Figure~\ref{fig:chiNN}
(appearing in the next subsection).
In the static limit where $m\rightarrow \infty$
(a standard approximation utilized in nuclear physics consisting of 
neglecting nucleonic kinetic energies),
evaluation of the transition amplitude associated with this diagram
leads to the expression of the standard OPE potential, which in momentum space reads
\begin{equation}
\label{eq:v_ope}
v_{\pi}({\bf k})=-\frac{g_A^2}{F_{\pi}^2}\,{\bm \tau}_i\cdot{\bm \tau}_j\,
\frac{{\bm \sigma}_i\cdot{\bf k}\,{\bm\sigma}_j\cdot{\bf k}}{\omega_k^2}\  .
\end{equation}
In the equation above, the potential is obtained in the center-of-mass frame
where the nucleons' initial and final relative momenta are ${\bf p}$ and ${\bf p}^{\prime}$,
respectively. We also define ${\bf k} = {\bf p}^{\prime} -{\bf p}$, then
the energy of the exchanged pion is $\omega_k= \sqrt{k^2+m_{\pi}^2}$,
while ${\sigma}_i$ and ${\tau}_i$ are the nucleonic spin and isospin Pauli matrices, respectively. 
The $\pi NN$ coupling is given in terms of the 
nucleon axial coupling constant $g_A \simeq 1.27 $ and the pion decay amplitude
 $F_\pi \simeq 186$ MeV~\footnotemark[1]\footnotetext{Note that some authors use 
the convention $F_{\pi}/2$ to define the pion decay amplitude.}. 

Intermediate-- and short--range components of the NN interaction present 
a rich structure, which is expressed in terms of nucleonic momenta, spin and isospin operators. 
Realistic potentials involve a number of parameters that
are constrained to fit $pp$ and $np$ scattering data up to energies of $\simeq 350$
MeV in the laboratory frame, along with the binding energy of the deuteron.
Common conventional NN potentials describe intermediate-- and short--range parts
in terms of one-boson exchange contributions, as done in the
CD-Bonn potential~\cite{CDBONN}, or parameterizations in terms of operator structures with strengths
specified by special functions as done, for example, in the series of Argonne NN potentials whose latest implementation
is the Argonne-$v_{18}$ (AV18) version developed in Ref.~\cite{AV18}.  

Unlike NN potentials, 3N forces are not fixed to 3N scattering data yet.
Static properties of few-nucleon systems, such as binding energies,
and nuclear-matter saturation properties are instead exploited to constrain them. Then,
the values of the fitted parameters entering 3N forces necessarily depend on the particular NN
potential utilized in the calculations. For example, 3N forces constructed in combination with
the AV18 NN interaction belong to the Urbana series~\cite{Pudliner95}, whose latest version
is called Urbana IX (UIX), and to the more recently developed Illinois series~\cite{IL,IL7},
whose latest implementation is called Illinois-7 (IL7). Both these 3N interactions
include the Fujita-Miyazawa term~\cite{Fujita57a,Fujita57b}---a two-pion exchange
contribution involving the excitation of a virtual $\Delta$-isobar---and a short-ranged repulsive
phenomenological term~\cite{Pudliner95}. The Illinois interaction adds to the Urbana one
the contributions due to an S-wave two-pion exchange term plus so called ring
diagrams which involve the exchanges of three pions combined
with excitations of one virtual $\Delta$-isobar~\cite{IL,IL7}.
When used with the AV18 NN and IL7 3N potentials,
the final GFMC results reproduce the experimental ground- and excited-state energies
for $A\le12$ nuclei very well~\cite{pieper2001,Pieper02,Pieper04,Pieper04a,IL7}.

Nuclear e.m.~charge ($\rho$) and current (${\bf j}$) operators---that is the time and vector components
of the four-vector current $j^{\mu}=(\rho, {\bf j})$---are also expressed as
an expansion in many-body operators that act on nucleonic degrees of freedom
\begin{eqnarray}
\label{eq:fourvec}
\rho({\bf q})    &=& \sum_i {\rho}_i({\bf q}) + \sum_{i<j} {\rho}_{ij}({\bf q}) + \dots\ , \\
\nonumber
{\bf j}({\bf q}) &=&  \sum_i {\bf j}_i({\bf q}) + \sum_{i<j} {\bf j}_{ij}({\bf q}) +\dots \ .
\end{eqnarray}
Here,  ${\bf q}$ is the momentum associated with the external photon field.
Calculations in IA performed using one-body operators only are based on 
the simplified picture in which nuclear
properties can be expressed in terms of those associated with free nucleons, {\it i.e.},
the probing photon interacts with individual nucleons. This kind of
contributions is schematically represented by disconnected diagrams such as the
one illustrated in panel $(a)$ of Figure~\ref{fig:chi_cnt} (appearing in the next subsection). 

One-body e.m.~current operators are obtained from the non-relativistic expansion of
the covariant single-nucleon current operator~\cite{Carlson98}.
At leading order in this $Q/m$ expansion (where $Q$ denotes a generic nucleon momentum),
the charge operator, here given in momentum-space, reads
\begin{equation}
\label{eq:charge_IA0}
\rho_i^{\rm NR} ({\bf k}_i) = e\, e_{N,i}   \ ,
\end{equation}
where ${\bf k}_i={\bf p}_i^\prime-{\bf p}_i$ is momentum transferred to nucleon
$i$ (${\bf p}_i$  and ${\bf p}_i^\prime$ are the initial and final momenta of nucleon $i$),
$e>0$ is the electric charge, and  $e_{N,i}=( 1+\tau_{i,z} )/2$ is the proton projection operator.
To simplify the notation in the equation above and in the following ones we drop 
 terms proportional to the $\delta$-function $\delta({\bf k}_i-{\bf q})$, enforcing momentum
conservation. Consequently, the ${\bf k}_i$-dependence on the left hand side of Eq.~(\ref{eq:charge_IA0})
becomes equivalent to the ${\bf q}$-dependence introduced in Eq.~(\ref{eq:fourvec}).

Implicit in the expression above is the assumption that nucleons are point-like objects,
an approximation which holds in the (long-wave) limit of ${\bf q}\rightarrow 0$. However,
nucleons' charge and magnetization internal distributions cannot be neglected as the spatial
resolution of the probe increases. In order to account for these nucleonic structure effects,
form factors that depend on the four-momentum transferred
%\footnotemark[1]\footnotetext{We are approximating here the nucleon form factors 
%with on-shell form factors.} 
$Q^2_\mu = -q^\mu q_\mu = -\omega^2 + q^2$
are folded in the  expressions of the e.m.~currents for point-like nucleons.
In particular, the charge operator of Eq.~(\ref{eq:charge_IA0}) becomes
\begin{equation}
 \label{eq:charge_IA}
\rho_i^{\rm NR}({\bf k}_i) = e\, e_{N,i}(Q^2_\mu)  \ ,
\end{equation}
where now
 \begin{eqnarray}
e_{N,i}(Q^2_\mu) &=& \frac{G_E^S(Q^2_\mu)+G_E^V(Q^2_\mu)\, \tau_{i,z}}{2}\ , \nonumber \\ 
\label{eq:e_ff}
\end{eqnarray}
and $G^{S/V}_E$ denote the isoscalar/isovector combinations of the proton and neutron electric
form factors, normalized as $G^S_E(0)=G^V_E(0)=1$~\cite{Sachs62}.

The non-relativistic one-body e.m.~current operator consists of two terms
namely the convection current, generated by the motion of charged nucleons, and the
spin-magnetization current associated with the nucleonic spins, and in momentum representation it reads
\begin{eqnarray}
\label{eq:cnt_IA}
{\bf j}_i^{\rm NR} ({\bf k}_i,{\bf K}_i)&=&\frac{e}{2\, m}
\left[ \,2\, e_{N,i}(Q^2_\mu) \, {\bf K}_i
+i\,\mu_{N,i}(Q^2_\mu)\, {\bm \sigma}_i\times {\bf q }\,\right]  \ ,
 \end{eqnarray}
where ${\bf K}_i=\left({\bf p}_i^\prime+{\bf p}_i\right)/2 $, and 
\begin{eqnarray}
 \mu_{N,i}(Q^2_\mu) &=&\frac{G_M^S(Q^2_\mu)+G_M^V(Q^2_\mu)\, \tau_{i,z}}{2} \ ,
\label{eq:m_ff}
\end{eqnarray}
with the isoscalar/isovector combinations of the proton and neutron magnetic
form factors $G^{S/V}_M$~\cite{Sachs62} normalized as $G^S_M(0)=0.880
\, \mu_N$, and $G^V_M(0)=4.706\, \mu_N$ in units of the nuclear magneton $\mu_N$.
Note that by Fourier transforming  Eqs.~(\ref{eq:charge_IA}) and~(\ref{eq:cnt_IA})
to obtain a representation in terms of the nucleon coordinates ${\bf r}_i$, the familiar form
of the charge and current operators (as given, {\it e.g.}, in the Ref.~\cite{Carlson98}) is restored.

Relativistic effects are accounted for by retaining higher order terms in the
$Q/m$ expansion of the covariant single-nucleon e.m.~currents~\cite{Carlson98}. 
Inclusion of the first non-vanishing relativistic correction (RC) leads to IA
operators which are suppressed by a $(Q/m)^2$ factor with respect to the corresponding
non-relativistic ones, and their formal expressions can be found, {\it e.g.},
in Ref.~\cite{Piarulli12}.  

The IA description of nuclei is improved  by accounting for the effects
of NN and 3N interactions onto the e.m.~currents associated with nucleon
pairs and triples, respectively. MEC follow naturally
once meson-exchange mechanisms are invoked to describe the interactions among
nucleons. For example, at large inter-particle distances, where the NN interaction
is mediated by the exchange of one pion, two-body e.m.~currents of one-pion range emerge.
They result from photons hooking up with exchanged pions, as  shown
by the seagull and pion-in-flight diagrams illustrated in panels $(b)$ and $(c)$,
respectively, of Figure~\ref{fig:chi_cnt} (see next subsection). Direct evaluation
of the transition amplitudes associated with these diagrams leads to the OPE current
commonly used in the literature~(see, {\it e.g.}, Ref.~\cite{ericson88}),
which in momentum representation reads
\begin{eqnarray}
 {\bf j}_\pi&=& -i\, e\frac{g^2_A}{F^2_\pi}
 ({\bm \tau}_i \times {\bm \tau}_j)_z 
 \left( {\bm \sigma}_i -{\bf k}_i\,\frac{{\bm \sigma}_i\cdot {\bf k}_i} {\omega_{k_i}^2}  \right)
 \frac{{\bm \sigma}_j\cdot {\bf k}_j}{\omega^2_{k_j}} + i \rightleftharpoons j \ ,
 \label{eq:cnt_OPE} 
 \end{eqnarray}
where $\omega_{k_i}^2=k_i^2+m_\pi^2$.

The explicit connection between many-body potentials and
many-body current operators is provided by the continuity equation
imposed by the gauge invariance of the theory:
\begin{equation}
{\bf q}\cdot {\bf j}=\left[H ,\, \rho \,\right ] \ ,
\label{eq:continuity}
\end{equation}
where $H$ is the nuclear Hamiltonian, whose many-body operatorial decomposition is given in
Eq.~(\ref{eq:hamiltonian}), and $[\dots,\dots]$ denotes a commutator. Neglecting relativistic
effects, the equation above can be re-casted in a set of equations of the form
\begin{eqnarray}
\label{eq:cont_1}
 {\bf q}\cdot {\bf j}_i   &=&\left[t_i ,\, \rho^{\rm NR}_{i} \,\right ] =  
                             \left[\frac{{\bf p}_i^2}{2\,m} ,\, \rho^{\rm NR}_{i} \,\right ]\ , \\
 {\bf q}\cdot {\bf j}_{ij}&=&\left[v_{ij},\, \rho^{\rm NR}_{i} + \rho^{\rm NR}_{j}\,\right ] \ , \label{eq:cont_2}
\end{eqnarray}
where $\rho^{\rm NR}_{i}$ is the non-relativistic IA charge operator defined in
Eqs.~(\ref{eq:charge_IA0})-(\ref{eq:charge_IA}). Similar equations hold in the
presence of 3N forces and beyond. In particular, Eq.~(\ref{eq:cont_1}) is satisfied by
the non-relativistic IA current operator ${\bf j}^{\rm NR}_{i}$
given in Eq.~(\ref{eq:cnt_IA}), while the pion-in-flight and seagull e.m.~currents of
Eq.~(\ref{eq:cnt_OPE}) satisfy Eq.~(\ref{eq:cont_2}) with the
OPE potential given in Eq.~(\ref{eq:v_ope})~(see, {\it e.g.}, Ref.~\cite{ericson88}).

The conventional approach exploits the continuity equation and the meson-exchange theoretical
insight to consistently construct e.m.~two-body currents from the given $v_{ij}$ potential utilized
in the {\it ab-initio} calculations (see Refs.~\cite{Carlson98,Marcucci05,Marcucci08} and
references therein).
These currents are, in the literature, referred  to as ``model independent'' in that they are
completely constrained by the NN interaction via gauge invariance.  The dominant terms are
isovector in character, and they satisfy, by construction, the continuity equation with the
static part of the NN potential. The latter is assumed to be due to the exchange of `effective'
pseudoscalar (PS or ``$\pi$-like'') and vector (V or ``$\rho$-like'') mesons. 
MEC currents are then due to the exchange of these `effective' PS- and V-mesons,
and are constructed utilizing PS- and V-meson propagators projected out of the static
part of $v_{ij}$. This ensures that the short- and intermediate-range behavior of
the MEC is consistent with that of the NN potential.
Additional ``model independent'' currents of short-range follow by minimal substitution in the
momentum-dependent part of the potential~\cite{Marcucci05}. They have both isoscalar and
isovector terms, and lead to contributions which are typically much smaller
(in magnitude) than those generated by the PS and V currents.  At large
inter-nucleon separations, where the NN potential is driven by the OPE
mechanism, the ``model independent'' current coincides with the standard seagull and pion-in-flight
OPE currents given in Eq.~(\ref{eq:cnt_OPE}) and diagrammatically illustrated in 
panels $(b)$ and $(c)$, respectively, of Figure~\ref{fig:chi_cnt} (appearing in the next subsection).
Inclusion of two-body terms in the e.m.~current operator, particularly the $\pi$-like exchange
contribution, is crucial to improve the agreement with the experimental data.

Models of 3N currents that satisfy the continuity equation with conventional 3N potentials
have been most recently developed in Ref.~\cite{Marcucci05}. In general, 3N MEC currents, are
found to provide small corrections to photo- and electronuclear
process~\cite{Viviani96,Schiavilla89,Carlson90,Marcucci98,Marcucci05}. 

\begin{figure}[htb]
\centering
\includegraphics[width=8cm,clip]{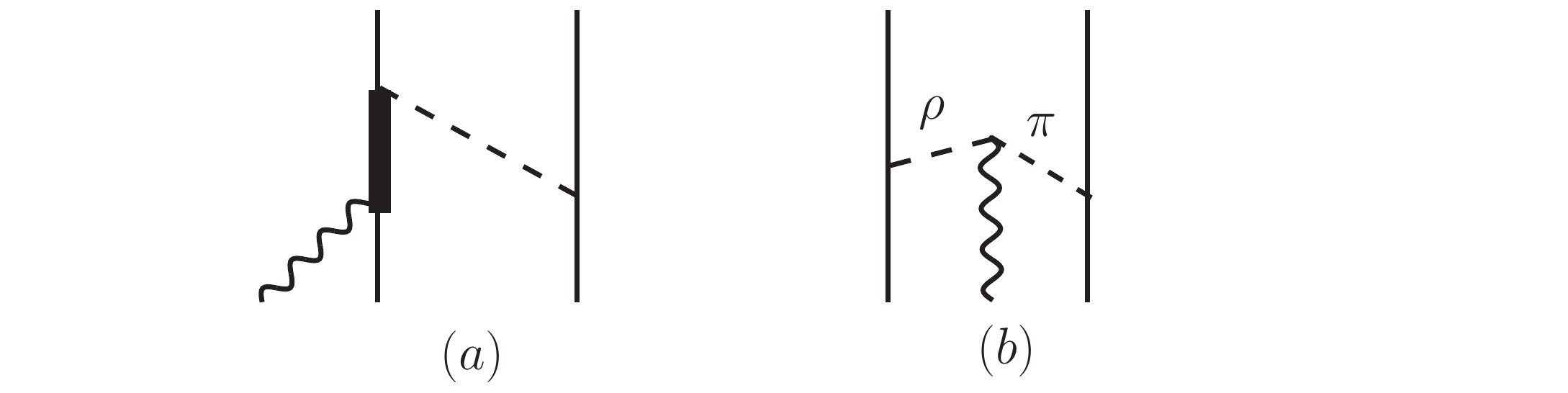}
\caption{Diagrams illustrating conventional ``model dependent'' two-body currents. 
See text for explanation. Nucleons, $\Delta$'s, mesons, and photons are denoted by solid,
thick, dashed, and wavy lines, respectively.}
\label{fig:delta_cnt}
\end{figure}

The continuity equation does not  constrain components of the e.m.~currents
which are orthogonal to the momentum ${\bf q}$ carried by the photon field.
Currents of this nature, that are purely transverse currents, are, in the literature,
referred to as ``model dependent''  to distinguish them from the ``model independent'' ones
which are completely determined by the nuclear potentials. Possible photon
interactions with nucleon pairs involve, for example, its coupling to mesons 
being exchanged among the nucleons, or virtual nucleon excitations, {\it e.g.},
$\Delta$ resonances.  The dominant two-body ``model dependent'' term is found
to be associated with excitations of intermediate $\Delta$
isobars~\cite{Fabian75, Weber78,Carlson98,Marcucci08}.
A contribution belonging to this class is schematically represented by the
diagram shown in panel $(a)$ of Figure~\ref{fig:delta_cnt}. 
Additional (and numerically small) ``model dependent'' currents arise from the isoscalar
$\rho\pi\gamma$---schematically illustrated in panel $(b)$ of Figure~\ref{fig:delta_cnt}---and isovector
$\omega\pi\gamma$ transition mechanisms~\cite{Carlson98,Marcucci08}.
Conventional e.m.~currents do not involve free-parameters in the sense that they
are determined by the many-body potentials entering the nuclear Hamiltonian.
However, they are not uniquely determined because the continuity equation does not constrain
their transverse component. Their range of applicability extends, in some
cases, beyond the pion-production threshold, where isobar currents become
more important~\cite{Carlson98, Schwamb2010}.

Conventional two-body e.m.~charge operators are entirely ``model dependent'' in that they cannot be
derived from the NN interaction~\cite{Carlson98}. Two-body contributions commonly utilized
in {\it ab-initio} studies include $\pi$-, $\omega$-, $\rho$-meson exchange charge operators,
as well as $\rho\pi\gamma$ and $\omega\pi\gamma$ transition charge operators~\cite{Carlson98}.
In order to reduce the model dependence due to a poor knowledge of meson-NN coupling
constants, the $\pi$- and $\rho$-meson propagators entering the associated meson-exchange amplitudes
can be replaced by the `effective' PS- and V-meson propagators projected out of the static NN potential,
much in the same fashion implemented to construct the ``model dependent'' two-body current operators.
The dominant contribution to the two-body charge operator is provided by the pion-exchange operator,
which is seen to be one order of magnitude larger than that of the remaining two-body
contributions~\cite{Carlson98}. Studies on charge form factors of light nuclei~\cite{Carlson98}
show that two-body effects in the charge operators become important above values of the momentum transfer $q\simeq 3$ fm$^{-1}$,
while below those the IA picture is sufficient to reach agreement with the experimental data. 

\subsection{The chiral effective filed theory approach}
\label{subsec:chi}

We now turn our attention to the $\chi$EFT formulation of nuclear forces and e.m.~currents.
Chiral Lagrangians are arranged in classes characterized by
the power $\nu$ of a small parameter $Q$ associated with the pion momentum, and so are the nuclear 
operators obtained from them. The scheme by which nuclear operators are organized according
to their scaling in $Q$ is called power counting. Here, we present the hierarchic arrangement
of nuclear many-body operators emerging from the application of the counting scheme introduced by
Weinberg in Refs.~\cite{Weinberg90,Weinberg91,Weinberg92}. 
\begin{figure}[bthp]
\centering{
\includegraphics[width=4in]{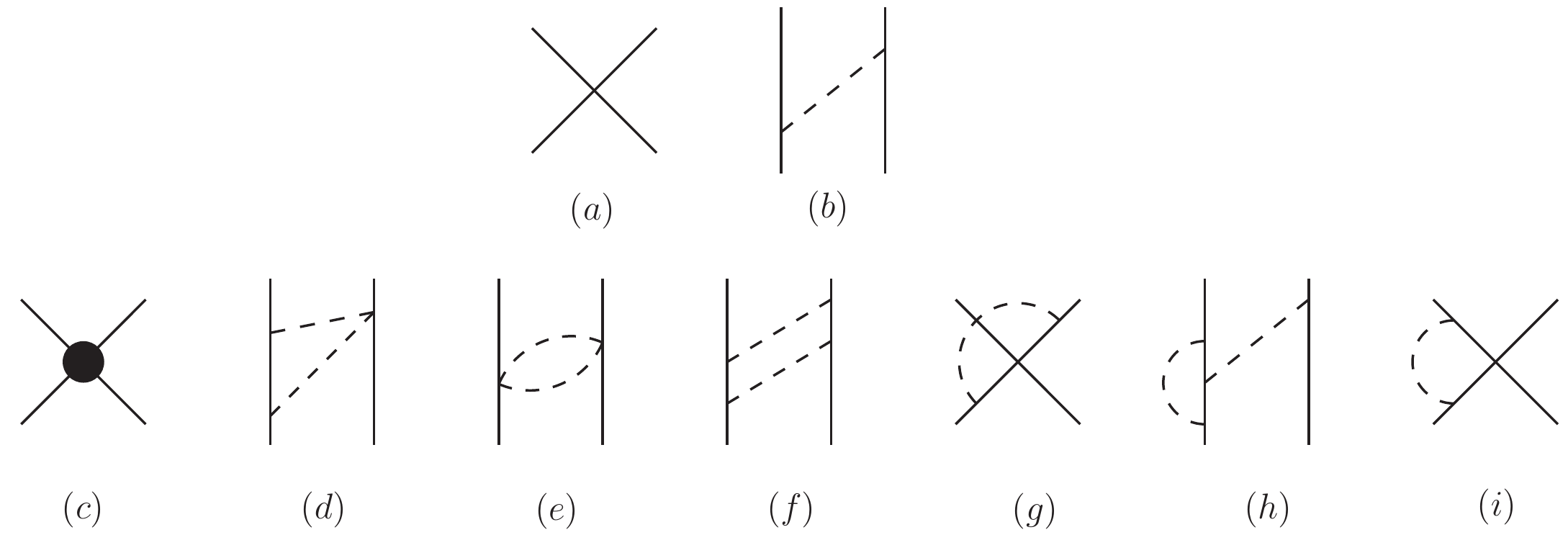}}
\caption{
Diagrams illustrating contributions to the chiral NN potential
entering at LO ($\nu=0$), panels $(a)$ and $(b)$, and NLO ($\nu=2$), panels $(c)$-$(i)$.
The solid circle in panel $(c)$ indicate contact interactions of order $Q^2$
that involve two nucleons' momenta (see text for explanation).
Notation is as in Figure~\ref{fig:delta_cnt}, but for dashed lines that
here represent pions.}
\label{fig:chiNN}
\end{figure}

\subsection*{Chiral many-body potentials}

Nuclear $\chi$EFT potentials are expressed in terms of multiple-pion exchange contributions,
which describe the long- and intermediate-range components of the interaction,
along with many-nucleons contact terms (CT) that encode short-range dynamics~\cite{Epelbaum09,Epelbaum12,Machleidt11}.  
The chiral nuclear force involves a number of unknown LECs which need to be determined.
LECs entering the NN chiral potential are fitted to the available NN scattering data,
as it is done for conventional potentials. The chiral NN potential at LO ($Q^{\nu=0}$) consists
of the  static OPE potential given in Eq.~(\ref{eq:v_ope}) plus two four-nucleon contact
interaction terms involving two unknown LECs, $C_S$ and $C_T$. The contact contribution
of order $\nu=0$ is represented by the diagram illustrated in panel $(a)$ of Figure~\ref{fig:chiNN},
and the complete chiral potential at LO (${\nu=0}$) reads
\begin{equation}
v^{\nu=0}= v_\pi + v_{\rm CT}=-\frac{g_A^2}{F_{\pi}^2}\,{\bm \tau}_i\cdot{\bm \tau}_j\,
\frac{{\bm \sigma}_i\cdot{\bf k}\,{\bm\sigma}_j\cdot{\bf k}}{\omega_k^2} + C_S+C_T\,{\bm \sigma}_i\cdot{\bm \sigma}_j \  .
\label{eq:ct0}
\end{equation}
The NLO\footnotemark[1]\footnotetext{Note that in Ref.~\cite{Pastore08} the NN potential of order
$Q^{\nu=2}$ is counted as being an N2LO instead of an NLO correction. To avoid confusion
induced by the use of different conventions, we always specify
the power $\nu$ of $Q$ when we use the `N$n$LO' notation.} contribution scales as $Q^{\nu=2}$
(there are no $\nu=1$ contributions to the NN interaction)  and, referring
to Figure~\ref{fig:chiNN}, it includes the one-loop diagrams illustrated in panels $(d)$--$(i)$,
plus seven additional contact terms, illustrated in panel $(c)$. The latter involve two nucleon
momenta and are written in terms of seven unknown LECs. At this order, there are also 3N
contributions of one-pion range which, however, are seen to vanish~(see, {\it e.g.},
Ref.~\cite{vanKolck94}). At NLO ($\nu=2$), all the spin-isospin operator structures necessary
to describe the NN interaction have been generated, however one needs to incorporate
higher order contributions in order to accurately fit NN data. The N2LO (${\nu=3}$) nuclear force
consists of additional one-loop terms contributing to the NN sector of the interaction,
along with the first non-vanishing contribution to the 3N interaction~\cite{Epelbaum09,Epelbaum12,Machleidt11}. We note that the
leading 3N forces (${\nu=3}$) are suppressed by two orders with respect to the
leading NN potential (${\nu=0}$). Currently, phase-shift
equivalent chiral NN potentials  include up to
N3LO ($\nu=4$) corrections consisting of two-loop contributions
and 15 additional contact terms associated with 15 unknown LECs to be fitted to
NN scattering data. Chiral NN potentials up to N3LO ($\nu=4$) have been been developed
by Epelbaum and collaborators in Refs.~\cite{Epelbaum98,Epelbaum00,Epelbaum05},
and by Entem and Machleidt in Refs.~\cite{Entem01,Entem03}. 

So far, we discussed NN potentials that are  constructed in the center-of-mass frame of the two nucleons.
However, in Ref.~\cite{Girlanda10a} it has been shown that, by requiring Poincar\'e covariance
of the theory, one is forced to introduce additional terms with fixed LECs,
in order to describe the two-nucleon potential in reference frames other than the center-of-mass frame.
These terms depend on the total momentum of a nucleon pair which can be set to be equal to zero in
the center-of-mass frame of a two-nucleon system, but it is in general different from zero in systems
with mass number $A>2$, and the effects of these corrections have not yet been quantitatively studied within
$\chi$EFT formulations~\cite{Girlanda10a}.

The 3N forces~\cite{Epelbaum09,Epelbaum12,Machleidt11} enter at N2LO ($\nu=3$) and involve two unknown LECs.
Strong observables that can be used to pin them down are, for example, the binding energies of $A=3$
and $4$ nuclei. An alternative strategy, implemented by Gazit and collaborators in Ref.~\cite{Ga_beta_triton},
is based on the observation that one of the strong LECs enters also the chiral contact
two-nucleon weak current at N2LO ($\nu=2$)~\cite{Park93}. Therefore, experimental data
of electroweak processes induced by this weak operator can also be used to fix the LEC.
In Ref.~\cite{Ga_beta_triton}, the 3N force at N2LO (${\nu=3}$) is constrained so as to reproduce
the binding energies of the $A=3$ nuclei and the empirical value of the Gamow-Teller matrix element in triton $\beta$-decay.
Of course, the values of the LECs entering the 3N potential are obtained in combination with the chiral
NN interaction utilized in the calculations. 3N forces at N3LO ($\nu=4$) are  presently being investigated
by the theoretical community~\cite{lenpic}. 
Chiral 3N forces, if included, are taken at N2LO in the results presented in this review. 
%
%In all the results shown in this review that are comprehensive 
%of 3N forces the latter are used at N2LO.
 
Once the chiral potentials are constructed, standard {\it ab-initio} techniques
may be used to solve the  many-body Schr\"odinger equation.
Calculations of the energy
spectrum of light nuclei that use NN interactions at N3LO and 3N forces at N2LO 
compare very well with the experimental data
(see, {\it e.g.}, Refs.~\cite{roth2011a,Barrett13}).
More recently, computational techniques developed in the field
of lattice QCD have been implemented to solve the many-body problem of nuclear
physics in combination with chiral potentials. The method is known as lattice EFT and currently 
it has been applied to structure studies of nuclei with mass number up
to $A=28$~\cite{Epelbaum09c,Epelbaum11_hoyle,Lahde13}, while selected 
e.m.~transitions have been investigated in $^{12}$C~\cite{Epelbaum12_hoyle_BE2} and $^{16}$O~\cite{Epelbaum13paa}.

Few comments are now in order. Nuclear potentials and current operators present ultraviolet divergences which 
need to be removed by a proper regularization procedure. There are two
kinds of regularizations which are implemented. The first one concerns
the regularization of the loop integrals entering loop contributions (for example,
those illustrated in diagrams $(d)$--$(i)$ of Figure~\ref{fig:chiNN}). This is
accomplished via one of the usual schemes commonly adopted in quantum filed theory,
{\it e.g.}, dimensional regularization, and it is followed by a renormalization procedure
in which divergences, isolated by the regularization scheme, are reabsorbed
by the LECs of the theory.  The second kind of regularization follows from the fact that
nuclear operators, obtained by evaluating transition amplitudes, are then used in the Schr\"odinger
equation  to construct nuclear wave functions and to evaluate matrix elements. 
Chiral operators have a power law behavior at large momenta that needs to be cut off to
avoid infinities. This is  accomplished by multiplying nuclear operators with a short-range momentum cutoff.
The latter is usually taken to be of the form $exp[- (Q/\Lambda)^{2\,n}]$,
where the choice of $n$ is contingent to the accuracy one aims to reach. That is,
for small values of $Q$, where the cutoff behaves as $\sim 1-(Q/\Lambda)^{2\,n}+\dots$,
spurious contributions $\propto (Q/\Lambda)^{2\,n}$ generated by the insertion
of the cutoff need to be much smaller than the order at which the calculation is performed.
In a theory that explicitly includes up to three-pion-exchange contributions, as it is done 
in the case of the chiral NN potential at N3LO ($\nu=4$), it is reasonable to set $\Lambda \sim 500$
MeV $\sim 3\,m_\pi$. This way, the cutoff eliminates short-range contributions of
four-pion range and beyond, that are not explicitly included in the theory,
but are rather subsumed in the LECs.  We remark that the discussion
on what is the proper renormalization scheme (and power counting) to be adopted is still
open~\cite{Epelbaum09,Epelbaum12,Machleidt11,Nogga05,PavonValderrama05,Epelbaum06,Phillips13,Epelbaum13}.
We will not address this issue further, and simply keep in mind that calculations with
chiral operators depend on an additional parameter, {\it i.e.}, the momentum cutoff $\Lambda$ defined above.  
For comprehensive reviews on $\chi$EFT nuclear forces we refer to the review articles
by Epelbaum {\it et al.}~and by Machleidt and Entem given in Refs.~\cite{Epelbaum09,Epelbaum12},
and  Ref.~\cite{Machleidt11}, respectively. 

\subsection*{Chiral many-body electromagnetic current operator}

One of the great advantages of the $\chi$EFT formulation is that e.m.~currents
are naturally constructed with nuclear forces in a consistent way. Gauge invariance
is one among the fundamental symmetries the theory is required to satisfy.
Hence, e.m.~fields are coupled to nuclear currents which satisfy the continuity
equation, order by order, with chiral potentials.  
Chiral two-body currents ${\bf j}$ were first investigated within
$\chi$EFT by Park, Min, and Rho in Ref.~\cite{Park96} and constructed  up to include one-loop corrections by using covariant perturbation theory.
In recent years, the two-body operators for both $\rho$ and ${\bf j}$
have been derived within two different implementation of time-ordered perturbation theory
up to include TPE corrections. The JLab-Pisa group~\cite{Pastore08,Pastore09,Pastore11,Piarulli12}
used standard time-ordered perturbation theory, while the Bochum-Bonn group~\cite{Kolling09,Kolling11}
used the method of the unitary transformation, which was also utilized to 
construct the N3LO ($\nu=4$) two-body potential developed in Refs.~\cite{Epelbaum98,Epelbaum00,Epelbaum05}. 
Differences between these e.m.~operators have been discussed at length in
Refs.~\cite{Pastore08,Pastore09,Pastore11,Piarulli12,Kolling11}. Here,
we will qualitatively describe the hierarchy of the e.m.~currents and
charge operators that emerges from the chiral expansion,
and refer to the aforementioned references for details and formal
expressions of the operators. We point out that a proper renormalization
of the e.m.~OPE operators has been carried out only within the unitary
transformation formalism (see Ref.~\cite{Kolling11}). 

\begin{figure}[htb]
\centering
\includegraphics[width=9cm,clip]{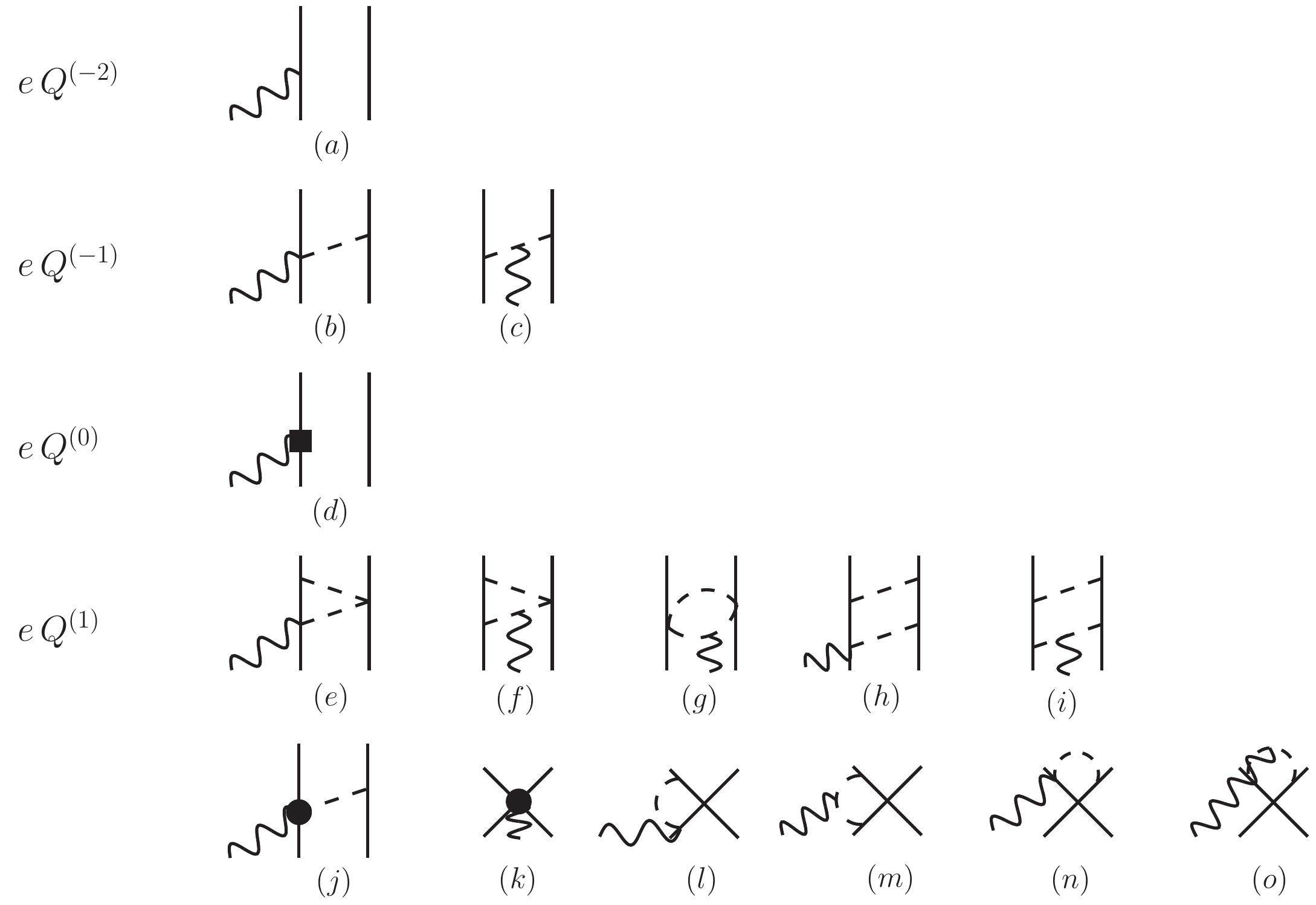}
\caption{Diagrams illustrating one- and two-body chiral e.m.~current operators entering at LO ($\nu=-2$)
[panel $(a)$], NLO ($\nu=-1$) [panels $(b)$ and $(c)$], N2LO ($\nu=0$) [panel $(d)$],
and N3LO ($\nu=1$) [panels $(e)$--$(o)$]. The LO operator corresponds to the non-relativistic
IA operator of Eq.~(\ref{eq:cnt_IA}). The NLO seagull and pion-in-flight contributions lead
to the current operator of Eq.~(\ref{eq:cnt_OPE}).  The square in panel $(d)$ represents 
the $(Q/m_N)^2$, or $(v/c)^2$, relativistic correction to the LO one-body current 
operator [or IA(RC)], whereas the solid circle in the tree-level
diagram illustrated in panel $(j)$ is associated with a $\gamma \pi N$ coupling of order $e\, Q^2$ (see text for explanation).
The solid circle in panel $(k)$ represent a 
%$\gamma$4N 
vertex of order  $e\, Q$.
Notation is as in Figure~\ref{fig:chiNN}. }
\label{fig:chi_cnt}
\end{figure}

We start off with the e.m.~current operator ${\bf j}$, whose contributions are diagrammatically listed
in Figure~\ref{fig:chi_cnt}. In Table~\ref{tb:scalingQ}, we also report the scaling in $e\,Q^\nu$
(where $e$ is the electric charge brought in by the e.m.~coupling) and the range of the operators
at each order. Note that the LO term here scales as $e\,Q^{\nu = -2}$,
which follows from direct application of the power counting~\cite{Park96} to the
disconnected diagram illustrated  in panel $(a)$ of Figure~\ref{fig:chi_cnt}. In particular,
the $\gamma$NN vertex scales as the single nucleon e.m.~current given in Eq.~(\ref{eq:cnt_IA}),
that is $e\,Q$. In addition, disconnected contributions, such as those illustrated in panels 
$(a)$ and $(d)$ of this figure, involve an extra $\delta$-function in the initial and final momenta
of the spectator nucleon---that is $\delta({\bf p}_j^\prime-{\bf p}_j)$, and they are thus enhanced
by a factor $Q^{-3}$ with respect to connected diagrams. Then, the power counting of diagram $(a)$
is $e\,Q\times Q^{-3}=e\,Q^{-2}$. 
Other conventions for the LO's scaling have also been utilized in the literature. Ultimately,
what matters is the suppression factor of a given operator $\propto e\,Q^{\nu}$ with respect to
the LO term $\propto  e\,Q^{\nu_{\rm LO}}$, that is $e\,Q^{(\nu-\nu_{\rm LO})}$, which is independent
of the choice made for the LO term scaling.  Translation from one notation to the other is easily achieved
keeping that in mind\footnotemark[1]\footnotetext{We point out that there are also slightly different power counting
schemes adopted in the literature. These differences will be highlighted where needed.}.

Referring to Figure~\ref{fig:chi_cnt}, the LO ($\nu=-2$)
operator is represented by the diagram illustrated in panel $(a)$ and corresponds to the standard 
non-relativistic IA current given in Eq.~(\ref{eq:cnt_IA}). 
A strict chiral expansion of the nuclear current operators does not guarantee the convergence
of the calculated observables to the experimental data. One has to account for nucleonic structure 
effects via suited form factors. The latter can also be derived within $\chi$EFT, however, as it has been
observed in Refs.~\cite{Walzl01,Phillips03,Phillips07} in the case of the deuteron charge form factor,
the description of nuclear observables is limited by the difficulty of $\chi$EFT in describing 
nucleonic form factors for values of momentum transfer larger than $q\sim 1.5$ fm$^{-1}$. 
Therefore,  it is customary also within $\chi$EFT formulations
to account for nucleonic structure effects via form factors taken from fits to elastic
electron scattering data on deuteron and proton.
As discussed above, relativistic corrections to the LO one-body current generate a one-body operator
which is suppressed by a $(Q/m)^2$ factor with respect to the LO one. This single-nucleon
term, also denoted with IA(RC), 
is represented by the disconnected diagram illustrated in panel $(d)$ and scales 
as $e\,Q^{\nu=0}$ (N2LO). Nuclear two-body effects appear at NLO ($\nu=-1$) with the OPE
currents represented by the diagrams illustrated in panels $(b)$ and $(c)$. These are the
well known seagull and pion-in-flight currents of Eq.~(\ref{eq:cnt_OPE}). At N3LO ($\nu = 1$),
there are currents of one- and two-pion range as well as contact currents. In particular, pure
TPE one-loop contributions are illustrated in panels $(e)$--$(i)$, while one-loop short-range currents are
represented in panels $(l)$--$(o)$. One-loop corrections at N3LO ($\nu = 1$)---that is those illustrated in 
panels $(e)$--$(i)$ and $(l)$--$(o)$---and OPE currents at NLO ($\nu=-1$) lead to purely isovector operators.

\begin{center}
\begin{table}[bth]
\centering{
\begin{tabular}{c|c|c|c|c|c}
\hline
\hline
 {\rm Operator}     &  {\rm LO}    & {\rm NLO}   & {\rm N2LO}  & {\rm N3LO} & {\rm N4LO}  \\
\hline
 ${\bf j}$          &  $\nu= -2$   & $\nu= -1$   &   $\nu= 0$  & $\nu= 1$                     &  \\
                    &  {\rm IA(NR)} & {\rm OPE}   & {\rm IA(RC)} & {\rm \,\,\,OPE(LECs)}    &  \\
                    &              &             &             & {\rm \,\,\,TPE{\color{white}(LECs)} } & \\
                    &              &             &             & {\rm CT(LECs)} & \\
\hline
$\rho$              &  $\nu= -3$   & $\nu= -2$   &   $\nu= -1$  & $\nu= 0$  & $\nu= 1$ \\
                    &  {\rm IA(NR)} & ---         & {\rm IA(RC)}  & {\rm OPE} &  {\rm TPE} \\
                    
\hline
\hline
\end{tabular}
\caption{Scaling in $e\,Q^\nu$ up to $\nu=1$ and ranges of the chiral e.m.~current
and charge operators. The ${\bf j}$ operator at N4LO $(\nu=2)$ are have not been derived yet.
The acronyms stand for OPE = one-pion-exchange,
TPE = two-pion-exchange, CT = contact term, IA = impulse approximation, NR = non-relativistic,
and RC = relativistic correction.}
\label{tb:scalingQ}}
\end{table}
\end{center}

We note that, so far, the e.m.~operators involve known LECs, namely the axial coupling constant
$g_A$ and the pion decay amplitude $F_\pi$. Unknown LECs enter the N3LO ($\nu = 1$) tree-level
and contact currents illustrated  in panels $(j)$ and $(k)$, respectively. We distinguish between
`minimal' and `non-minimal' contact currents. The former are linked to the contact potential at NLO
($\nu=2$)---schematically illustrated in panel $(c)$ of Figure~\ref{fig:chiNN}---via the continuity equation.
That is, the diagram %$\gamma 4N$ vertex
 of panel $(j)$ in Figure~\ref{fig:chi_cnt} is determined from the %$4N$ 
nucleon vertex in panel $(c)$ of Figure~\ref{fig:chiNN} via minimal substitution in the nucleon momentum,
${\bf p}\rightarrow{\bf p} -i\,e\,{\bf A}$, where ${\bf A}$ is the vector photon field.
Therefore `minimal' currents involve the same LECs that enter the contact NN potential at NLO ($\nu=2$),
and these can be constrained by fitting $np$ and $pp$ elastic scattering
data and the deuteron binding energy. Additional contact currents at N3LO ($\nu = 1$) are of
`non-minimal' nature and follow from the coupling of the e.m.~field tensor
$F_{\mu\nu} = (\partial_\mu A_\nu-\partial_\nu A_\mu)$. In particular, there are two `non-minimal'
contact currents, one is isoscalar and the other one isovector. These contact
currents involve unknown LECs, which need to be fixed so as to reproduce e.m.~observables.

The tree-level current at N3LO ($\nu = 1$), represented by the diagram illustrated
in panel $(j)$, results from a $\gamma\pi N$ coupling
of order $e\,Q^2$ (indicated by a solid circle) and involves three unknown LECs. Two of them multiply
isovector operators and the remaining one multiplies an isoscalar operator.
In the present $\chi$EFT formulation, with pions and nucleons
as relevant degrees of freedom, LECs subsume interactions involving heavy-mesons or 
nucleon resonances integrated out from the theory. In fact, the isovector part of the 
tree-level current at N3LO ($\nu = 1$) has the same structure as the current involving the $N \Delta$-excitation
that is illustrated in panel $(a)$ of Figure~\ref{fig:delta_cnt}. The latter 
is known to provide the major contribution to the ``model dependent'' conventional currents~\cite{Carlson98}. 
Similarly, the isoscalar component of the tree-level current at N3LO ($\nu = 1$)
simulates the $\rho\pi\gamma$  transition current~\cite{Carlson98},
illustrated in panel $(b)$ of Figure~\ref{fig:delta_cnt}. 
A strategy often implemented to reduce the number of unknown LECs is to `saturate' them
with the couplings entering the $\Delta$-resonance and/or the $\rho\pi\gamma$ 
transition currents illustrated in panels $(a)$ and $(b)$ of Figure~\ref{fig:delta_cnt},
respectively (see Ref.~\cite{Park96}). Once the five LECs listed so far, two associated with
contact currents and three belonging to the tree-level current at N3LO ($\nu=1$), are
determined (via saturation and/or via fits to e.m.~observables) the current operator can
then be used predictively. 
 
\begin{figure}[htb]
\centering
\includegraphics[width=10cm]{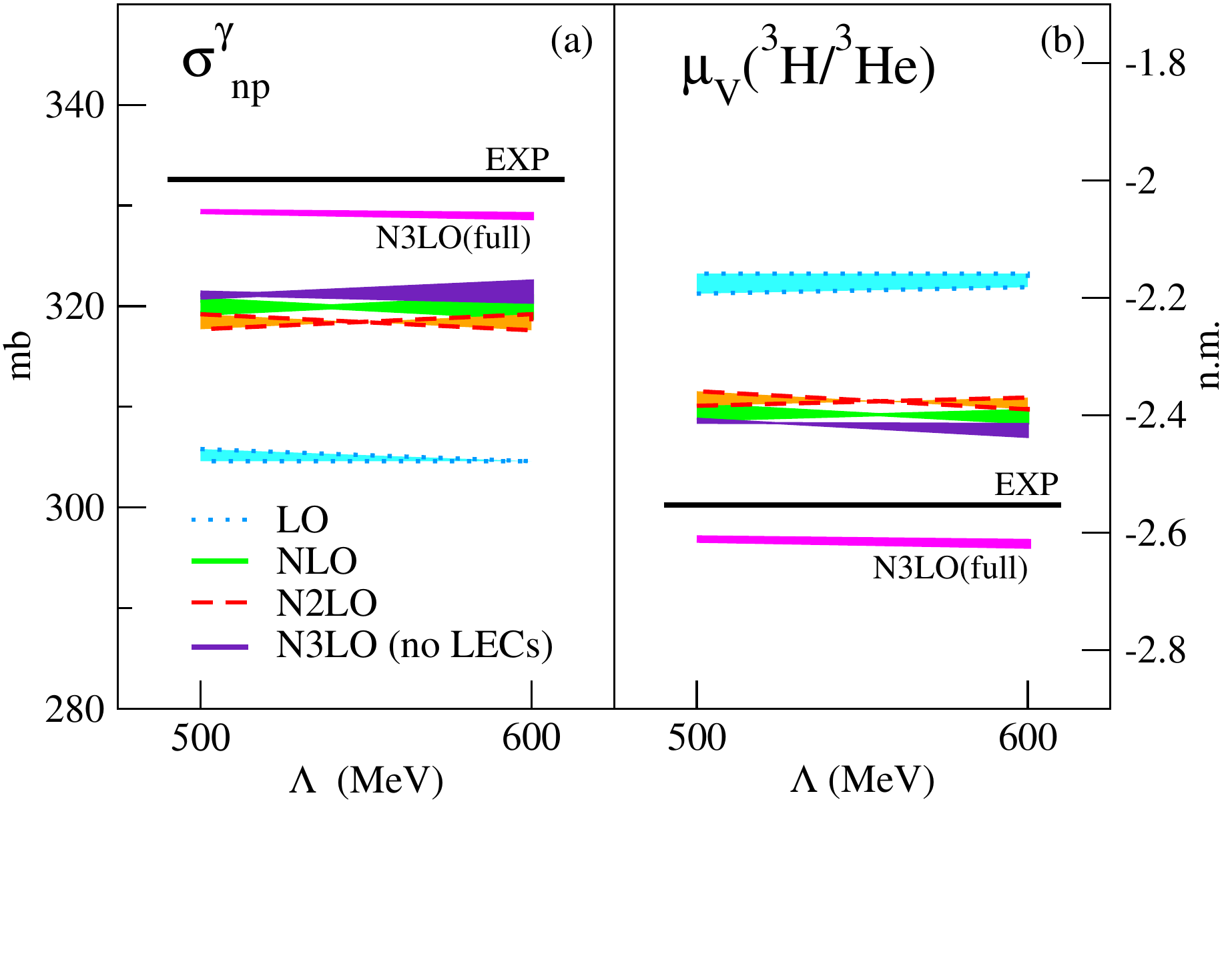}
\vspace*{-1.5cm}
\caption{ (Color online)
Cumulative contributions to the $n+p\rightarrow d+\gamma$ cross section
at thermal neutron energies $\sigma^\gamma_{\rm np}$ (left panel) and trinucleon isovector magnetic $\mu^{\rm V}$($^3$H/$^3$He)
(right panel).  The bands represent the sensitivity of the results to
the two nuclear Hamiltonians utilized, namely the NN(N3LO)+3N(N2LO) and AV18+UIX. 
The experimental values for $\sigma^\gamma_{\rm np}$  and $\mu^{\rm V}$($^3$H/$^3$He)
are 332.6 $\pm$ 0.7  and −2.553 $\mu_N$, respectively.
Calculations are from Ref.~\cite{Piarulli12}.
}
\label{fig:npdg_trimm}
\end{figure}

As an example, we briefly discuss  calculations~\cite{Piarulli12}
of the $n+p\rightarrow d+\gamma$ cross section at thermal neutron energy $\sigma^\gamma_{\rm np}$
(left panel of Figure~\ref{fig:npdg_trimm}) and the isovector combination of the $A=3$ nuclei's
magnetic moments $\mu^{\rm V}$($^3$H/$^3$He) (right panel of Figure~\ref{fig:npdg_trimm}). These
observables are induced by the magnetic dipole operator associated with the
e.m.~nuclear currents. In these calculations, `minimal' LECs associated with the contact
currents at N3LO---see panel $(k)$ of Figure~\ref{fig:chi_cnt}---have been fixed to NN scattering
data~\cite{Entem03,Machleidt11}. The saturation of the $\Delta$-resonance has been exploited to
fix the two LECs entering the isovector component of the tree-level current at N3LO ($\nu = 1$)---see panel $(j)$ of
Figure~\ref{fig:chi_cnt}, while the two isoscalar LECs, entering the tree-level and contact currents
at N3LO ($\nu = 1$)---see panels $(j)$ and $(k)$ of Figure~\ref{fig:chi_cnt},
respectively---are determined so as to reproduce the deuteron  as well as the isoscalar
combination of the trinucleon magnetic moments. Then, only one unknown isovector LEC is left.
For the calculation in the left (right) panel this LEC is obtained by reproducing the experimental
value of the isovector combination of the trinucleon magnetic moments ($n+p\rightarrow d+\gamma$
cross section at thermal neutron energy) for each value of the momentum cutoff $\Lambda=500,600$ MeV.
The results are obtained including cumulatively contributions at LO ($\nu=-2$), NLO ($\nu=-1$),
N2LO ($\nu=0$), and at N3LO ($\nu=1$). The cumulative contribution at N3LO ($\nu=1$) is separated
into the results labeled `N3LO(no-LECs)' and those labeled `N3LO(full)'. The former include only
contributions of minimal nature. That is the loop-corrections illustrated in panels $(e)$--$(i)$ and $(l)$--$(o)$ of
Figure~\ref{fig:chi_cnt}, and the `minimal' contact currents  represented in panel $(k)$ of the same figure.
The magenta band labeled `N3LO(full)' is obtained using the full e.m.~current operator up to N3LO ($\nu=1$).
The thickness of the bands represents variations due to the use of different nuclear Hamiltonians
to generate the nuclear wave functions, namely the AV18+UIX or the NN(N3LO)+3N(N2LO), 
where the two-- and three--body chiral interactions
are form Refs.~\cite{Entem03} and~\cite{Ga_beta_triton}, respectively.   
The predictions indicated by the magenta bands are  within 1\% for $\sigma^\gamma_{\rm np}$ and 3\% for
$\mu^{\rm V}$($^3$H/$^3$He) of the experimental values~\cite{Piarulli12}, which are represented by the
black bands in the figure. We note that the leading two-body correction at NLO ($\nu=-1$) given by
the long-ranged OPE seagull and pion-in-flight currents enhances the LO ($\nu=0$) prediction (or IA prediction) for
$\sigma^\gamma_{\rm np}$ [$\mu^{\rm V}$($^3$H/$^3$He)] by $\sim$4\% ($\sim$9\%). 
We also note that the N3LO ($\nu = 1$) contributions due to the tree-level and `non-minimal' contact
currents are `large' and crucial to bring the theory in agreement with the experimental data. This is
an indication that explicit inclusion of the $\Delta$-resonance as a fundamental degree of freedom
of the theory should improve the convergence of the chiral expansion. 

An important check on the chiral e.m.~currents  is to verify that they
satisfy the continuity equation with the chiral potential order by order in the chiral expansion.
Of course, up to NLO ($\nu=-1$) the current is conserved since it just involves the non-relativistic
IA and OPE seagull and pion-in-flight operators of Eqs.~(\ref{eq:cnt_IA}) and~(\ref{eq:cnt_OPE}),
respectively, which are known to satisfy the continuity equation with kinetic term and OPE potential
entering the many-body nuclear Hamiltonian. A simple counting of the powers of momentum entering the
l.h.s. and r.h.s. of Eq.~(\ref{eq:cont_2})\footnotemark[1]\footnotetext{Note that the commutator on the r.h.s. of
Eq.~(\ref{eq:cont_2}) brings in an extra factor of $Q^3$
due to the implicit momentum integrations. Then, for example, for two-body currents at N3LO
($\nu=1$) the l.h.s. of Eq.~(\ref{eq:cont_2}) scales as $e\, Q^2$. The non-relativist IA charge
operator entering the r.h.s. scales as $e\,Q^{-3}$. Therefore, in order to equate the l.h.s. with the r.h.s.
the two-body potential $v_{ij}$ has to scale as $Q^2$.}, indicates that `minimal' N3LO ($\nu=1$) e.m.~currents,
illustrated in Figure~\ref{fig:chi_cnt}, must satisfy the continuity equation with the NN potential
of order $Q^2$, illustrated in Figure~\ref{fig:chiNN}. This statement has been explicitly verified
in Refs.~\cite{Pastore08,Kolling09}.

A natural question to ask is whether at the order we are considering, that is N3LO ($\nu=1$), there
are three-body currents. Above, we mentioned that 3N forces appear for the first time at NLO ($\nu=2$),
however they are seen to vanish. Therefore, in order to satisfy the continuity equation, three-body
currents at N3LO ($\nu=1$) must either vanish, or be transverse to the photon field. They are, in fact,
found to vanish~\cite{Girlanda09}, thus the current operator up to N3LO ($\nu=1$) includes only 
one- and two-body terms. Leading three-body e.m.~currents appear at N4LO ($\nu=2$)~\cite{Park96},
and they have not been derived yet.

\subsection*{Chiral many-body electromagnetic charge operator}

The e.m.~charge operator has been first investigated from a $\chi$EFT perspective by 
Walzl {\it et al.} in Ref.~\cite{Walzl01} and Phillips in Refs.~\cite{Phillips03,Phillips07}.
Phillips derived it up to N3LO (that is, up to $\nu=0$ in our 
counting~\footnotemark[1])\footnotetext{Note that in the counting utilized by Phillips 
the IA charge operator at LO is taken to scale as $e\,Q^0$
as opposed to $e\,Q^{-3}$ as it is done here, therefore N3LO$_{\rm Phillips}=Q^3$.}, 
while the first derivation of one-loop corrections,
entering at N4LO ($\nu=1$), has been carried out in Ref.~\cite{Kolling09} by K\"olling {\it et al.}
using the unitary transformation method.
A time-ordered perturbation theory calculation has subsequently appeared in Ref.~\cite{Pastore11}.
Within this framework, the construction of the charge operator up to one-loop necessarily requires
the study of non-static contributions to the chiral OPE and TPE potentials.
These corrections that go beyond the static limit are not uniquely determined
off-the-energy-shell, therefore the specific form of the N3LO ($\nu=0$) and N4LO ($\nu=1$) corrections 
of one- and two-pion range are found to depend on the off-the-energy-shell prescriptions
adopted for non-static terms in the OPE and TPE potentials, respectively~\cite{Pastore11}.
The ambiguity in the non-static potential and charge operators is of no consequence, since different
forms are related to each other by a unitary transformation~\cite{Pastore11}, a finding that
was first unraveled by Friar~\cite{Friar77} for non-static potentials and charge
operators of one-pion range. 

\begin{figure}[htb]
\centering
\includegraphics[width=8cm,clip]{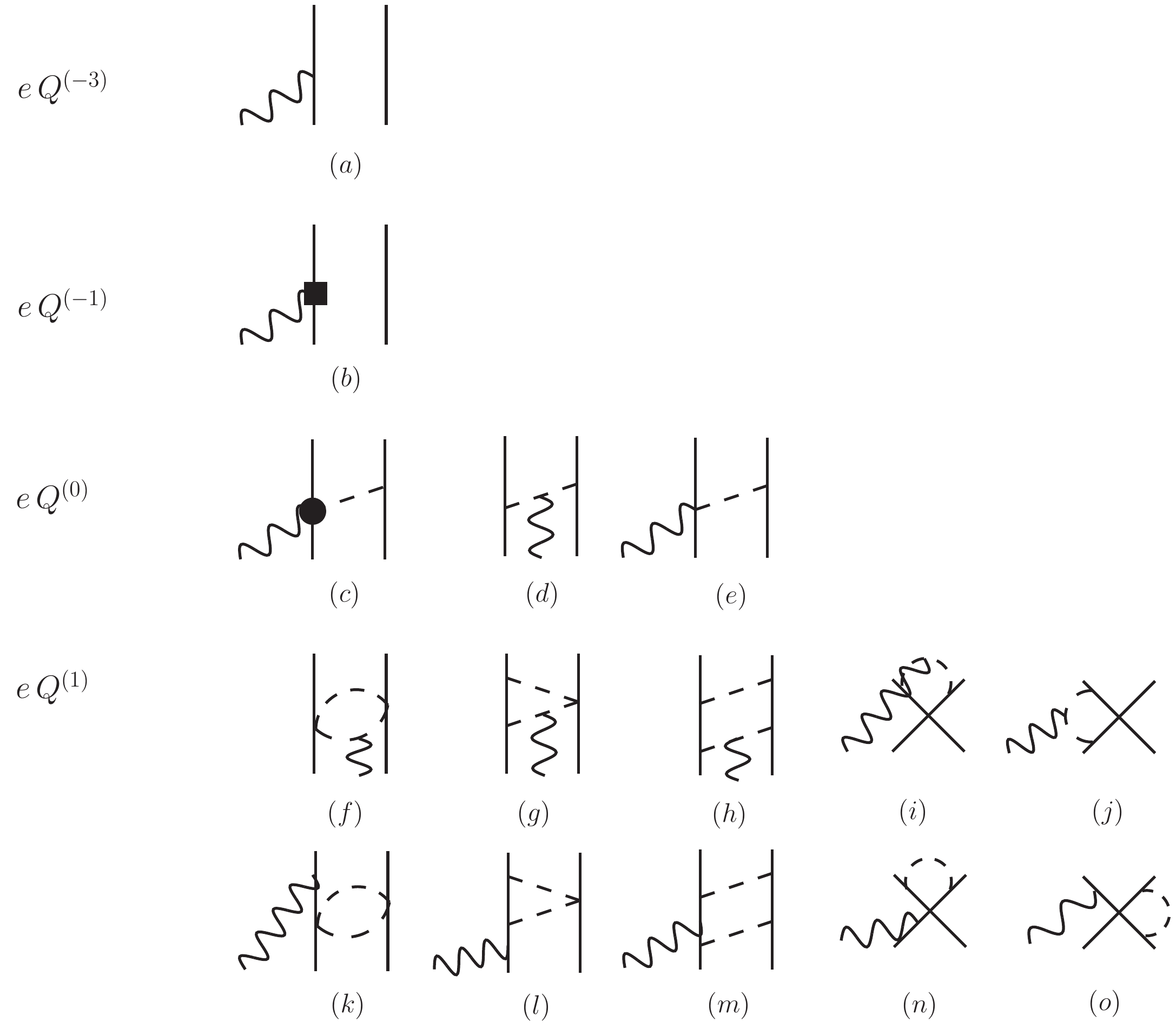}
\caption{Diagrams illustrating one- and two-body charge operators entering at LO ($e\, Q^{-3}$), panel $(a)$,
N2LO ($e\, Q^{-1}$), panels $(b)$, N3LO ($e\, Q^{0}$), panels $(c)$--$(e)$,
and N4LO ($e\, Q^{1}$), panels $(f)$--$(o)$.  The square in panel $(b)$ represents the
$(Q/m)^2$, or $(v/c)^2$, relativistic correction to the LO one-body charge operator [or IA(RC)],
whereas the solid circle in
panel $(c)$ is associated with a $\gamma \pi N$ charge coupling of order $e\, Q$ (see text for explanation). 
 Notation is as in Figure~\ref{fig:chiNN}.}
\label{fig:chi_charge}
\end{figure}

In what follows, we refer to Figure~\ref{fig:chi_charge} and list the various contributions
to the charge operator up to corrections of order $e\,Q^{(\nu=1)}$. The LO ($\nu=-3$)
contribution is represented by the diagram illustrated in panel $(a)$ and corresponds to
the non-relativistic IA operator given in Eq.~(\ref{eq:charge_IA}). In principle, at NLO
($\nu=-2$) there are contributions of one-pion range which, however, are seen to vanish in the static limit.
At N2LO ($\nu=-1$) there is the one-body operator illustrated
in panel $(b)$ corresponding to the relativistic correction to the leading IA operator [or IA(RC)].
Two-body contributions appear at N3LO ($\nu=0$) and are illustrated in panels $(c)$, $(d)$ and
$(e)$. In particular, the contribution of panel $(c)$ involves a $\gamma\pi$NN vertex---denoted
by a solid circle---obtained from a chiral Lagrangian of order $e\,Q$~\cite{Phillips03,Phillips07},
while operators associated with the diagrams of panels $(d)$ and $(e)$ result from accounting
for non-static corrections, and are found to be suppressed with respect to the contribution
of panel $(c)$~\cite{Piarulli12}. An operator that has the same structure as the operator
resulting from the evaluation of the transition amplitude associated with the diagram of
panel $(c)$, has been reported in the late eighties in the review article by Riska~\cite{Riska89},
however its derivation within a $\chi$EFT context has been carried out in recent years
by Phillips~\cite{Phillips03}. Finally, one-loop corrections enter at N4LO ($\nu=1$) and are
represented by the diagrams of panels $(f)$--$(o)$. Of course, at N3LO ($\nu=0$) and N4LO ($\nu=1$)
one should also include relativistic corrections to the OPE at NLO ($\nu=-2$) and IA at N2LO ($\nu=-1$),
respectively. Note that, due to charge conservation, contributions beyond the IA or LO term 
vanish at $|{\bf q}|=0$.
 
We emphasize that the structure of the charge operator is quite different from that of the current
operator. To begin with, the e.m.~current operator at LO ($\nu=-2$) is suppressed by a factor of
$Q$ with respect to the LO ($\nu=-3$) charge operator. Secondly, two-body effects of one-pion range
enter the charge operator at N3LO ($\nu=0$). They instead constitute the NLO ($\nu=-1$) and therefore the
major correction to the IA picture in the case of the e.m.~current operator. In addition,
OPE corrections to the charge operator are $1/m$ terms that vanish in
the static limit of $m\rightarrow \infty$.
Finally, the charge operator does not involve unknown LECs.

\subsection{Conventions and notations}

Here, we introduce and define some nomenclature.
We will use the word `conventional' to describe calculations carried out
within the conventional approach described in Sec.~\ref{subsec:SNPA}.
With `$\chi$EFT calculations' we generically denote calculations
that use  potentials and currents derived within $\chi$EFT formulations,
briefly outlined in Sec.~\ref{subsec:chi}. We stress that such $\chi$EFT
calculations are not always strictly consistent, in that, for example,
$(i)$ they may use different regulators for the chiral NN and 3N potentials
and currents, $(ii)$ they may use two- and three-body forces which
are not evaluated at the same order in the chiral expansion, $(iii)$ they may
use chiral e.m.~currents which satisfy the continuity equation with only part
of the chiral potential.
We denote with `hybrid calculations' calculations in which nuclear wave functions
are obtained from conventional nuclear Hamiltonians while e.m.~current operators
are derived from $\chi$EFT formulations. 

Throughout the course of this review, we try, where possible, to consistently
implement the following color scheme: we indicate with black symbols experimental data;
cold colors ({\it e.g.}, cyan, blue) indicate calculations in IA;
warm colors ({\it e.g.}, magenta, red) indicate calculations with two-body currents.
The combination (cyan, magenta) is mostly used for $\chi$EFT calculations,
while the (blue, red) one refers to conventional or hybrid calculations.
Similarly, for calculations in which the effect of 3N forces is investigated,
warm colors are assigned to final results inclusive of 3N forces' effects, while
cold colors are associated with results that use only NN potentials.  

In the figures, nuclear models are indicated using the convention $v_{ij}+V_{ijk}$.
For example, calculations that use the conventional AV18 two-nucleon and
IL7 three-nucleon interactions are indicated with AV18+IL7. For calculations
that use chiral potentials we use the notation NN(N$n$LO)+3N(N$n^\prime$LO),
where $n$ and $n^\prime$ specify the chiral orders of the two-nucleon and three-nucleon
potentials, respectively. Conventional electromagnetic charge and current operators are denoted
using the convention $\rho^{\rm 1-body + 2-body}$ and ${\bf j}^{\rm 1-body + 2-body}$.
We use `IA' to denote the non-relativistic operators given in Eqs.~(\ref{eq:charge_IA0}),
(\ref{eq:charge_IA}), (\ref{eq:cnt_IA}), while relativistic corrections of order $Q^2/m^2$
to the (non-relativistic) IA operator are denote with `IA(RC)'. Conventional two-body e.m.
currents are indicated with `MEC'. For example, a calculations that uses conventional
one- and two-body vector currents are indicated with ${\bf j}^{\rm IA + MEC}$, or with
${\bf j}^{\rm IA + IA(RC) + MEC}$, if relativistic correction to the IA operator are included.
For chiral e.m. charge (current) operators
we use the notation $\rho^{{\rm N}n{\rm LO}}$ (${\bf j}^{{\rm N}n{\rm LO}}$), where $n$ specifies
the cumulative contributions up to N$n$LO corrections included. Of course, chiral 
e.m. charge and current operators at LO are equivalent to 
conventional non-relativistic operators in IA, {\it i.e.}, $\rho^{\rm LO}\equiv\rho^{\rm IA}$ 
and ${\bf j}^{\rm LO}\equiv{\bf j}^{\rm IA}$.   
When we want to indicate both the current operator and the nuclear potential, we use the following notation: 
e.m. operator/$v_{ij}+V_{ijk}$.

\section{Electron scattering reactions}
\label{sec:electron}

\begin{center}
\begin{figure}[htb]
\centering
\includegraphics[width=5cm,clip]{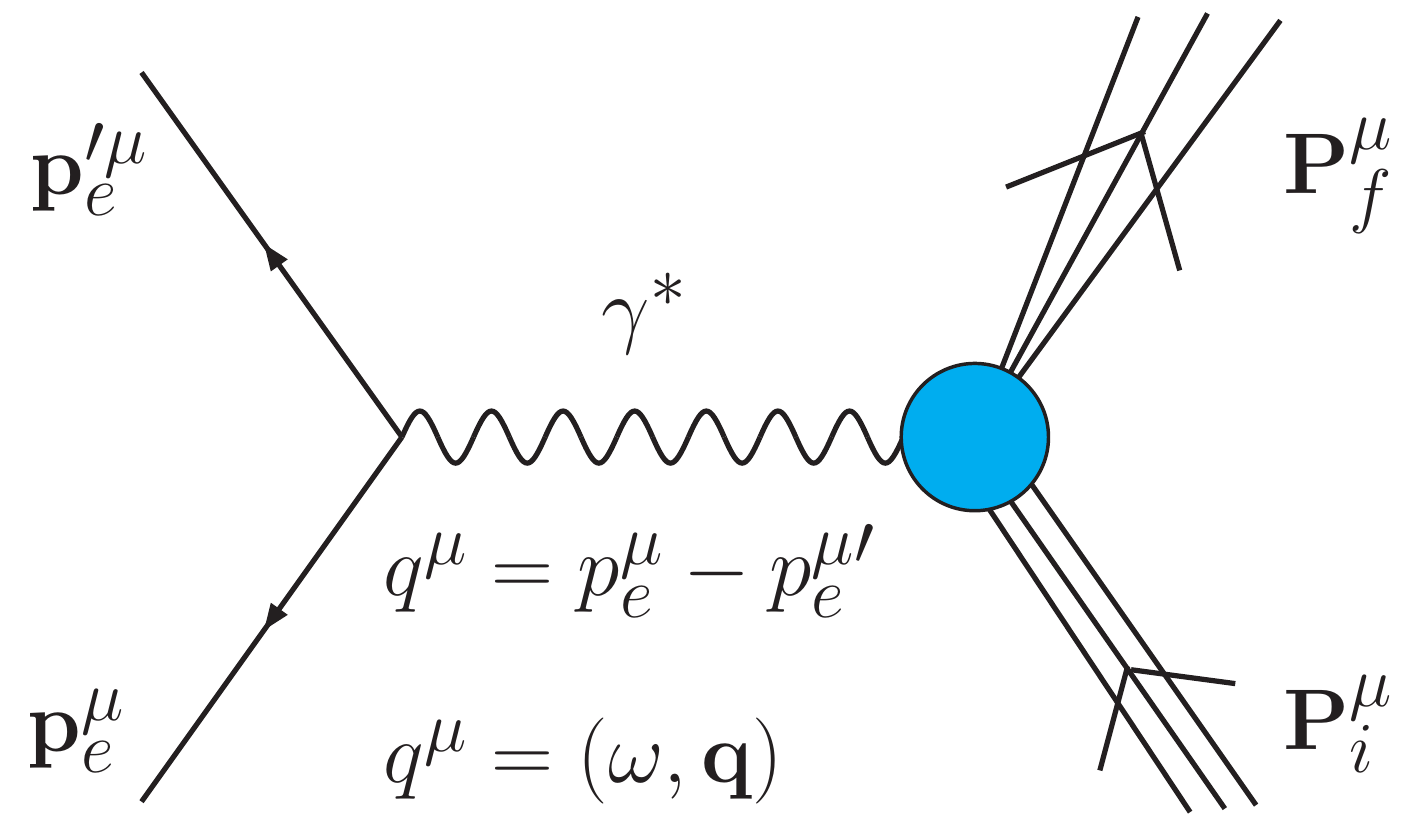}
\caption{(Color online) Diagram illustrating the electron scattering off a nucleus in the one-photon-exchange approximation.
A virtual photon $\gamma^*$ (wavy line) is exchanged between an electron (single solid lines) of initial and final four-momenta 
${\rm p}_e^{\mu}$ and ${\rm p^\prime}_e^{\mu}$, respectively, and a nucleus (triple solid lines) with initial and final
momenta ${\rm P}^{\mu}_i$ and ${\rm P}^{\mu}_f$, respectively.}
\label{fig:electron}       % Give a unique label
\end{figure}
\end{center}

In Born approximation, the electron scattering off a
nucleus~\cite{DeForest66,Donnelly75,Donnelly84,Eisenberg76,Ring80, Boffi96,SICK2001}
occurs via the exchange of one virtual photon between the probing electron and the nuclear target,
as it is schematically shown  in Figure~\ref{fig:electron}.
Elastic electron scattering is and ideal probe to study nuclear properties. In fact, due to the weakness of the e.m.~interaction with respect to the strong one, responsible of defining
the main nuclear structure features, electrons are noninvasive probes,
in that they barely perturb the targets. In the light nuclei of interest here,
the proton number $Z$ is small, thus incoming and scattered electrons are accurately
described by plane-waves, {\it i.e.}, they are assumed to be free particles unaffected by
the Coulomb field of the nucleus. This approximation is referred to as plane wave
Born approximation, and corrections beyond it are accounted for through
the evaluation of successive higher-order terms in the $Z\alpha$ expansion~\cite{Eisenberg76}.   

\begin{center}
\begin{figure}[htb]
\centering
\includegraphics[width=10cm,clip]{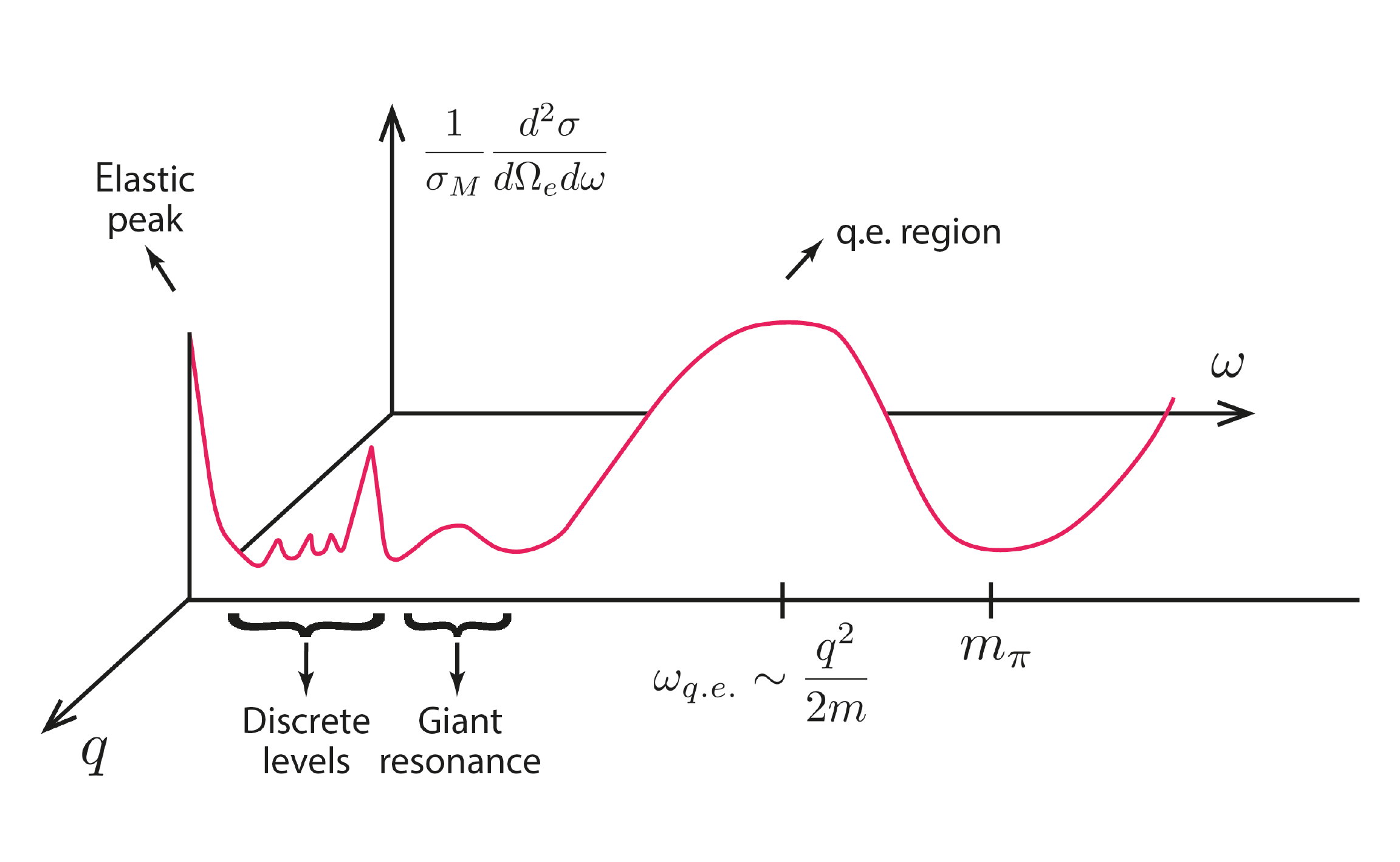}
\caption{(Color online) Schematic representation of the double differential cross section at fixed value of momentum transfer.}
\label{fig:xsec}       % Give a unique label
\end{figure}
\end{center}

In Figure~\ref{fig:electron}, an electron with momentum ${\rm p}_e^{\mu}=(\epsilon,{\bf p}_e)$
is scattered to a state with momentum ${{\rm p}^\prime}_e^{\mu}=(\epsilon^\prime,{\bf p}^\prime_e)$.
The virtual photon transfers a momentum $q^{\mu}=(\omega, {\bf q})$ to the nucleus, which transitions
from an initial state $\left| \Psi_i  \right\rangle$ with momentum  ${\rm P}^{\mu}_i=(E_i, {\bf P}_i)$ to
a final state $\left| \Psi_f  \right\rangle$ with momentum  ${\rm P}^{\mu}_f=(E_f, {\bf P}_f)$,
and momentum conservation implies $q^{\mu}=p_e^{\mu}-{p_e}^{\prime\mu}={\rm P}^{\mu}_f-{\rm P}^{\mu}_i$.
Furthermore, the interaction proceeds through the exchange of a space-like virtual photon,
for which $q_{\mu}^2=\omega^2-{\bf q}^2 < 0$\footnotemark[1]\footnotetext{The four-vector squared $q_{\mu}q^{\mu}$ is
here denoted with $q_{\mu}^2$.}.
In electron-induced reactions $\omega$ and ${\bf q}$ can vary independently (provided that
$|{\bf q}| > \omega$), as opposed to reactions induced by real photons where $|{\bf q}| = \omega$.
In elastic reactions $\omega=0$ (neglecting the recoil
of the nucleus), which implies $\left| \Psi_i  \right\rangle=\left| \Psi_f  \right\rangle$.
Reactions in which $\omega\neq0$ are instead called inelastic. To different values of $\omega=E_f-E_i$,
correspond different excitation energies of the nucleus. As $\omega$ increases to a few MeV,
low-lying (discrete) nuclear excited states can be accessed. For energies transferred of the order of
$\sim 10-30$ MeV, giant resonance modes in the continuum spectrum of the nucleus are excited,
while for values of $\omega_{q.e.}\sim q^2/(2 m)$ quasi-elastic effects dominate, in which the reaction is
in first approximation well described as if electrons were scattered off single nucleons. 
Beyond the quasi-elastic energy region, meson production can be observed. 
A schematic representation of the double differential cross section for electron scattering 
at a fixed value of momentum transfer $q$ is provided in Figure~\ref{fig:xsec}.

Because in inelastic electron scattering $\omega$ and ${\bf q}$ can vary independently,
for each value of excitation energy $\omega$, one can study the matrix elements' behavior 
as a function of the momentum transfer. In particular, by varying ${\bf q}$ one changes the
spatial resolution of the electron probe, which is $\propto 1/|{\bf q}|$. At low values of
momentum transfer, electron scattering reactions probe long ranged dynamics, while at higher
values of momentum transfer shorter distance phenomena are tested, where dynamics from heavier mesons and baryons become relevant.  

Cross sections for elastic scattering and scattering
to discrete excited states, for which the transferred energy $\omega$ is fixed,
are expressed in terms of longitudinal (or charge) and transverse (or magnetic)
form factors, which are functions of the momentum transferred $q=|{\bf q}|$,
and provide information on the e.m.~charge and current spatial distributions inside
the nucleus. 
The double differential cross section for inclusive processes,
in which only the scattered electron is detected, is expressed in terms of the longitudinal 
and transverse  response functions. These represent deviations from
the Mott cross section, $\sigma_M$, associated with electron scattering off point-like nuclei, therefore
they contain the nuclear structure information. In particular, longitudinal responses
are obtained from a multipole expansion of the nuclear e.m~charge operator $\rho$, while the
transverse ones are obtained from matrix elements of the nuclear e.m.~current operator
${\bf j}$. Thus, electron scattering processes are particularly suited to test models for
both nuclear Hamiltonians, which are used to generate the nuclear wave functions, and e.m.~current
operators. 
                            
In the next subsections we present an overview of recent progress in calculations 
of electron scattering observables for light nuclei, with emphasis on latest applications 
of $\chi$EFT nuclear and e.m. operators.

\subsection{Formalism}
\label{sec:formalism}

In plane wave Born approximation, the formal expression of the electron scattering
cross section for a process in which a nucleus transitions from an initial state $\left| \Psi_i \right\rangle$
to a final bound state $\left| \Psi_f \right\rangle$ reads~\cite{DeForest66,Donnelly75,Donnelly84,Carlson98}, in the laboratory frame\footnotemark[1]\footnotetext{Note that the convention used for the electric charge is $\frac{e^2}{4\pi\hbar c}$. }, 
\begin{equation}
\frac{d \sigma}{d\Omega_e}=4\pi \sigma_M f^{-1}_{rec}\left[\frac{Q_\mu^4}{q^4} F^2_L(q )
+\left(\!\frac{Q_\mu^2}{2q^2}+\tan^2{\frac{\theta_e}{2}}\! \right)F^2_T(q)\right]
\label{cross_el}
\end{equation}
where $ Q_\mu^2=-q_{\mu}^2$ and $q=|{\bf q}|$. The Mott cross section $\sigma_M$ is given by
\begin{equation}
\nonumber
\sigma_M=\bigg\{\frac{\alpha \cos{\theta_e/2}}{2\epsilon \sin^2{\theta_e/2}}\bigg\}^2,
\label{eq:mott}
\end{equation} 
and the recoil term is
\begin{equation}
 f^{-1}_{rec}=1+\frac{2\epsilon}{M_A}\sin^2(\theta_e/2)\ ,
\end{equation}
where $M_A$ is the rest mass of the target nucleus with mass number $A$,   $\theta_e$ and $\epsilon$
are the electron scattering angle and initial energy.  

The experimental longitudinal and transverse form factors for elastic and inelastic
transitions, $F_L(q)$ and $F_T(q)$, respectively, are extracted from the measured cross section
via  Rosenbluth separation, a procedure that allows, by varying $\theta_e$ at fixed values
of $\omega$ and $q$, to separate the longitudinal from the transverse form factor. 
The multipole expansion of the charge and current operators allows to write  form 
factors in terms of reduced matrix elements of Coulomb ($T_J^C$), magnetic ($T^M_J$),
and electric ($T^E_J$) multipole operators~\cite{DeForest66,Donnelly75,Donnelly84,Carlson98}:
\begin{eqnarray}
\label{eq:f-long}
F_L^2(q)\!=\!\frac{1}{2\,J_i+1} \!\! \sum_{J=0}^\infty |\!\!\left< J_f, E_f|| T_J^C(q)||J_i, E_i\right>\!\!|^2 \ ,  \\
 F_T^2(q)\!= \!\frac{1}{2\,J_i+1}\!\! \sum_{J=0}^\infty \!|\!\!\left< J_f, E_f|| T^M_J(q)||J_i, E_i\right>\!\!|^2 \!
+\!|\!\!\left< J_f, E_f|| T^E_J(q)||J_i, E_i\right>\!\!|^2 ,
\label{eq:f-tran} 
\end{eqnarray}
where $J_i$ ($J_f$) is the initial (final) total angular momentum of the nucleus.
The Coulomb multipoles ($CJ$) contributing to the longitudinal form factor are obtained from matrix
elements of the charge operator $\rho$, while the electric ($EJ$) and magnetic ($MJ$) multipoles
contributing to the transverse form factor are obtained from matrix elements of the current
operator ${\bf j}$ using standard formulae~\cite{DeForest66,Donnelly75,Donnelly84,Carlson98}.
Parity and time-reversal conservation laws impose selection rules. In particular,
in elastic scattering (where $\omega=0$, and $J_i=J_f$)
only Coulomb multipoles with even $J$ contribute starting from $J=0$, while electric multipole
cannot contribute due to time-reversal invariance.  Thus, only magnetic multipole operators
enter the transverse form factor and they carry an odd value of $J$, starting from $J=1$ since a
transverse photon carries unit helicity. In addition, conservation of angular momentum
requires that $J \leq 2 J_i$.
 
In the  limit of $q\rightarrow 0$, where the wavelength of the radiation
is large compared to the nuclear radius, the elastic Coulomb and magnetic
reduced matrix elements are proportional to static charge and magnetic
moments~\cite{DeForest66,Donnelly75,Donnelly84,Ring80,Carlson98}. In particular,
the Coulomb monopole ($C0$) contribution
is proportional to the proton number $Z$, while  for scattering off ground states
(thus $J_i=J_f=J_0$), the Coulomb quadrupole ($C2$) and magnetic
dipole ($M1$) reduced matrix elements
are proportional to the ground-state quadruple and magnetic dipole moments,
respectively. They are obtained from the evaluations of the following
ground-state expectation values~\cite{DeForest66,Donnelly75,Donnelly84,Ring80}
\begin{eqnarray}
 Q & = & \langle J_0 \,\,M_0 \!= \! J_0 \,\, |  \,\,\hat{Q} \,\, |  \,\, J_0 \,\,M_0 \!= \! J_0 \rangle \ , \\
 \mu & = & 2\,m \langle J_0\,\, M_0 \!= \! J_0 \,\, |  \,\, \hat{\mu}_z \,\, |  \,\, J_0 \,\,M_0 \!= \!J_0  \rangle \ ,
\end{eqnarray}
where the quadrupole and magnetic dipole operators are defined as
\begin{eqnarray}
\label{eq:q2-mom}
\hat{Q}&=& \int {\rm d}{\bf x} \,\rho({\bf x}) (3z^2-{\bf x}^2)\ , \\
\hat{{\bm \mu}} &=&\frac{1}{2}\int {\rm d}{\bf x} \left[{\bf x}\times{\bf j}(\bf x)\right] \ .
\label{eq:mag-mom}
\end{eqnarray}
and $\rho({\bf x})$ and ${\bf j}(\bf x)$ are the charge and current density operators~\cite{ericson88}. 
Therefore, probing electrons are sensitive to
static e.m. properties of nuclei as $q\rightarrow 0$.

In inelastic transitions, all kinds of multiple operators with $|J_i-J_f|\leq J\leq J_i+J_f$
contribute to the form factors, which are in this case referred to as transition
form factors. In addition, parity conservation requires  that 
$\pi_i \pi_f = (-1)^J$ [$\pi_i \pi_f = (-1)^{J+1}$] for Coulomb and electric (magnetic)
transitions, where $\pi_{i,f}$ are the parities of the initial
and final nuclear states.  In the long-wave limit, the transverse electric
multiple operator $EJ$ can be written in terms of the Coulomb operator $CJ$.
In this limit, one replaces the longitudinal current entering the $EJ$
transition operator with the charge operator $\rho$ using the continuity 
equation (this is, in fact, the Siegert theorem~\cite{Siegert}). Therefore,
in the long-wave limit, the transverse electric multipole operator
$EJ$ is obtained from matrix elements of the charge operator $\rho$, {\it i.e.}, from Coulomb $CJ$ multipoles. 
This is a powerful approximation that allows to implicitly include in $EJ$ multipoles
the effect of MEC corrections (of the three-vector e.m.~currents) without explicitly
including them in the calculations.  

Reduced matrix elements entering the form factors for elastic and inelastic
scattering to discrete excited nuclear states also enter the transition rate for 
photo-emission reactions (which will be discussed in Sec.~\ref{sec:e2m1}).
Because in this kind of reactions the emitted photon is real,
the process is induced by transverse multipoles. For natural parity states, the dominant transverse
contributions are the $M1$ and $E2$ transitions operators. With the use of standard formulas~\cite{Ring80}
the reduced transition probabilities $B(E2)$ and $B(M1)$ are obtained from them as follows
\begin{eqnarray}
\label{eq:be2}
 B(E2)&=&\frac{1}{2 J_i +1} |\langle J_f||\hat{Q} ||J_i \rangle|^2 \ , \\
 B(M1)&=&\frac{1}{2 J_i +1} |\langle J_f||\hat{\mu}_z||J_i\rangle|^2 \ ,
\label{eq:bm1}
\end{eqnarray}
where $\hat{Q}$ and ${\hat{\bm \mu}_z}$ are the operators defined in Eqs.~(\ref{eq:q2-mom}) and~(\ref{eq:mag-mom}),
respectively\footnotemark[1].\footnotetext{Note that Eq.~(\ref{eq:be2}) is obtained in the long-wave limit with the use of the Siegert theorem.}

Below, we focus on ground-state properties which are inferred from elastic electron scattering.
In particular, Section~\ref{sec:gs-mag-mom} is devoted to discuss static e.m.~nuclear moments,
while in Section~\ref{sec:gs-ffs} we present a number of calculations of elastic form factors.
Strictly speaking, ground-state static electric and magnetic moments are accessed via
many different experimental procedures including, in some cases, elastic
electron scattering. Despite the fact that this section is devoted solely to
electron scattering processes, we discuss the theoretical calculations of light nuclei
e.m.~moments here. We find this arrangement to be convenient because the discussion on
nuclear static e.m. properties is complementary to the discussion on nuclear elastic form
factor that follows. A presentation of the main features of the double differential cross
section for inelastic scattering and associated observables is deferred to Section~\ref{sec:e-inelastic}.

\subsection{Ground-state properties: electromagnetic moments}
\label{sec:gs-mag-mom}

Static e.m.~properties of light nuclei play an important role in 
testing the validity of nuclear models. For example, the experimental
evidence of a non-zero deuteron quadrupole moment pointed to the
necessity of introducing tensor components in the NN interaction. 
Similarly, static magnetic moments of nuclear ground states have been
determinant to establish the role of MEC. 
Indeed, the seminal studies on MEC effects were focused on evaluating their contributions to
nuclear magnetic moments~\cite{Villars47,Miyazawa51}. Since then,
modern and highly sophisticated conventional MEC have been successfully utilized in
a number of calculations of nuclear e.m.~properties~\cite{Carlson98}. 
In recent  calculations, based on wave functions
obtained with the HH method, trinucleon magnetic moments are found to
be within less than $1\%$ of the experimental
values~\cite{Marcucci98}, when the AV18+UIX %~\cite{AV18,Pudliner95}
nuclear Hamiltonian and consistent MEC currents are used. MEC  
provide a $\sim 16\%$ correction to the total calculated values~\cite{Marcucci98}.

\begin{table*}[bth]
\caption{GFMC results from Ref.~\cite{Pastore12} for $A\leq9$ nuclear states'
energies, dipole magnetic [$\mu(\rm IA)$] and quadrupole ($Q$) moments,
compared to experimental values~\cite{Amroun94,Tilley02,Tilley04,Audi03,Purcell10}.
%,Shiner95,Nortershauser11,Nortershauser09}. 
Numbers in parentheses are statistical errors for the GFMC calculations
or experimental errors; errors of less than one in the last decimal place
are not shown.}
\label{tb:energies}
\begin{center}
 \begin{tabular}{c|l|l|l|l|l|l}
\hline
\hline
    $^AZ(J^\pi,T)$                &
   \multicolumn{2}{c}{$E$ [MeV]} &
   \multicolumn{2}{c}{$\mu$(IA) [n.m.]} &
   \multicolumn{2}{c}{$Q$ [fm$^2$]}
   \\ \cline{2-3} \cline{4-5} \cline{6-7} 
                            &
   \multicolumn{1}{c}{GFMC} &
   \multicolumn{1}{c}{Expt.}&
   \multicolumn{1}{c}{GFMC} &
   \multicolumn{1}{c}{Expt.}&
   \multicolumn{1}{c}{GFMC} &
   \multicolumn{1}{c}{Expt.}
   \\ \hline
   \nuc{2}{H}$(1^{+},0)$                    &  $-$2.225   &  $-$2.2246&  {\color{white}$-$}0.847      & {\color{white}$-$}0.8574 & {\color{white}$-$}0.270   &  {\color{white}$-$}0.286
   \\
   \nuc{3}{H}$(\frac{1}{2}^+,\frac{1}{2})$  &  $-$8.50(1) &  $-$8.482 &   {\color{white}$-$}2.556      & {\color{white}$-$}2.979  &         &
   \\
   \nuc{3}{He}$(\frac{1}{2}^+,\frac{1}{2})$ &  $-$7.73(1) &  $-$7.718  & $-$1.743     &$-$2.127  &         &
   \\
   \nuc{6}{Li}$(1^+,0)$                     & $-$31.82(3) & $-$31.99 &  {\color{white}$-$}0.817 &  {\color{white}$-$}0.822  &$-$0.20(6) &$-$0.082(2)
   \\
   \nuc{7}{Li}$(\frac{3}{2}^-,\frac{1}{2})$ & $-$39.0(1)  & $-$39.24 &   {\color{white}$-$}2.87  &  {\color{white}$-$}3.256  &$-$4.0(1)  &$-$4.06(8)
   \\
   \nuc{8}{Li}$(2^+,1)$                     & $-$41.5(2)  & $-$41.28 &   {\color{white}$-$}1.16(2)&  {\color{white}$-$}1.654 & {\color{white}$-$}3.3(1) & {\color{white}$-$}3.14(2)
   \\
   \nuc{8}{B}$(2^+,1)$                      & $-$37.5(2)  & $-$37.74 &   {\color{white}$-$}1.45(1)&  {\color{white}$-$}1.036 & {\color{white}$-$}5.9(4) & {\color{white}$-$}6.83(21)
   \\
   \nuc{9}{Be}$(\frac{3}{2}^-,\frac{1}{2})$ & $-$58.1(2)  & $-$58.16 & $-$1.18(1)&$-$1.178 & {\color{white}$-$}5.1(1) & {\color{white}$-$}5.29(4)
   \\
   \nuc{9}{Li}$(\frac{3}{2}^-,\frac{3}{2})$ & $-$45.2(3)  & $-$45.34 & {\color{white}$-$}2.66(3)& {\color{white}$-$}3.439 &$-$2.3(1) & $-$3.06(2)
   \\
   \nuc{9}{C}$(\frac{3}{2}^-,\frac{3}{2})$  & $-$39.7(3)  & $-$39.04 & $-$0.75(3)&$-$1.391 &$-$4.1(4)
   \\
\hline
\hline
\end{tabular}
\end{center}
\end{table*}

Hybrid studies on $M1$ static properties of $A\leq3$ nuclei, based on the 
N3LO ($\nu=1$) $\chi$EFT e.m.~currents of Ref.~\cite{Park93}, have been
first carried out by Song and collaborators in Refs.~\cite{Song07,Song09}.
The model dependence of the hybrid predictions has been studied
by utilizing different realistic nuclear Hamiltonians, including Hamiltonians
derived within $\chi$EFTs. Using resonance saturation
arguments~\cite{Song07,Song09}, the authors
have reduced the number of LECs entering the e.m.~currents to two.
These LECs multiply contact operators at N3LO---see panel $(k)$ of Figure~\ref{fig:chi_cnt}---and 
they have been fixed to the experimental trinucleon magnetic moments ($np\rightarrow d\gamma$ cross
section at thermal neutron energies and the deuteron magnetic moment) when used to predict
the $np\rightarrow d\gamma$ cross section at thermal neutron energies and the
deuteron magnetic moment (trinucleon magnetic moments). In Ref.~\cite{Song07},
it has been found that the isoscalar and isovector combinations of trinucleon
magnetic moments agree with the experimental value at the $\sim 2-3\%$ level (a finding that
was confirmed in the studies of Ref.~\cite{Piarulli12} summarized in Figure~\ref{fig:npdg_trimm}),
and it has been conjectured that  additional corrections from three--body e.m.~currents
entering at N4LO, along with the use of improved nuclear wave functions,
could reduce the present gap between theoretical and experimental values.

Moving towards larger nuclei, we find a number of {\it ab-initio} calculations
of magnetic moments carried out in IA. A recent summary on the current
status for nuclei up to  $A\sim15$ can be found, {\it e.g.}, in Ref.~\cite{Maris13}.
Here, we limit ourself to report in Table~\ref{tb:energies} a set of GFMC calculations
of energies and static e.m.~moments in IA of nuclei with $A\leq9$. In fact, we want to emphasize
calculations
which account for two-body effects in the e.m.~currents,  as most recently
reported in Ref.~\cite{Pastore12}. Magnetic moments for $A\leq 7$ nuclei, comprehensive
of two-body corrections, have been first evaluated in Ref.~\cite{Marcucci08} using GFMC
computational techniques %~\cite{Carlson87,Wiringa91,pudliner1997,Wiringa00,pieper2001,Pervin07},
and conventional MEC operators~\cite{Marcucci05,Marcucci08},
in combination with the AV18+IL2 %~\cite{AV18,IL} 
nuclear Hamiltonian.
In that work, it has been found that MEC corrections increase the $A=3,7$ isovector magnetic
moments by up to $16\%$, bringing them into very good agreement with experimental data.
That  study  has been recently extended in Ref.~\cite{Pastore12} to include larger nuclei
and improved nuclear wave functions obtained from the AV18 %+IL7 %~\cite{AV18,IL7} 
and an updated version of the 3N interaction, {\it i.e.}, the IL7. %~\cite{IL7}.
In addition, $\chi$EFT e.m.~current operators of Refs.~\cite{Pastore08,Pastore09,Piarulli12}
have been tested in hybrid calculations which use the same nuclear wave functions
obtained from the AV18+IL7
%~\cite{AV18,IL7} 
interaction.

\begin{center}
\begin{figure}
\centering
\includegraphics[height=.40\textheight,angle=270,keepaspectratio=true]{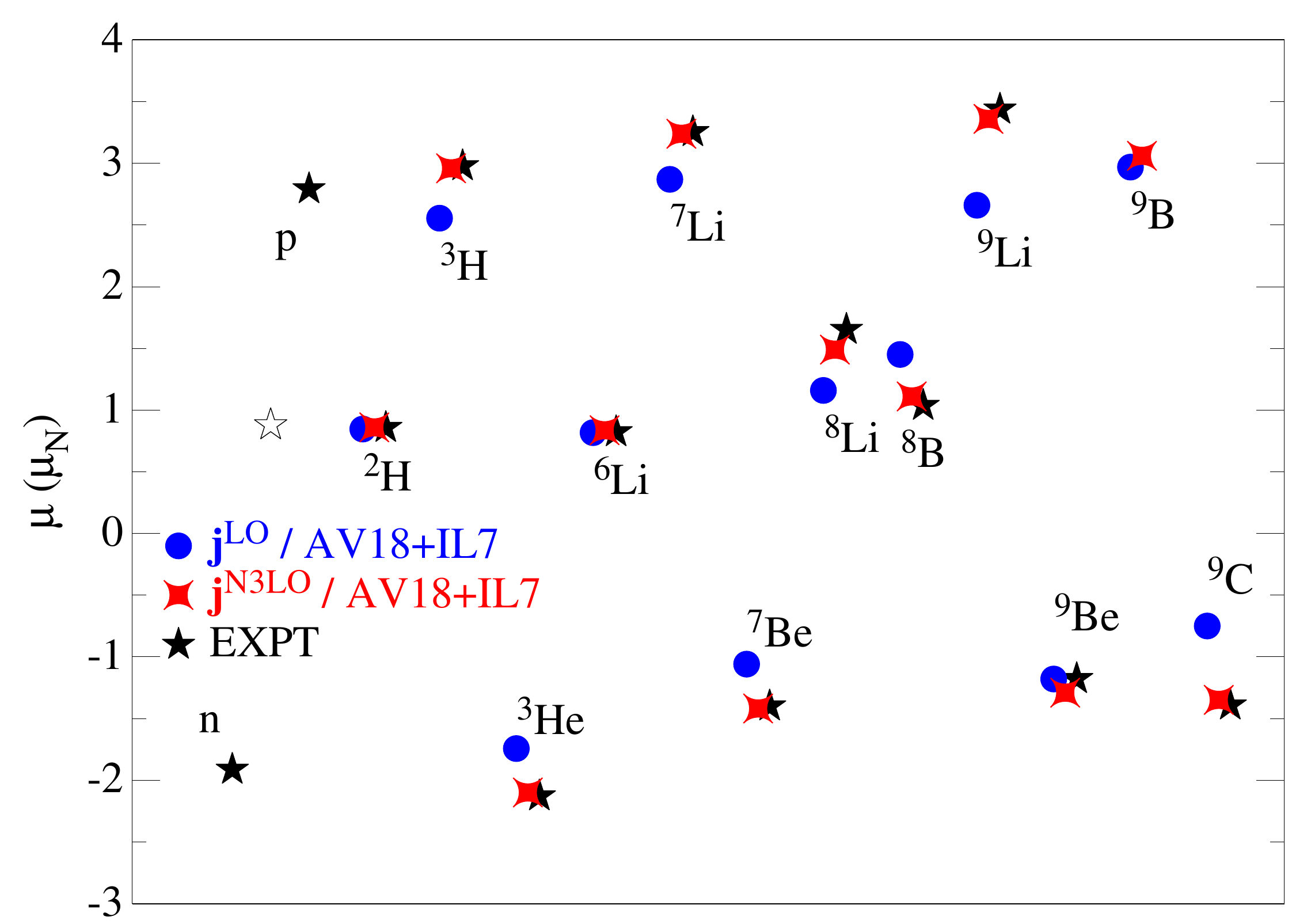}
%\includegraphics[height=.40\textheight,keepaspectratio=true]{figures_eps/fig8.eps}
%\vspace*{-0.5in}
\caption{(Color online) Magnetic moments in nuclear magnetons for $A\leq9$ nuclei from Ref.~\cite{Pastore12}.
           Black stars indicate
           the experimental values from Refs.~\cite{Borremans05,Tilley02,Tilley04},
           while blue dots (red diamonds)  represent GFMC calculations which include the LO or IA
           (N3LO) e.m.~currents from $\chi$EFT.  Predictions are for nuclei with $A>3$. }
\label{fig:mm}
\end{figure}
\end{center}

GFMC results for nuclear magnetic moments are summarized
in Figure~\ref{fig:mm}, where black stars represent the experimental
data~\cite{Borremans05,Tilley02,Tilley04}---there are no data
for the $^9$B magnetic moment. For completeness,
experimental values for the proton (p) and neutron (n) magnetic moments,
as well as their sum (empty star), which corresponds to the magnetic moment of
an S-wave deuteron, are shown.  The static nuclear magnetic dipole operator induced by the IA
current of Eq.~(\ref{eq:cnt_IA}) coincides with the LO in $\chi$EFT and reads 
\begin{equation}
M1= {\bm \mu}({\rm IA}) =  \sum_i \left(e_{N,i}\, {\bf L}_i + \mu_{N,i}\,
  {\bm \sigma}_i\right)  \ ,
\label{eq:M1_IA}
\end{equation}
where  ${\bf L}_i$ is the orbital angular momentum of nucleon $i$ (only protons contribute
to the convection e.m.~current), and the $e_{N,i}$ and $\mu_{N,i}$ expressions are given in
Eqs.~(\ref{eq:e_ff}) and~(\ref{eq:m_ff}), respectively, and should be taken in the limit
of $Q^2_\mu\rightarrow0$. Results obtained with the operator
given above are represented by blue dots of Figure~\ref{fig:mm}, and they reproduce the bulk properties
of the considered nuclear magnetic moments. The magnetic moment associated with the protons'
convection current is found to be small if compared with proton and neutron spin magnetization
terms~\cite{Pastore12}. Nuclear magnetic moments in IA
are driven by those associated with valence nucleons.  In particular, the magnetic moment of
an odd-even nucleus is driven by that of the unpaired proton, and
lines up in the upper part of Figure~\ref{fig:mm}, similarly, the magnetic moment of an
even-odd nucleus is driven by that of the unpaired neutron, and sits in the bottom part of the figure.
Magnetic moments for odd-odd nuclei are instead driven either by a proton-neutron 
or by a triton-neutron ($^3$He-proton) cluster. 
\begin{center}
\begin{figure}%[h!]
\centering
\includegraphics[height=3.5in]{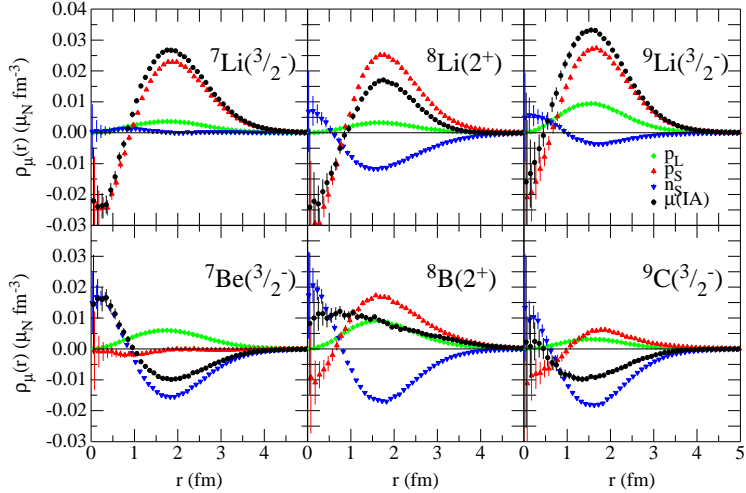}
\vspace*{-0.8cm}
\caption{(Color online)  One-body magnetic density in nuclear magnetons per fm$^3$ for 
selected nuclei (see text for explanation). }
\label{fig:mag_density}
\end{figure}
\end{center}
These general features can be best appreciated
by looking at the IA magnetic densities represented in Figure~\ref{fig:mag_density}.
Here, the red upward-pointing triangles are the contribution from
the proton magnetic moment weighted by the difference between the spatial
distributions for proton with spin up and down, $\mu_p[\rho_{p\uparrow}(r) - \rho_{p\downarrow}(r)]$, 
while the blue downward-pointing triangles are the analogous contribution 
but from the neutron magnetic moments. The green diamonds are the proton orbital
(convection current) contribution, and the black circles are the sum.
The integrals of the black curves over $d^3r$ give the total magnetic moments
of the nuclei in IA. For example, for the odd-even $^7$Li and $^9$Li nuclei,
the neutrons are paired up, and give only a small  contribution, so the total IA
magnetic moment is close to the sum of the
proton spin and orbital parts.  However, $^8$Li has one unpaired neutron which
acts against the proton and significantly reduces the overall IA values.
Similar considerations apply to the bottom panels of Figure~\ref{fig:mag_density}.

GFMC results of nuclear magnetic moments, which account for two-body effects and relativistic corrections in the
e.m.~current operator, are represented  in Figure~\ref{fig:mm} by red stars.
The main features of the $\chi$EFT e.m.~currents~\cite{Pastore08,Pastore09,Piarulli12} utilized in 
the calculations have been discussed in Section~\ref{subsec:chi}. As mentioned there, LECs of minimal
nature, associated with N3LO contact currents of the type illustrated in panel $(k)$ of Figure~\ref{fig:chi_cnt},
can be fixed to NN scattering data. In this work, they have been assigned the values obtained from the 
phase-shifts analysis carried out in Refs.~\cite{Entem01,Entem03,Machleidt11}. Instead, referring to
Figure~\ref{fig:chi_cnt}, LECs entering the N3LO operator illustrated in panel $(j)$ and those of non-minimal nature
associated with the contact currents of panel $(k)$, have been determined  so as to reproduce the magnetic moments
of $A=2$ and $3$ nuclei~\cite{Piarulli12}. Thus, only the results for $A>3$ nuclei 
with this $\chi$EFT e.m.~current are predictions.   Referring to Figure~\ref{fig:mm}, corrections beyond the
leading IA operator always increase the IA results in the direction of
the experimental data, except for $^6$Li and $^9$Be, where the contributions from
two-body e.m.~currents, while being small, make the predictions slightly worse. The effect
of two-body components in the e.m.~current operator is particularly pronounced in the
$^9$C and $^9$Li nuclei, where they provide up to $\sim 40\%$ and $\sim 20\%$, respectively,
of the total predicted magnetic moments' values. Two-body effects, while being
significant for the $^9$C and $^9$Li magnetic moments, have been found to be 
negligible for those of $^9$Be and $^9$B. This behavior can be explained  considering
the dominant spatial symmetries of the nuclear wave functions for these $A=9$ systems.
For example, the dominant spatial symmetry of $^9$Be ($^9$B) corresponds to an $[\alpha, \alpha, n (p)]$
structure~\cite{Wiringa06}. Therefore, the unpaired nucleon outside the $\alpha$ clusters
feels no OPE potential---see panel $(a)$ Figure~\ref{fig:chiNN}.
As a consequence, two-body currents of one-pion range entering at NLO---see panels $(b)$ and $(c)$
of Figure~\ref{fig:chi_cnt}---produce a negligible correction.
On the other hand, the dominant spatial symmetry of $^9$C ($^9$Li) corresponds to
an [$\alpha$, $^3$He ($^3$H), $pp$ ($nn$)] structure, and NLO OPE e.m.~currents 
contribute in both the trinucleon clusters and in between the trinucleon 
clusters and the valence $pp$ ($nn$) pair.

In Ref.~\cite{Pastore12}, also conventional MEC of Refs.~\cite{Marcucci05,Marcucci08} have been used to
calculate nuclear magnetic moments. It has been found that GFMC results obtained with $\chi$EFT
and conventional currents are qualitatively in agreement, particularly for isovector combinations
of magnetic moments.
% This is mainly due to the fact that both models describe the long-range
%behavior of the e.m.~operator in terms of OPE contributions---see panels $(b)$ and $(c)$ of
%Figure~\ref{fig:chi_cnt} and Eq.~(\ref{eq:cnt_OPE}). These isovector two-body currents
%constitute the major correction (occurring at NLO in the chiral expansion of the e.m.~current operator)
%to the IA current.
Quantitative differences between the two models have been found
to be more pronounced in isoscalar combinations of magnetic moments\footnotemark[1]\footnotetext{We note that 
the isoscalar, $\mu({\rm IS})$, and isovector, $\mu({\rm IV})$, combinations of nuclear magnetic 
moments are defined as $\mu(T,T_z)=\mu({\rm IS})+\mu({\rm IV})T_z$, where $\mu(T,T_z)$ is the magnetic moment
of a nucleus with total isospin, $T$, and total isospin $z$-projection, $T_z$. }.
 In general, $\chi$EFT e.m.~currents are
found to provide results which are in a better agreement with the experimental data for the
considered nuclear magnetic moments.

\subsection{Ground-state properties: elastic form factors}
\label{sec:gs-ffs}

Elastic form factors of light nuclei are key observables to test nuclear Hamiltonians and current operators. 
Their longitudinal (L) and transverse (T) components are given in Eqs.~(\ref{eq:f-long}) and~(\ref{eq:f-tran}),
respectively, to be taken with $J_f=J_i=J_0$. Elastic longitudinal and transverse form factors 
reflect the charge and magnetic spatial distributions, respectively.
In this section we provide an overview of the present status of elastic form factors in
{\it ab-initio} calculations, with emphasis on studies that include two-body currents
and/or 3N forces.  We organized the results in subsections devoted to different nuclei
with increasing mass number, ranging from the deuteron to $^{12}$C.
When possible, we compare theoretical results obtained by different groups for the same
observables and check them against experimental data.

\subsubsection{The deuteron}
\label{gs-deuteron}

$-\, \, \,\,$ Electron-deuteron elastic scattering reactions have been intensively investigated
both from an experimental and a theoretical point of view. There exists a vast
literature on this topic, for which we refer to the review
articles of Refs.~\cite{SICK2001,Garcon01,Gilman}.
Conventional calculations based on realistic NN interactions and consistent
e.m.~two-body MEC provide a satisfactory description of the available
experimental data~\cite{Carlson98,AV18,Arenhovel00,Schiavilla02}.
Currently, there  are also quite a few calculations of deuteron e.m.~properties
that are based on $\chi$EFT 
formulations~\cite{Phillips00,Walzl01,Phillips03,Phillips05,Phillips07,Valderrama07,Kolling12,Piarulli12}.
Within the $\chi$EFT framework, a general good agreement between the calculated and experimental
deuteron form factors is observed, provided that the e.m.~structure of
the nucleons is accounted for via suited nucleonic form factors.

Traditionally, the charge and magnetic spatial distributions of the
deuteron are studied in terms of the charge ($G_{ C}$),
quadrupole ($G_{\rm Q}$), and magnetic  ($G_{M}$) form factors,
and they are related to the Coulomb, $T^C_J$, and magnetic transverse, $T^M_J$,
multipole operators discussed in Sec.~\ref{sec:formalism}
via~\cite{Carlson98}
\begin{eqnarray}
&& \sqrt{\frac{4\pi}{3}} T^C_0(q) = (1+\eta)G_{C}(q) \ , \\
&& \sqrt{\frac{4\pi}{3}} T^C_2(q) = \frac{2\sqrt{2}}{3}\eta(1+\eta)G_{Q}(q) \ , \\
&&\!\!\!\!\!\!\!\!\!\!\!\!-\,i\,\sqrt{\frac{4\pi}{3}} T^M_1(Q) =\frac{2}{\sqrt{3}}\sqrt{\eta(1+\eta)}G_{C}(q) \ , 
\end{eqnarray}
where $\eta=(q/2\,M_d)^2$ and $M_d$ is the deuteron mass. 
These form factors are normalized as
\begin{equation}
\label{eq:norm0}
 G_C(0)=1\ ,\,  G_M(0)=(M_d/m_N)\, \mu_d\ ,\,  G_Q(0)=M_d^2\, Q_d \ ,
\end{equation}
where $\mu_d$ and $Q_d$ are the deuteron magnetic moment
(in units of $\mu_N$) and quadrupole moment, respectively.
The deuteron form factors are extracted from the measured 
structure functions $A$ and $B$, and tensor polarization
$T_{20}$. The expressions relating these measured quantities with
the form factors can be found in Ref.~\cite{Carlson98}.

\begin{figure}
\begin{minipage}[t]{0.35\textheight}
\includegraphics[height=.24\textheight,angle=0,keepaspectratio=true]{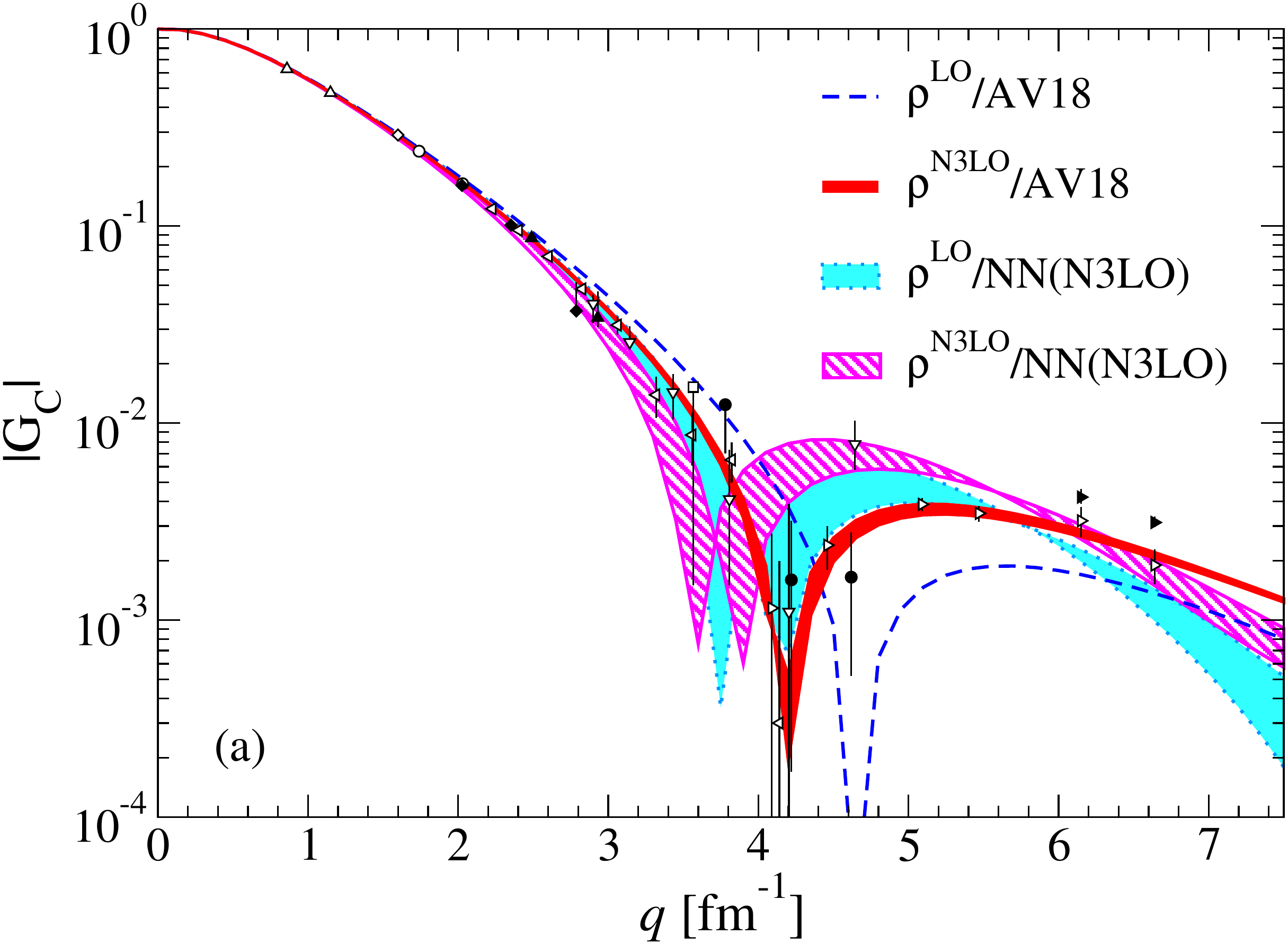}
\end{minipage}
\begin{minipage}[t]{0.35\textheight}
\includegraphics[height=.24\textheight,angle=0,keepaspectratio=true]{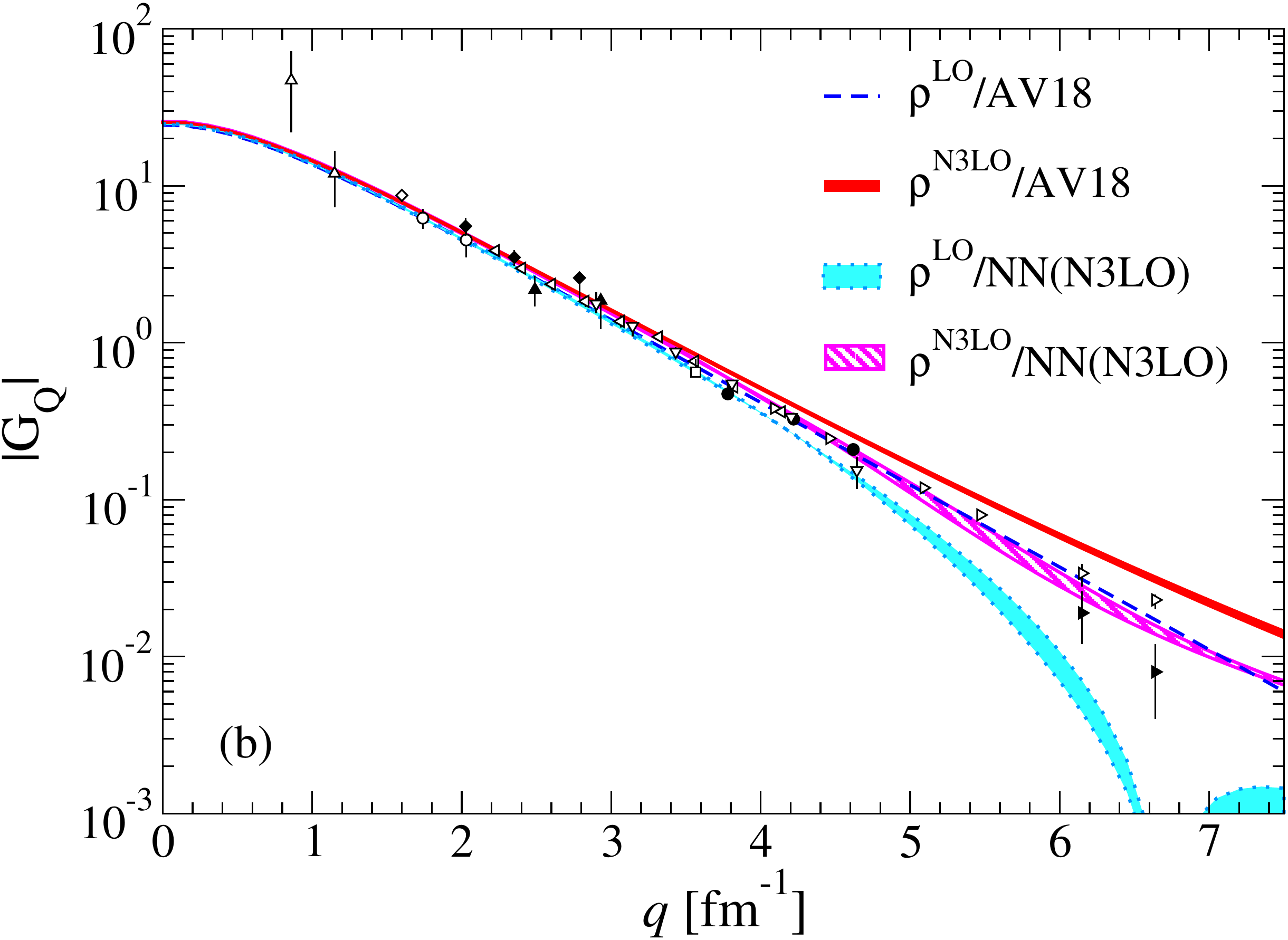}
\end{minipage}
\begin{minipage}[t]{0.35\textheight}
\includegraphics[height=.24\textheight,angle=0,keepaspectratio=true]{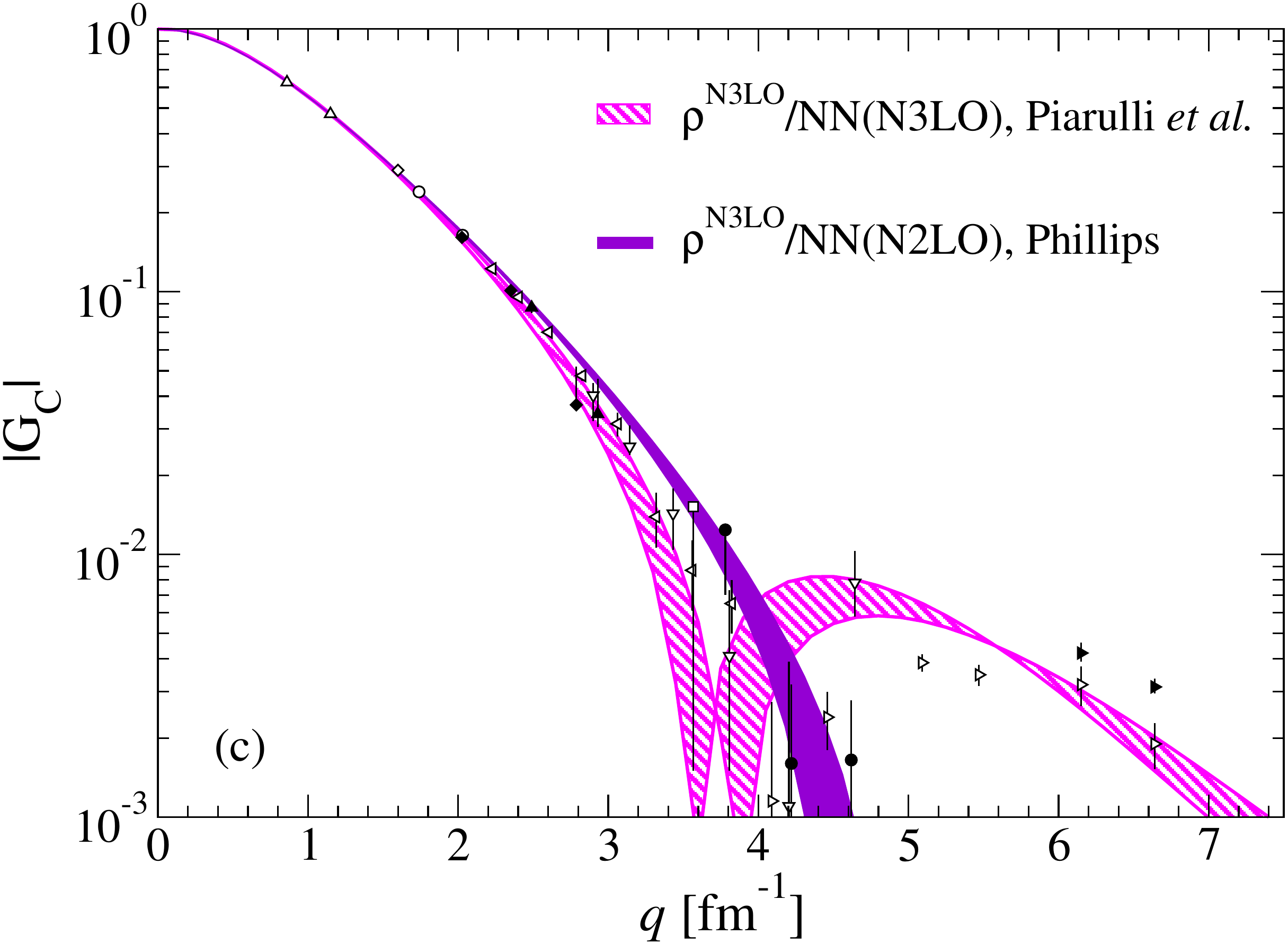}
\end{minipage}
\begin{minipage}[t]{0.35\textheight}
\includegraphics[height=.24\textheight,angle=0,keepaspectratio=true]{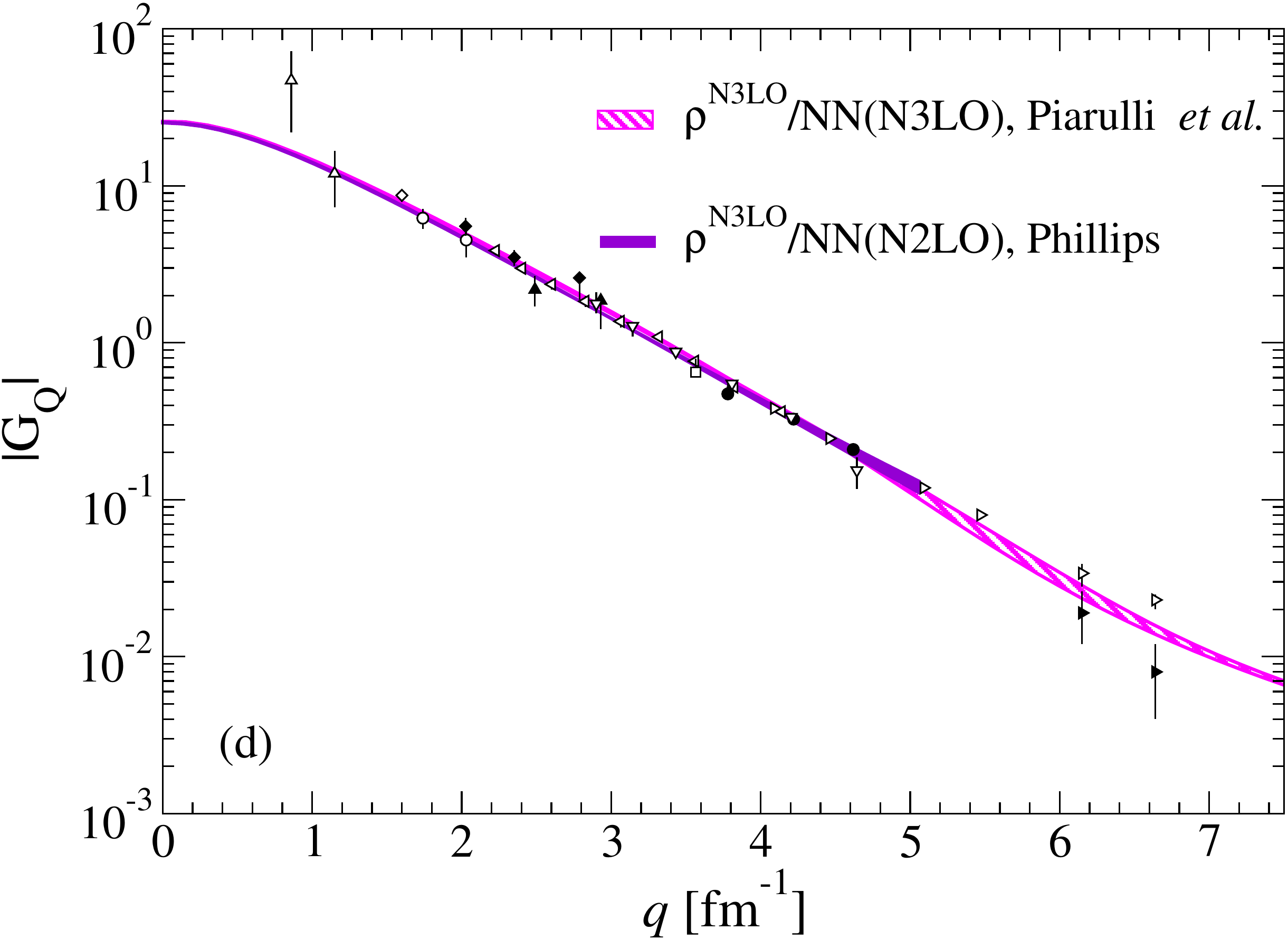}
\end{minipage}
\caption{ (Color online) Deuteron charge, panel $(a)$, and quadrupole, panel $(b)$, form factors from Ref.~\cite{Piarulli12}
compared with experimental data from
Refs.~\cite{STANFORD65,ORSAY66,CEA69,DESY71,MONTEREY73,SLAC75,
MAINZ81,BONN85,SACLAY90,JLABHALLA99,JLABHALLC99,BATES84,BATES91,
BATES2011,VEPP85,VEPP86,VEPP90,VEPP2003,
BONN91,NIKHEF96,NIKHEF99,JLAB2000}.
Cutoffs $\Lambda$ in the range 500--600 MeV are displayed by the bands.
In panels $(c)$ and $(d)$, results from Ref.~\cite{Piarulli12} (magenta hatched bands) are compared with
those obtained in Ref.~\cite{Phillips07} (purple solid bands). See  text for more details.} 
\label{fig:gc-d}
\end{figure}

In Figure~\ref{fig:gc-d}, the calculated deuteron charge and quadrupole form factors 
from Ref.~\cite{Piarulli12} are compared with the experimental data in panels 
$(a)$ and $(b)$, respectively. In these figures, results from both a $\chi$EFT and
a hybrid calculation based on the $\chi$EFT charge operator at LO ($\nu=-3$) and N3LO
($\nu=0$) are shown. The cyan dotted band (blue dashed line) and magenta hatched (red solid)
band are obtained using deuteron wave functions from the N3LO ($\nu=4$) chiral NN potential by Entem and Machleidt~\cite{Entem03}
(AV18 potential), in combination with the charge operator at LO ($\nu=-3$) and N3LO ($\nu=0$), respectively.
%The blue dashed line and red solid band represent hybrid calculations obtained with
%the AV18 potential and with the charge operator at LO ($\nu=-3$) and N3LO ($\nu=0$), respectively.
The thickness of the bands indicates the sensitivity of the results with respect to two values
of the regularization cutoff corresponding to $\Lambda=500$ and $600$ MeV (used consistently
in the chiral NN potential and e.m. currents, for the $\chi$EFT calculation). The IA and IA(RC)
operators at LO and N2LO---see panels $(a)$ and $(b)$ of Figure~\ref{fig:chi_charge}---are
cutoff independent, therefore  hybrid results at LO are represented by a line,
while the thickness of the cyan dotted band associated with the $\chi$EFT results at LO 
is due to variations in the cutoff utilized to regularize
the chiral NN potential.

The charge and quadrupole form factors are determined by the isoscalar
component of the charge operator. Therefore, the N4LO ($\nu=1$) loop-corrections,
illustrated in panels $(f)$--$(o)$ of Figure~\ref{fig:chi_charge}, do not contribute to
these observables, as they lead to isovector operator structures. 

We emphasize that the results discussed here include e.m.~nucleonic
form factors---that is the $G_{E/M}^{S/V}$ introduced in Eqs.~(\ref{eq:e_ff})
and~(\ref{eq:m_ff}). As noted in  Ref.~\cite{Walzl01},
a simple chiral expansion of the charge operator provides  poor agreement with
the experimental data for $q\gtrsim 1.5$ fm$^{-1}$.
The inclusion of e.m.~nucleonic form factors taken from experimental data ensures a reasonable fall-off
of the calculated deuteron form factors at increasing values of $q$. 
In doing so, a very good agreement between the full theory and experiment is observed
up to values of $q\sim3$ fm$^{-1}$ in the case of the charge form factor,
and of $q\sim6$ fm$^{-1}$ in the case of the quadrupole form factor.
Hybrid and $\chi$EFT results are in very good agreement for low values of momentum 
transfered ($q\lesssim 3$ fm$^{-1}$), however, 
as $q$ increases these observables become sensitive to
the wave functions utilized in the computations, 
which display some differences (see, {\it e.g.}, Figure 16 of Ref.~\cite{Machleidt11}).
This can be appreciated by comparing the calculations based on the 
N3LO ($\nu=0$) e.m.~charge operators,  with the chiral N3LO ($\nu=4$)~\cite{Entem03} 
(magenta hatched band) and the AV18~\cite{AV18} (solid red band) NN potentials.
The diffraction region of $G_C$ is better reproduced by the AV18 calculation, 
while the $\chi$EFT results for $G_Q$ are in better agreement with the experimental 
data for $q\gtrsim 3$ fm$^{-1}$.

In the bottom panels of Figure~\ref{fig:gc-d}, 
we compare the full $\chi$EFT results by Piarulli {\it et al.}~\cite{Piarulli12} (magenta hatched band) 
with those obtained by Phillips in Refs.~\cite{Phillips03,Phillips07} (purple solid band).
The latter employ the chiral N2LO ($\nu=3$) NN interaction from Ref.~\cite{Epelbaum05}
and the charge operator at N3LO ($\nu=0$)\footnotemark[1]\footnotetext{Note that the power counting
used in Ref.~\cite{Phillips07} is slightly different from the one we use.
In particular, in the counting of Ref.~\cite{Phillips07},
relativistic corrections to the LO charge operator lead to a one-body operator that is of order N4LO,
which is, therefore, neglected.}. These
calculations are in good agreement at low values of momentum transfer and exhibit a similar
cutoff dependence. However, as $q$ increases the predicted 
form factors exhibit a more noticeable sensitivity to the particular NN potential 
utilized to solve the Schr\"odinger equation (a finding that confirms the conclusions
of previous studies by Phillips~\cite{Phillips03,Phillips07}).

We also report that the $\chi$EFT (AV18) calculated deuteron quadrupole moment, $Q_d$, inclusive of MEC,
is found to be within $1\%$ ($2\%$) of the experimental value, and to show a 
$\sim 1\%$ (negligible) variation in going from $\Lambda=500$ to $600$ MeV~\cite{Piarulli12}.

We now turn our attention to the deuteron magnetic form factor. 
Because the deuteron is isoscalar, only isoscalar components in the chiral e.m.~current
contribute to its magnetic form factor. Therefore, OPE and TPE currents
at NLO ($\nu=-1$)---see panels $(b)$ and $(c)$ of Figure~\ref{fig:chi_cnt}---and N3LO ($\nu=1$)---see
panels $(e)$--$(i)$ and $(l)$--$(o)$ of the same figure---do not  contribute to this observable.
The isoscalar e.m.~current operator involves only two e.m.~LECs both entering at N3LO ($\nu=1$);
one accompanies a short-ranged contact interaction---see panel $(k)$ of Figure~\ref{fig:chi_cnt}---and
the other one multiplies the isoscalar part of the tree-level current illustrated in panel $(j)$
of Figure~\ref{fig:chi_cnt}.  We note that isoscalar contributions
of one-pion range are suppressed by three powers of $Q$ with respect to the LO ($\nu=-2$) or IA
term.

\begin{figure}
\begin{minipage}[t]{0.35\textheight}
\includegraphics[height=.24\textheight,angle=0,keepaspectratio=true]{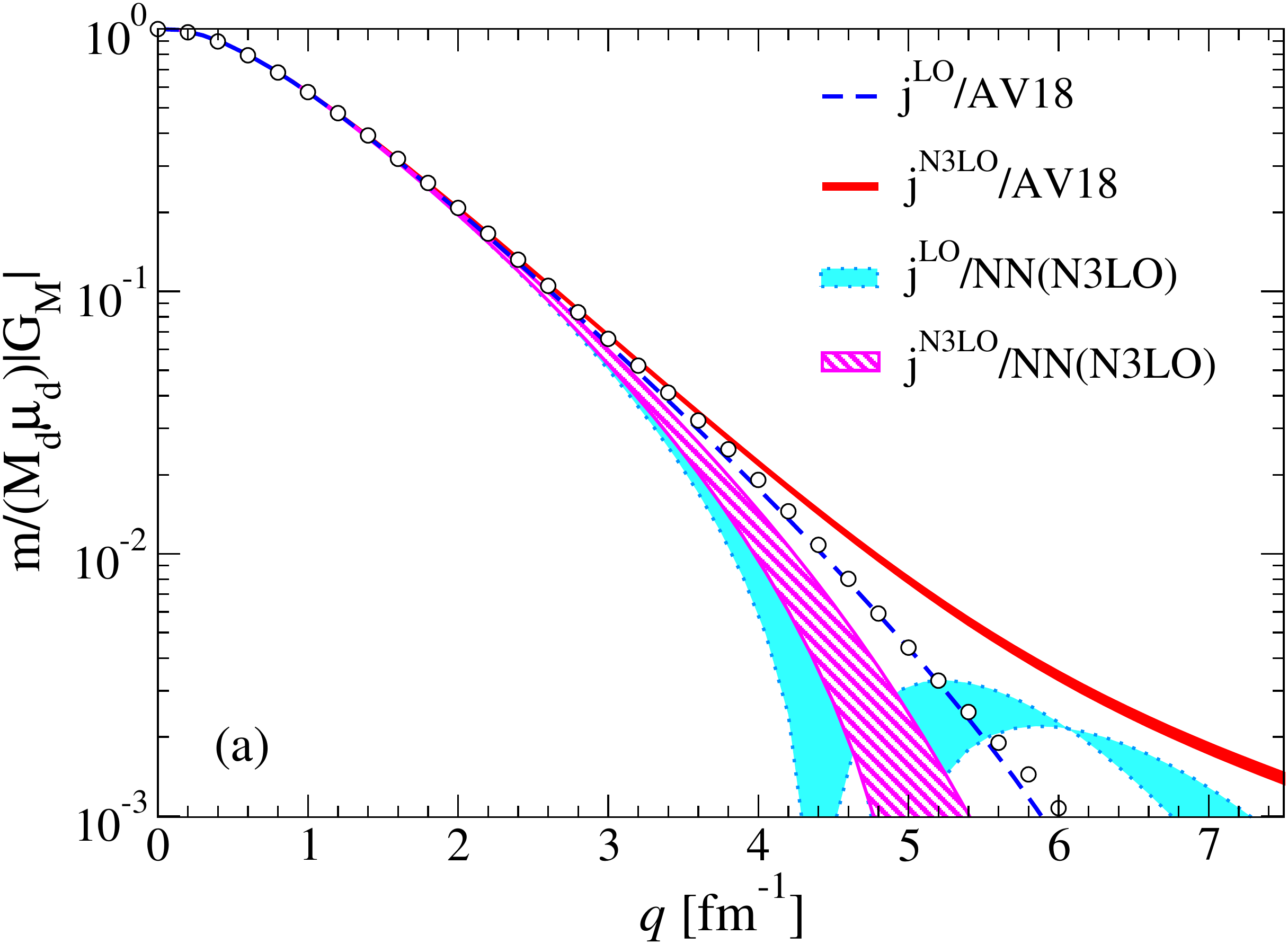}
\end{minipage}
\begin{minipage}[t]{0.35\textheight}
\includegraphics[height=.24\textheight,angle=0,keepaspectratio=true]{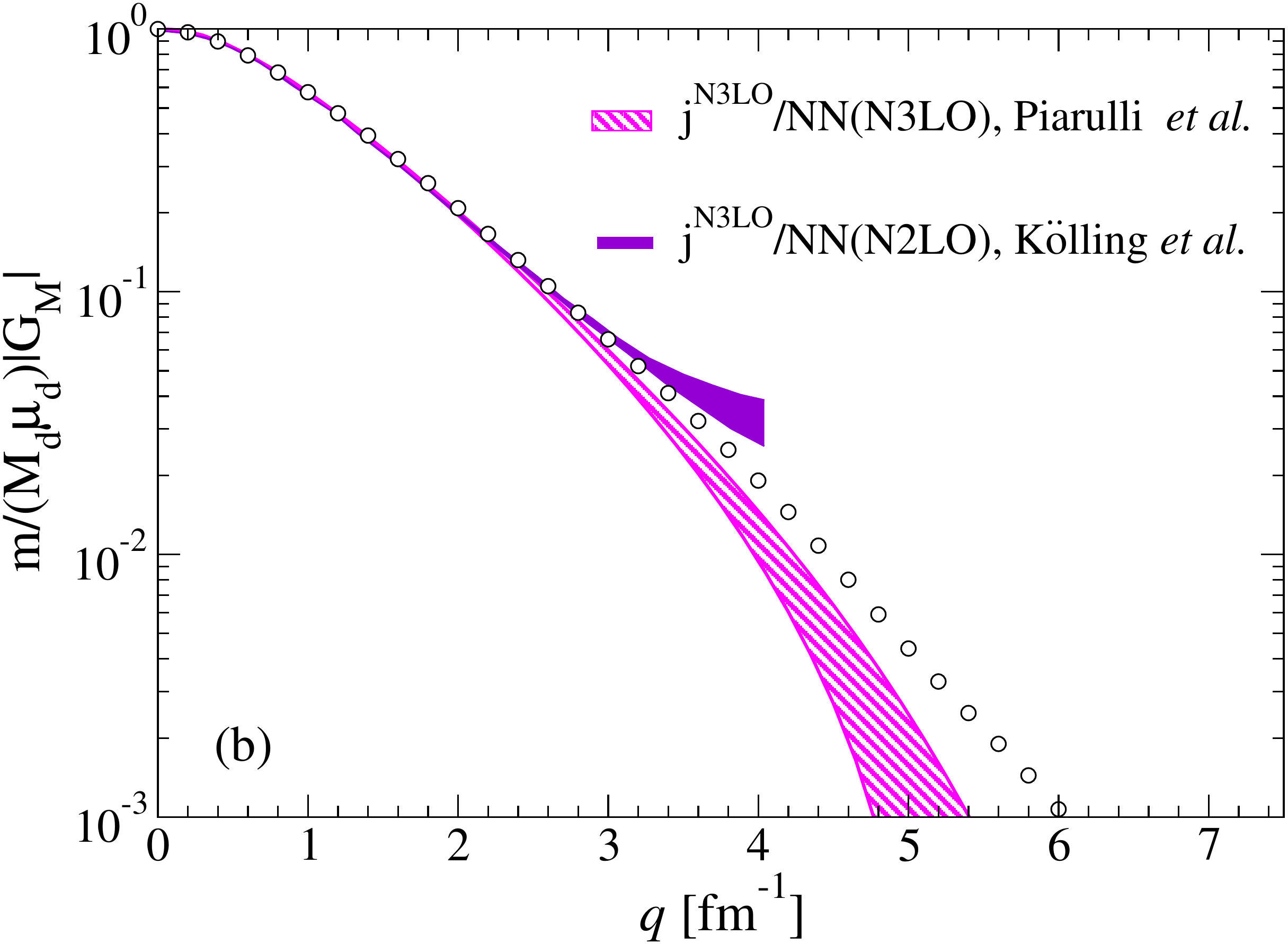}
\end{minipage}
\caption{(Color online) Panel $(a)$: The deuteron magnetic form factor $G_M(q)$ from Ref.~\cite{Piarulli12}
%, obtained at leading order (LO) and
%with inclusion of current operators up to N3LO, based on the N3LO NN potential from Refs.~\cite{Entem03}
 compared with the experimental data
from Refs.~\cite{STANFORD65,MAINZ81,SACLAY85,BONN85,SLAC90,SICK2001}.
Color code as in Figure.~\ref{fig:gc-d}.
 Panel $(b)$: Calculated deuteron magnetic form factor from
Ref.~\cite{Piarulli12} (magenta hatched band) compared with
that obtained in Ref.~\cite{Kolling12} (purple solid band), based on the N2LO NN interaction
from Ref.~\cite{Epelbaum05} and current operators at N3LO.} \label{fig:gm-d}
\end{figure}

In panel $(a)$ of Figure~\ref{fig:gm-d}, we show the calculated deuteron magnetic form 
factor from Ref.~\cite{Piarulli12}, with the same color code as in Figure~\ref{fig:gc-d}.
Here, the e.m.~LECs entering the isoscalar current at N3LO ($\nu=1$)
have been constrained so as to reproduce the deuteron magnetic moment, $\mu_d$,
as well as the isoscalar combination of the trinucleon magnetic moments, $\mu^S$,
defined in Eq.~(\ref{eq:muSmuV}). Therefore, there is no prediction for the deuteron
static magnetic moment with this chiral e.m.~current. LECs of minimal nature that multiply
contact operators at N3LO ($\nu=1$) have been taken from the phase-shift analysis
carried out in Refs.~\cite{Entem03,Machleidt11}.
We observe a rather good agreement with the experimental data for values
of $q\simeq 2$ fm$^{-1}$. However, at larger values of momentum transferred the hybrid and $\chi$EFT
results are quite different. In particular, calculations based on the AV18 interaction and e.m.~operators 
at LO ($\nu=-2$) do not show the diffraction pattern observed in the $\chi$EFT results at LO ($\nu=-2$). This
is due to the different deuteron wave functions generated by the two corresponding NN potentials.
We remark that the N3LO ($\nu=1$) e.m.~current is conserved with the NN potential at NLO ($\nu=2$),
therefore, the $\chi$EFT calculation based on the NN potential at N3LO ($\nu=4$) would require
an e.m.~current of order $\nu=3$ in order to strictly fulfill the continuity equation. 

In panel $(b)$ of Figure~\ref{fig:gm-d}, we compare the results by Piarulli 
{\it et al.}~\cite{Piarulli12} (hatched magenta band) with
the fully consistent $\chi$EFT calculations by K\"{o}lling {\it et al.}~\cite{Kolling12} 
(solid purple band) based on the chiral NN potential at N2LO ($\nu=3$)~\cite{Epelbaum05} 
and chiral e.m.~currents at N3LO ($\nu=1$) ~\cite{Kolling09,Kolling11}, both derived within 
the unitary transformation method.
In that work, the short-ranged e.m.~LEC---see panel $(k)$ of Figure~\ref{fig:chi_cnt}---is fixed
to the deuteron magnetic moment, while the isoscalar OPE e.m. LEC---see panel $(j)$ of Figure~\ref{fig:chi_cnt}---is 
fitted to the measured B structure function~\cite{Kolling12}. 
The theoretical results for the deuteron magnetic form factor are in very good agreement
with each other and with the experimental data for values of momentum transferred $q\simeq 3$ fm$^{-1}$, 
and present a comparable cutoff dependence.
% Two-body effects at
%N3LO ($\nu=1$) are seen to become comparable to the LO ($\nu=-2$) prediction at $q\simeq3$ fm$^{-1}$,
%which is where the chiral expansion is conjectured to break down.
% A very good agreement
%between these calculations is observed, along with a comparable cutoff dependence.

Finally, we report on a very recent evaluation of deuteron $G_C$, $G_Q$, and $G_M$ form factors
carried out in a fully consistent and cutoff independent $\chi$EFT formulation~\cite{Epelbaum13c}.
The calculation, based on the renormalizable approach of Ref.~\cite{Epelbaum12c}, uses e.m.~charge, e.m.~currents, 
and NN operators at LO, and predicts deuteron form factors which are in good agreement
with the experimental data, provided that the nucleonic e.m.~structure is folded in the
evaluation via appropriate nucleonic form factors.  

\subsubsection{The three-body nuclei}
\label{sec:a3-ffs}

$-\, \, \,\,$ Currently, the only complete study on elastic form factors of nuclei with mass number $A=3$
that uses $\chi$EFT potentials at N3LO ($\nu=4$), N3LO ($\nu=1$) e.m.~currents and N4LO ($\nu=1$)
charge operators has been carried out by Piarulli {\it et al.} in Ref.~\cite{Piarulli12}.
In that work, the calculations of the relevant matrix elements
have been performed using wave functions obtained with the 
HH expansion method carried out in momentum space. 
%developed in Refs.~\cite{Kievsky97,Kievsky01,Viviani05,Kievsky08}.
The e.m.~current and charge operators of Refs.~\cite{Pastore08,Pastore09,Pastore11,Piarulli12}
have been sandwiched in between wave functions obtained from the
NN(N3LO)+3N(N2LO) nuclear Hamiltonian, where the two-body interaction
is from Ref.~\cite{Entem03}, while the 3N force has been taken at N2LO ($\nu=3$)
with corresponding LECs constrained as in Ref.~\cite{Ga_beta_triton}. 
\begin{center}
\begin{figure}
\centering
\includegraphics[width=5.5in]{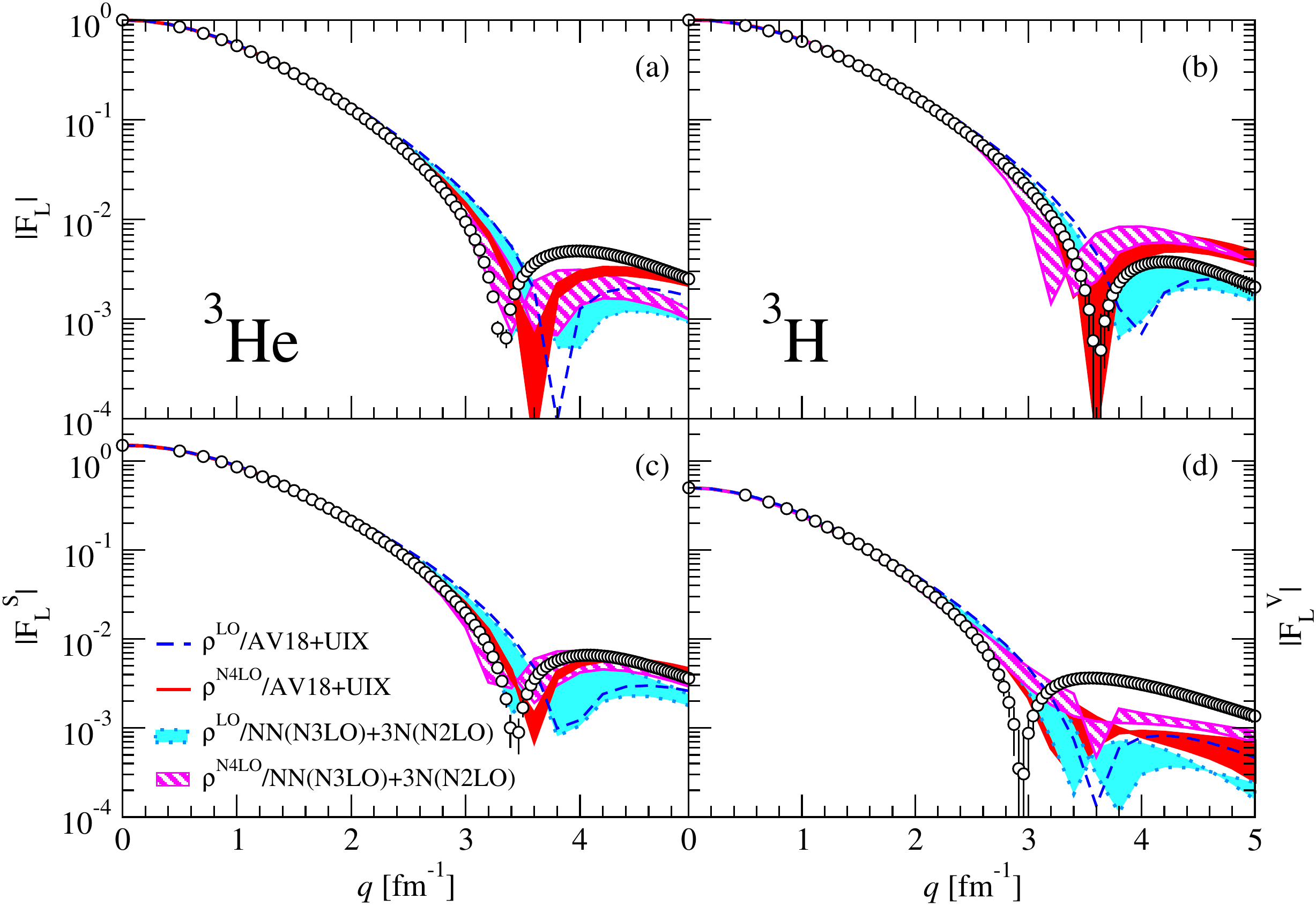}\\
\caption{(Color online) The $^3$He $(a)$ and $^3$H $(b)$ charge form factors,
and their isoscalar $(c)$ and isovector $(d)$ combinations from Ref.~\cite{Piarulli12},
%obtained with the LO charge operator, and with the inclusion of $\chi$EFT charge operators up to N4LO, are
compared with experimental data by Amroun {\it et al.}~\cite{AMROUN} (empty circles). Same color code as in Figure~\ref{fig:gc-d}.
 Predictions corresponding to cutoffs $\Lambda$ in the range (500--600) MeV are displayed by the bands.}
\label{fig:ff-charge-a3}
\end{figure}
\end{center}

In Figure~\ref{fig:ff-charge-a3},
the charge form factors ($F_C$) of $^3$He and $^3$H are represented in the 
upper panels, while the bottom panels show
the isoscalar ($F_C^S$) and isovector ($F_C^V$) combinations
defined as  
\begin{equation}
\label{eq:muSmuV}
 F^{S,V}_C(q) = \frac{1}{2} [2 F_C(q,^3{\rm He}) \pm  F_C(q,^3{\rm H})]  \ ,
\end{equation}
and normalized, respectively,  to 3/2 and 1/2 at $q=0$. The $\chi$EFT charge operator,
schematically illustrated in Figure~\ref{fig:chi_charge},
does not involve unknown LECs up to N4LO. Results obtained with the 
charge operator at LO (N4LO) are represented by 
the cyan dotted (magenta hatched) band, and their thickness represent the spread in the 
calculations corresponding to a variation of the regularization cutoff ($\Lambda=500$--$600$ MeV).
The latter has been varied consistently in both the potentials and the currents.
Hybrid results, obtained with wave functions from the AV18+UIX %~\cite{AV18,Pudliner95}
nuclear Hamiltonian,
are also given in Figure~\ref{fig:ff-charge-a3}. In particular,  those based on the charge operator
at LO (N4LO) are illustrated by the blue dashed line (red solid band). Note that one-body
operators at LO are cutoff independent, which is why hybrid results at LO
are given by a line. The calculated charge form factors
are in excellent agreement with the experimental data for values 
of $q<3$ fm$^{-1}$. However, the  positions of the zeros, as well as those 
of the maxima at $q\sim4$ fm$^{-1}$, are not well reproduced by the theory. 
The results displayed in the bottom panels of Figure~\ref{fig:ff-charge-a3}
suggest that larger (in magnitude) two-body isovector contributions to the e.m.~charge operator are
needed to reproduce the first zero in the isovector form factor, $F_C^V$.

\begin{center}
\begin{figure}
\centering
\includegraphics[width=5.5in]{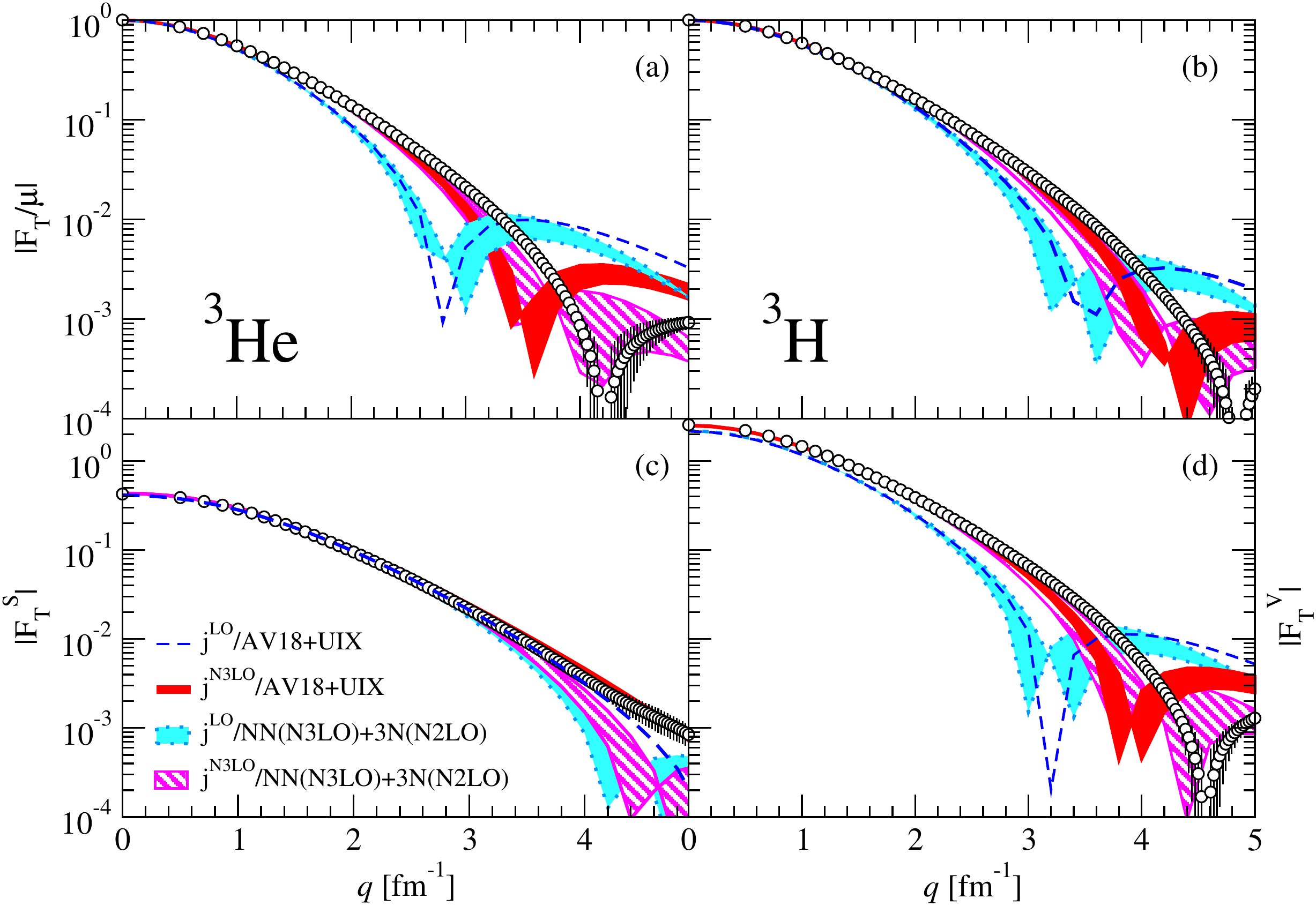}\\
\caption{(Color online) The $^3$He $(a)$ and $^3$H $(b)$ magnetic form factors,
and their isoscalar $(c)$ and isovector $(d)$ combinations from Ref.~\cite{Piarulli12},
obtained with the LO current operator,  
and with the inclusion of $\chi$EFT current operators up to N3LO, are compared with
experimental data by Amroun {\it et al.}~\cite{AMROUN}. 
 Same color code as in Figure~\ref{fig:gc-d}.
 Predictions relative to cutoffs
$\Lambda$ in the range (500--600) MeV are displayed by the bands.
}
\label{fig:ff-mag-a3}
\end{figure}
\end{center}

The trinucleon magnetic form factors are given in Figure~\ref{fig:ff-mag-a3}.
Minimal LECs entering the N3LO contact currents---see panel $(k)$ of
Figure~\ref{fig:chi_cnt}---have been taken from fits to NN scattering data~\cite{Entem03,Machleidt11}.
The LECs entering the isovector component of the $\chi$EFT e.m.~current at N3LO, 
illustrated in panel $(j)$ of Figure~\ref{fig:chi_cnt}, have been saturated by the 
$\Delta$-isobar. The LEC associated with the isoscalar component of the
same current, along with the LEC associated with the isoscalar N3LO contact current of non-minimal
nature (see diagram in panel $(k)$ of Figure~\ref{fig:chi_cnt}) have been determined by
reproducing the deuteron magnetic moment as well as the isoscalar combination of the
trinucleon magnetic moments, $\mu_S=[\mu(^3{\rm He})+\mu(^3{\rm H})]/2$.
The remaining LEC, which multiplies an isovector non-minimal contact current at N3LO (see 
panel $(k)$ of Figure~\ref{fig:chi_cnt}) is fixed so as to reproduce the isovector combination of
the trinucleon magnetic moments, $\mu_V=[\mu(^3{\rm He})-\mu(^3{\rm H})]/2$.
As before, in Figure~\ref{fig:ff-mag-a3}, the top panels show the  $^3$He and $^3$H magnetic
form factors, while the bottom ones show the isoscalar ($F_M^S$) and isovector ($F_M^V$) combinations
given by
\begin{equation}
 F^{S,V}_M(q) = \frac{1}{2} [\mu(^3{\rm He}) F_M(q,^3{\rm He}) \pm  \mu(^3{\rm H}) F_M(q,^3{\rm H})]  \ ,
\end{equation}
and normalized, respectively, to $\mu_S$ and $\mu_V$ at $q=0$. In Figure~\ref{fig:ff-mag-a3}, 
the color code is as in Figure~\ref{fig:ff-charge-a3}, the only difference being that the
vector e.m.~currents include up to N3LO ($\nu=1$) corrections.
As it is well know from studies based on the conventional approach~(see Ref.~\cite{Carlson98}),
two-body e.m.~currents are crucial to improve the agreement between the observed positions
of the zeros and the predicted ones at LO (or IA). Despite the excellent agreement between 
theory and experiment for $q\leq2$ fm$^{-1}$, the theory underpredicts the data at higher momentum transfers, while the
zeros are found at lower values of $q$ than observed. 
The theoretical description of the first diffraction region is still incomplete. This is a finding
that confirms the conclusions of previous studies based on the conventional approach~\cite{Carlson98,Marcucci98,Marcucci05,Golak05}. 
It is interesting to point out that, in Ref.~\cite{Piarulli12}, different LEC parameterizations
have been investigated. In particular, it has been found that if one fixes
the isovector LEC to the  $np$ radiative capture cross section at thermal neutron energies,
as opposed to $\mu_V$, the $\chi$EFT
calculations of the magnetic form factors would lead to significantly better agreement with
data over the whole range of momentum transfers, while overestimating the observed $\mu_V$ by $\simeq 3$\%
(see Figure~\ref{fig:npdg_trimm} and associated discussion).

\subsubsection{The $^4$He nucleus}

$-\, \, \,\,$  {\it Ab-initio} calculations of the $^4$He
elastic charge form factor have been performed by Schiavilla and collaborators~\cite{Schiavilla90}  
and reported in the 
review article of Ref.~\cite{Carlson98}. Since then, improved
nuclear wave functions 
from  conventional  nuclear interactions, as well as calculations
from  chiral potentials, 
have appeared in the literature.  

\begin{center}
\begin{figure}[htb]
\centering
\includegraphics[width=9cm,clip]{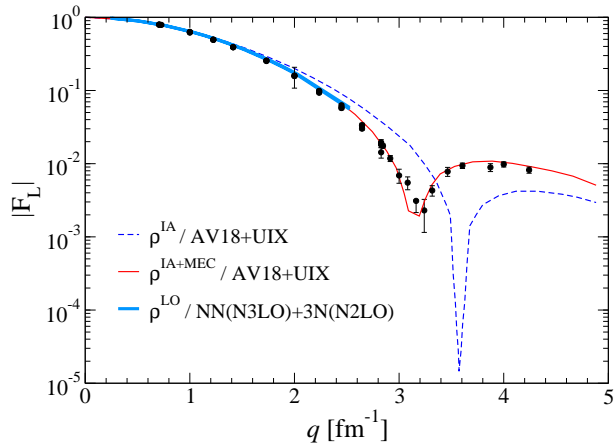}
\caption{(Color online) Calculated $^4$He elastic charge form factor in IA (blue dashed line) and with MEC (red solid line)
and nuclear wave functions from the AV18+UIX Hamiltonian from Ref.~\cite{Michele_He4_FF},
compared with the IA (or LO) calculation (cyan band) with the chiral NN(N3LO)/3N(N2LO) potential 
from Ref.~\cite{Bacca:2012xv}.   Experimental data are from Frosch {\it et al.}~\cite{FrM67}.}
\label{fig-el_he4}       
\end{figure}
\end{center}

Results from Viviani {\it et al.}~\cite{Michele_He4_FF} are presented in Figure~\ref{fig-el_he4}
along with the experimental data  by Frosch {\it et al.}~\cite{FrM67}.
Wave functions have been obtained using the
HH method with the AV18+UIX nuclear Hamiltonian and consistent conventional MEC.
The  calculation in IA (blue dashed line) deviates from 
the experimental data for values of momenta larger than $q\sim2$ fm$^{-1}$, while the
agreement with the experiment is restored once MEC are included (red solid line). The effect
of two-body currents becomes more pronounced as $q$ increases to larger values. In particular, 
two-body corrections lead to a very different diffraction minimum than that observed in the
IA results, and in agreement with experiment. 

In Ref.~\cite{Bacca:2012xv}, an IA calculation with the AV18+UIX Hamiltonian has been performed
using the EIHH method,  a different formulation of hyperspherical harmonics.
A perfect agreement between results from these calculations (not shown in Figure~\ref{fig-el_he4}, 
but displayed in Ref.~\cite{Bacca:2012xv}) and the IA results by
Viviani {\it et al.}~\cite{Michele_He4_FF} has been obtained, indicating that a high level of accuracy has been reached in
benchmarking few-body nuclei calculations, especially those of ground-state properties.

Also shown in Figure~\ref{fig-el_he4} is an IA (or LO) calculation (cyan band) from Ref.~\cite{Bacca:2012xv}
obtained from $\chi$EFT potentials within the EIHH method. 
The chiral Hamiltonians have been taken to consist of an NN force at N3LO ($\nu=4$)~\cite{Entem03} and a 3N force at N2LO
($\nu=3$), with LECs parameterized as in Ref.~\cite{Navratil07b} (choosing one LEC of natural value and fitting the other so as to reproduce the binding energy in $A=3$ nuclei) and as in Ref.~\cite{Ga_beta_triton} (fitting on the triton binding energy and empirical Gamow-Teller matrix element)
at fixed cutoff $\Lambda=500$ MeV. The curves corresponding to the two above mentioned chiral Hamiltonians overlap 
in the figure and give rise to the thin band.
At very low momentum,  the conventional and $\chi$EFT approaches agree with each other. 
This is not very surprising, because they both use the same charge operator. 
Furthermore, conventional and chiral Hamiltonians  give very similar results for the $^4$He radius,
which determines the shape of the  elastic form factor at low-$q$.
In fact, the point-proton radius is 1.432(2) fm~\cite{gazit2006} (1.464(2) fm~\cite{Bacca:2012xv})
when the AV18+UIX (NN(N3LO)+3N(N2LO)) nuclear Hamiltonian
is used. These calculated values are both close to the charged radius measured from elastic electron scattering, once 
the finite proton and neutron size are subtracted, yielding to
$1.463(6)$ fm  (see also Ref.~\cite{Brodeur} for more details). 

Finally, we remind that new JLab data have been recently published in Ref.~\cite{Camsonne13}, where both experiment and 
conventional theory have been extended up to higher values of momentum transfer.

\subsubsection{Nuclei with $A>4$}

$-\, \, \,\,$ There are few {\it ab-initio} calculations of elastic
form factors for $A>4$ nuclear systems. Elastic charge and magnetic 
form factors of $^6$Li have been calculated in Ref.~\cite{Schiavilla_Li6_PRL}
from VMC nuclear wave functions obtained using the AV18+UIX nuclear Hamiltonian,
and conventional MEC. 
In particular,
the charge form factor has been found to be in excellent agreement with
the experimental data, with conventional two-body corrections from the $\pi$-like 
charge operator improving on the IA results. The calculated transverse
form factor, instead, has been found to deviate from the experimental data 
beyond values of momentum transfer $q\sim 1$ fm$^{-1}$.  Appreciable MEC effects for
this isoscalar observable have been found for values of momentum transferred
larger than 3 fm$^{-1}$, that is beyond the range of the available experimental
data~\cite{Carlson98,Schiavilla_Li6_PRL}. The discrepancy with the experimental
data has been  attributed to a possible insufficient accuracy of the VMC nuclear wave functions,
which could be resolved using improved GFMC evolutions~\cite{Carlson98,Schiavilla_Li6_PRL}.
\begin{center}
\begin{figure}
\centering
\includegraphics[width=9cm]{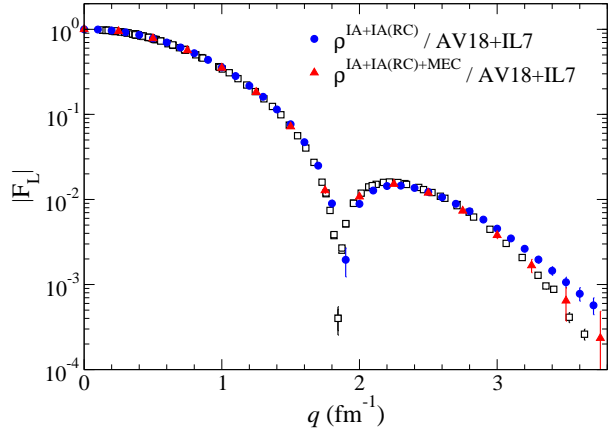}\\
\caption{(Color online) GFMC calculations from Ref.~\cite{Lovato13} of the  $^{12}$C elastic charge form
factor obtained with the AV18+IL7 Hamiltonian in IA, comprehensive of relativistic corrections,
(blue dots) and with MEC (red dots), compared with experimental data by de Vries {\it et al.}~\cite{DeVries87} 
(empty squares). Monte Carlo statistical errors are also shown. }
\label{fig:fel_C12}
\end{figure}
\end{center}

A comprehensive and computational demanding GFMC calculation of the $^{12}$C elastic
charge form factor has recently appeared in the literature~\cite{Lovato13}.
In that work, Lovato {\it et al.} have used the AV18+IL7 nuclear Hamiltonian
and conventional MEC. The calculated $^{12}$C values of the ground-state energy
and rms charge radius obtained with this Hamiltonian are -93.3(4) MeV and
2.46(2) fm, respectively, to be compared with the experimental values of
-92.16 MeV and 2.471(5) fm. 
In Figure~\ref{fig:fel_C12}, the  calculated $^{12}$C charge form factor~\cite{Lovato13} is shown along 
with the experimental data from Ref.~\cite{DeVries87}. Here, the one-body charge operator includes
relativistic corrections (RC) and the calculated form factor obtained from it is indicated by the blue circles,
which are in very good agreement with the experimental data for values of momentum
transferred $q \lesssim3$ fm$^{-1}$. Two-body charge operators used  in the calculations include 
$\pi$-like and $\rho$-like exchange terms, as well as a $\rho \pi \gamma$ transition current
(see Sec.~\ref{subsec:SNPA} for more details on conventional MEC operators).
The full calculation with MEC (red triangles) provides an improved description of the
data in the $q\ge3$ fm$^{-1}$ region.

Since the review by Carlson and Schiavilla~\cite{Carlson98}, thanks to 
the tremendous advancement of many-body computational techniques, accurate
nuclear wave functions are now available for up to $A\sim20$ nuclear
systems. This opens up the possibility of studying the charge and magnetic
spatial distributions of more complex nuclear systems. In addition, 
chiral potentials and e.m.~currents are now available and in the future studies,
such as those of Refs.~\cite{Schiavilla_Li6_PRL,Lovato13}, can be
performed with $\chi$EFT and extended to other nuclei.
 Effects of two-body e.m.~currents
have been proved to be significant and crucial for an accurate 
theoretical description of the available experimental data. It would also be
interesting to study their contributions in neutron-rich nuclei
that display halo features, such as, {\it e.g.}, the $^6$He nucleus.
Electron scattering experiments on unstable nuclei have not been carried out yet.
However, plans to measure charge distributions of neutron-rich nuclei
have been presented at both European (FAIR) and Japanese (RIKEN) facilities.
Such electron scattering reactions on unstable nuclei, to be performed
in storage rings, can potentially reveal new interesting nuclear structure
properties, for which the  community can provide theoretical guidance due
to the high level of reliability that is being reached by {\it ab-initio} calculations.

\subsection{Inelastic scattering}
\label{sec:e-inelastic}

The electron scattering off a nucleus is inelastic when both momentum ${\bf q}$
and energy $\omega$ are transferred to the target. In this case, the 
final state of the nucleus is different from the initial state
(typically the ground state), {\it i.e.}, $\left| \Psi_f  \right\rangle \ne \left| \Psi_i  \right\rangle$.
In the case that only the scattered electron is measured and no specific final hadronic state is selected,
one obtains an inclusive unpolarized double differential cross section.
In the usual one-photon-exchange approximation and
for ultrarelativistic electrons, such cross section in the laboratory is\footnotemark[1]\footnotetext{Note that, with respect to Eq.~(\ref{cross_el}), here a different convention for the charge is used, namely $\frac{e^2}{\hbar c}$.} 
\begin{equation}
\frac{d^2 \sigma}{d\Omega_e d\omega}=\sigma_M \left[\frac{Q_\mu^4}{q^4} R_L(\omega, {\bf q} )
+\left(\!\frac{Q_\mu^2}{2q^2}+\tan^2{\frac{\theta_e}{2}}\! \right)R_T(\omega, {\bf q})\right]\,,
\label{cross_inel}
\end{equation}
where the Mott cross section, $\sigma_M$, has been defined in  Eq.~(\ref{eq:mott}).
The dynamical quantities of interest are the longitudinal ($L$) and transverse ($T$) response functions, defined as
\begin{eqnarray}
\nonumber
R_L(\omega,{\bf q} )&= &\int \!\!\!\!\!\!\!\sum_f\left| \left\langle \Psi _{f}\right| {\rho}({\bf q})\left| \Psi _{0}\right\rangle \right| ^{2}\delta \left(E_{f}-E_{0}-\omega+ \frac{{\bf q}^2}{2M_A}\right)\\
R_T(\omega,{\bf q })&=&\int  \!\!\!\!\!\!\! \sum_f \!\! \sum_{\lambda=\pm1}\left| \left\langle \Psi _{f}\right| {J}_{\lambda}({\bf q})\left| \Psi _{0}\right\rangle \right| ^{2}\delta \left(E_{f}-E_{0}-\omega + \frac{{\bf q}^2}{2M_A} \right),
\label{lt_responses}
\end{eqnarray}
where the recoil energy in the $\delta$-function includes the mass of the target $M_A$. Here the symbol $\int \!\!\!\!\!\!\!\sum_f$ indicates a sum over the final states of the nucleus, including both the finite (sum) and continuum  (integral) ones. The longitudinal and transverse response functions
are obtained from the charge  and transverse current operator, respectively.
The longitudinal part can be disentangled from the transverse one by using the Rosenbluth separation
method~\cite{Rosenbluth}. Because this procedure is based on the one-photon-exchange assumption, 
it is valid only for light nuclei. Indeed, data from different laboratories (mostly Bates and Saclay),
extracted via the Rosenbluth separation,  agree fairly well with each other for light mass targets.
From the definitions of Eq.~({\ref{lt_responses}}),
it is clear that a comprehensive study of the nuclear response functions requires
not only the description of the initial state wave function and of the four-body current
$j^{\mu}$, but also of the final state wave functions, which could be in the 
continuum. 

Since $\omega$ and ${\bf q}$ can vary independently, one can study the responses as a function of the
energy $\omega$ while keeping the momentum $q=|{\bf q}|$ fixed or, vice versa, one can vary the momentum 
$q$ and keep  $\omega=E_f-E_i$ unchanged. 
In the last case, the response functions are called inelastic or transition form factors, 
since the nucleus undergoes a transition from the initial 
state $| \Psi_i \rangle $ to the specified final state $| \Psi_f \rangle$, and 
correspond to the quantities introduced in Eqs.~(\ref{eq:f-long}) and~(\ref{eq:f-tran}).

Below, we present the recent progress in {\it ab-initio} calculations of inelastic 
response functions and form factors and compare to the available experimental data. 
We divide the discussion into different subsections devoted to nuclei with increasing mass number, 
ranging from the deuteron to $^{12}$C.

\subsubsection{The deuteron and the $A=3$ nuclei}

$-\, \, \,\,$ In the case of $A=2$ and 3 nuclei, the final states  $| \Psi_f \rangle$  can be exactly 
evaluated with computational few-body techniques also in the continuum,  thus the longitudinal 
and transverse response functions, $R_L$ and $R_T$ of Eqs.~(\ref{lt_responses}), are obtained by 
summing over  all the final states. Alternatively, when specific break-up channels are selected, exclusive 
processes can be studied.

Results for the deuteron obtained within the conventional approach have been already discussed 
in earlier review articles (see, {\it e.g.}, Refs.~\cite{Carlson98,Arenhovel05}).  
Recently, Yang and Phillips performed a $\chi$EFT calculation of the deuteron longitudinal 
response function in Ref.~\cite{Yang13}. They have employed N2LO ($\nu=3$) wave functions 
from $\chi$EFT power counting and their subtractive renormalization method  \cite{Yang09}. 
By using the LO\footnotemark[1]\footnotetext{Note that the authors use a slightly different power counting, 
in particular they count $Q/m$ as $Q^2/\Lambda_\chi^2$ as opposed to $Q/\Lambda_\chi$. This 
demotes the relativistic one-body operator of panel (b) and the two-body operator of panel (c)  illustrated in
Figure~\ref{fig:chi_charge} to N4LO.} charge operator they obtained results which, 
at low energy and low momenta, are in agreement with conventional calculations by Arenh{\"o}vel {\it et al.} 
carried out in Ref.~\cite{Arenhovel92} using the Bonn potential~\cite{Bonnr}.

The inelastic responses of $^3$H and $^3$He with conventional potentials and currents in IA have been
discussed in the review by Carlson and Schiavilla~\cite{Carlson98}. Conventional two-body 
currents have been recently included in the studies of Refs.~\cite{Gloeckle:eA3,Viviani:2000},
and reviewed by Golak {\it et al.}~in Ref.~\cite{Golak05}, while $\Delta$-isobar excitations 
below pion threshold have been studied in Refs.~\cite{deltuva_electron, Yuan2011}.
For the inelastic electron scattering of the three-body nuclei not much has been 
done using the $\chi$EFT approach.

From the available calculations of $A=2$ and 3 nuclei, it is generally observed that
the effect of two-body operators is much larger on the transverse response, $R_T$, 
than in the longitudinal one, $R_L$. This can be easily understood in the $\chi$EFT language. 
In fact, two-body corrections appear  at NLO and at N3LO for the current and charge operator, 
respectively (see Table~\ref{tb:scalingQ}).  

\begin{center}
\begin{figure}
\centering
\includegraphics[width=14cm,clip]{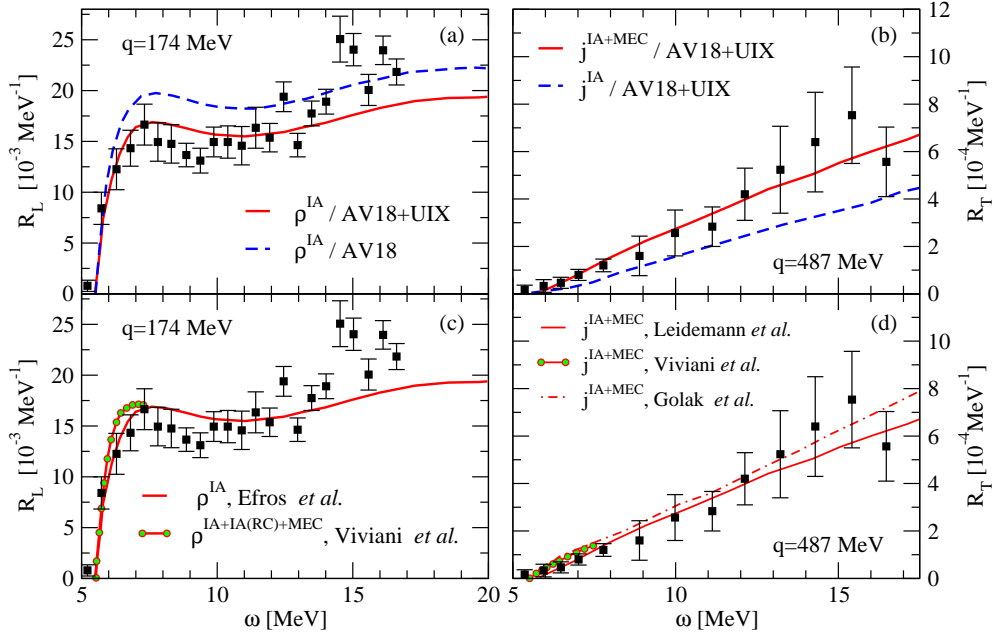}
\caption{(Color online) Calculations for the $^3$He longitudinal (left panels) and transverse (right panels) response functions at $q=174$  and $487$ MeV,  respectively, in comparison to data  by Retzlaff {\it et al.}~from Ref.~\cite{Retzlaff94} and from Ref.~\cite{Carlson02}, respectively. Panel $(a)$: calculations from Ref.~\cite{Efros04} with (red solid curve) and without (blue dashed curve) 3N forces. Panel $(c)$: $R_L$ calculations by Efros {\it et al.}~from Ref.~\cite{Efros04} compared to those  by Viviani {\it et al.}~\cite{Viviani:2000}  both obtained with the AV18+UIX Hamiltonians, but different charge operators (see text). 
Panel $(b)$: $R_T$ calculations from Ref.~\cite{Leidemann10} with (red solid curve) and without (blue dashed curve) MEC, obtained from the AV18+UIX Hamiltonian.
Panel $(d)$:  $R_T$ calculations by Leidemann {\it et al.}~from Ref.~\cite{Leidemann10} in comparison to those by  
  by Viviani {\it et al.}~\cite{Viviani:2000}  and by Golak {\it et al.}~\cite{Golak05}, where the same Hamiltonian (AV18+UIX) has been used,  but slightly different current operators, including MEC, have been implemented (see also text). 
\label{fig:InelasticA3}}
\end{figure}
\end{center}

Because the longitudinal response $R_L$ is  not  very sensitive
to two-body operators, for $A\ge3$ nuclei one can focus on studying the sensitivity 
of this observable to  3N forces, while keeping the charge operator in IA. 
The three-body nuclei have been studied for example in Ref.~\cite{Efros04} employing the 
LIT method. Conventional  3N forces  were found, {\it e.g.}, to reduce $R_L$ in the quasi-elastic peak 
by 5-10$\%$ in the momentum transfer regime between 174 MeV and about 400 MeV. However, 3N forces 
can also have  much larger effects. In Figure~\ref{fig:InelasticA3}$(a)$, 
the threshold behavior of $R_L$ is shown in the case of $^{3}$He  at $q=174$ MeV: 3N forces 
(red solid curve with the AV18+UIX potential) lead  to a  quenching of the response function with 
respect to the calculation without 3N forces (blue dashed curve with the AV18 potential) and they were 
proven to be necessary to obtain a satisfactory description of the experiment. In Figure~\ref{fig:InelasticA3}$(c)$,
a comparison with another calculation by Viviani {\it et al.}~\cite{Viviani:2000} is shown, where
 the same potentials were used and charge operators included MEC (green dots connected by red line). 
 In Ref.~\cite{Viviani:2000}, it was found that MEC have a negligible effect in this kinematical region, 
thus, the difference between the two calculations with the AV18+UIX force is  likely coming from the different 
numerical methods used. It is worth noticing that such theoretical difference is much smaller than the experimental error.
Furthermore, an even better agreement between the LIT calculations by Efros {\it et al.}~\cite{Efros04} and the  
variational calculations by Viviani {\it et al.}~\cite{Viviani:2000}   is found for larger momentum 
transfers, as shown in Ref.~\cite{Efros04}.

 In case of the transverse response function, $R_T$, it is generally observed that two-body 
currents have a substantial contribution on the total strength.
For the three-body nuclei, $R_T$ was calculated, for example, in 
Refs.~\cite{Viviani:2000, Golak05, Leidemann10} with the AV18+UIX force and  MEC and in 
Ref.~\cite{DellaMonaca08} with the  Bonn potential~\cite{Bonnr}, and the Tucson-Melbourne~\cite{TM} 
3N force and corresponding MEC. In the latter calculations, it has been shown that, 
for momentum transfer values of 174 MeV and 400 MeV, 
the two-body currents  enhance the quasi-elastic peak by about 10$\%$ and 6$\%$, respectively.
The largest MEC effects have been found away from the quasi-elastic peak and particularly close to threshold,
 where two-body currents can enhance the strength by up to 200$\%$. Generally, exchange currents contributions 
relative to the IA are enhanced when the difference $|\omega-\omega_{q.e.}|$ is larger. This is not unexpected 
as the quasi-elastic peaks correspond to the maximum contributions of the one-body current.

 In Figure~\ref{fig:InelasticA3}$(b)$ and $(d)$, we show the case of the $^{3}$He  transverse 
response at $q=487$ MeV and discuss the threshold results obtained in calculations that utilized the AV18+UIX Hamiltonian.
 In panel $(b)$, calculations by Leidemann {\it et al.}~from Ref.~\cite{Leidemann10} obtained with 
the LIT method point to the importance of MEC. The curve including  MEC (red solid) leads to a considerably
 improved agreement with experimental data with respect to the IA calculation (blue  dashed curve).  
In Figure~\ref{fig:InelasticA3}$(d)$, results by Leidemann {\it et al.}~from Ref.~\cite{Leidemann10}  
are compared to those by Viviani {\it et al.}~from Ref.~\cite{Viviani:2000} (green dots connected by red line) 
and by Golak {\it et al.}~from Ref.~\cite{Golak05} (dash-dotted curve). 
Besides using the same Hamiltonian, those calculations adopt slightly different implementations of the 
current operator. Calculations by  Leidemann {\it et al.}~include relativistic corrections to the one-body
operator and use the method devised by Arenh\"{o}vel and Schwamb~\cite{Arenhovel01} in coordinate space to derive the MEC. 
Calculations by Golak {\it et al.}~do not include Coulomb interactions  in the final state,
do not include relativistic corrections to the current operator and implement MEC obtained according to the 
momentum-space method of Riska~\cite{Riska85} (both methods for the MEC derivation lead to the same MEC contributions).
Calculations by Viviani {\it et al.}~instead employ more sophisticated MEC, which also include contribution 
from the $\Delta$-excitation. One can observe that all the three calculations nicely agree with data and with each other, 
proving that the dynamic of the few-body system is well under control.

At momentum transfer values of the order of 500 MeV and higher, relativity starts to play a role.
 Relativistic effects in $A=3$ due to the frame dependence were studied, {\it e.g.},in Refs.~\cite{Efros10,Efros11}.

While inclusive processes can help  planning for future exclusive experiments,
exclusive reactions are much richer in the information they provide about the nuclear dynamics.
In the  case of three-body nuclei, the exclusive $p-d$ break-up has been studied within the Faddeev approach
  and using conventional forces and currents~\cite{Gloeckle:eA3}. It has been observed that the final state interaction
 plays an important role, however not many experimental data are available to compare.

\subsubsection{The $^4$He nucleus}

$-\, \, \,\,$ For four-body systems the explicit calculation of all final states in the continuum becomes more cumbersome, 
thus often approximations of various kinds are introduced.
The simplest model for the response function is 
obtained under the assumption that the photon is absorbed by one nucleon only (via a one-body operator) 
and that the hadronic final state is just a free propagation (plane wave) of the knocked out nucleon that 
does not interact with the remaining spectator (A-1)-nucleus.
Such an approximation is named plane wave impulse approximation (PWIA). 
Under this assumption and neglecting excitations of the residual nucleus\footnotemark[1]\footnotetext{Note 
that, in principle, one may use the spectral function approach (see, {\it e.g.}, Ref.~\cite{Efros98})
to account for excitations of the residual nucleus.}, the longitudinal response function can be obtained from the ground-state nucleon 
momentum distribution $N({\bf p})$ as~\cite{Carlson98}
\begin{equation} 
  R_L^{\rm PWIA}(\omega,q)=\! \int d{\boldsymbol p}  \,N({\boldsymbol p})\, 
  \delta \left( \omega-\frac{( {\boldsymbol p} + {\boldsymbol q} )^2}{2
  m}-\frac{\boldsymbol{p}^2}{2(A-1)m} - E_s \right)\,, 
  \label{PWIA}
\end{equation}
 with $m$ being the mass of the struck nucleon (dominantly the proton) and  $E_s$ the proton separation energy.
The PWIA does not consider the antisymmetrization of the final state and 
ignores the effect of the final state interaction (FSI). It is thus expected to be better at high-momentum 
transfer, where the FSI is small. It also neglects the effect of two-body currents. In  the quasi-elastic regime, 
corresponding to momentum 
transfers of a few hundreds MeV and energies around  $\omega_{q.e.} \approx q^2/2m$, one can  envision that the 
electron has scattered elastically with a single nucleon. Thus, the PWIA can be considered the base line in a comparison with 
 more sophisticated calculations.
\begin{center}
\begin{figure}
\centering
\includegraphics[width=14cm,clip]{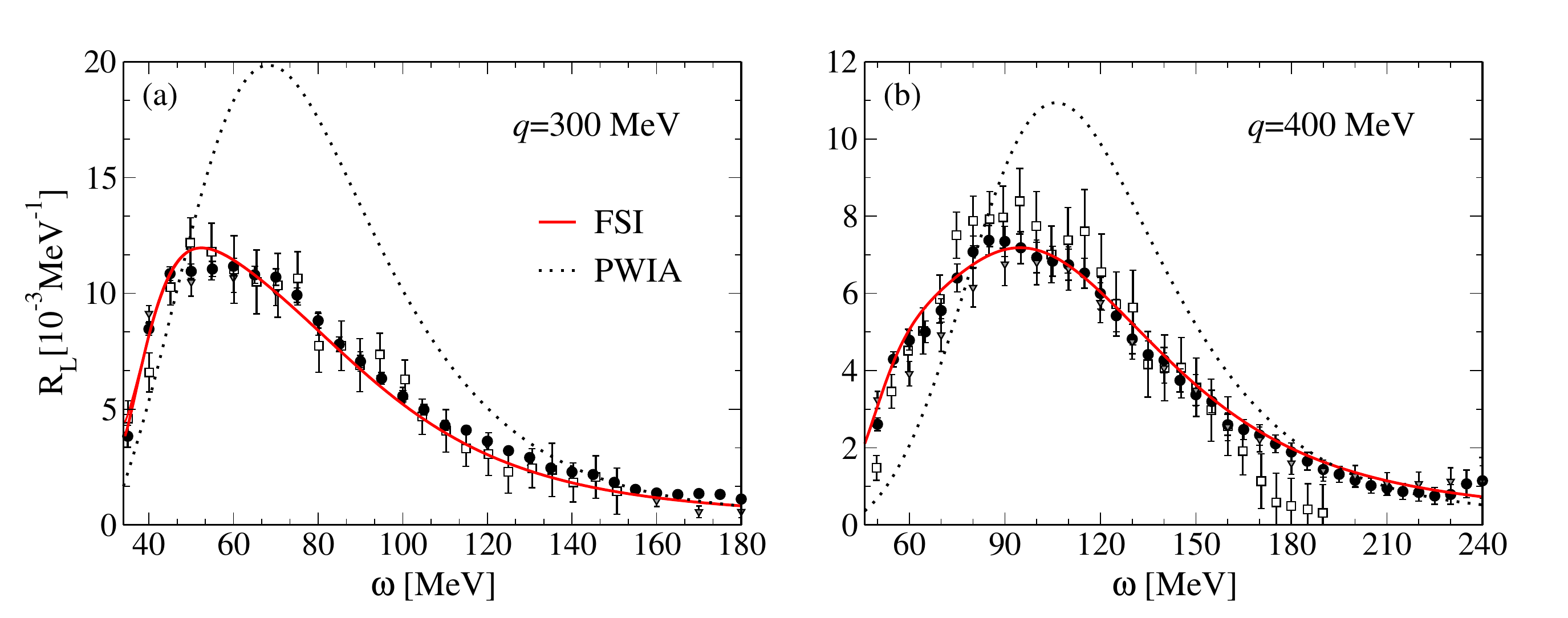}
\caption{(Color online) Longitudinal response function $R_L$ for $^4$He from Ref.~\cite{Bacca:2008tb} : 
PWIA (black dotted line) versus calculation with the FSI (red solid line)  with the AV18$+$UIX potential. 
The charge operator is used in IA. Experimental data are from Bates~\cite{Bates}
 (squares), Saclay~\cite{Saclay} (circles).}
\label{fig_FSI}
\end{figure}
\end{center}

For mass number  $A=2$ and 3 nuclei, the PWIA  fails to reproduce experimental data for $R_L$ below 
 momentum transfer $q\sim 500$ MeV, see, {\it e.g.}, Ref.~\cite{Carlson98}. In particular, it typically yields to too
 much strength near the quasi-elastic~peak and too little strength at energies below and above this peak. When the FSI is included, 
the agreement with experiment is restored. The same conclusions were drawn for the $^4$He nucleus, in the first calculation of $R_L$ 
with the Laplace transform approach~\cite{Carlson92, Carlson94} (already reported in the review~\cite{Carlson98})
and in a LIT calculation with a central NN potential model~\cite{EfL97}
 and then also confirmed  in a more
recent realistic calculation~\cite{Bacca:2008tb}.

In Figure~\ref{fig_FSI}, we show the longitudinal response function of $^4$He from Ref.~\cite{Bacca:2008tb} calculated with the 
AV18+UIX potential for two values of momentum transfer $q=300$ and $400$ MeV.
 In these calculations the charge density operator is used in IA.
 The latter is expanded in multipoles and then each multipole response function is separately calculated using the LIT method 
in conjunction with an EIHH expansion. 
 For these values of momentum transfer a numerical precision of 1$\%$ is reached when multipoles up to $J=6$ are 
summed up.
What is evident from  Figure~\ref{fig_FSI} is that  the PWIA (black dotted curve) is rather poor. When the FSI is included using 
the LIT method (red solid curve) the agreement with the experimental data is restored. 
The advantage of the LIT method over the Laplace transform  used in Ref.~\cite{Carlson94}
is that the first method allows for a stable inversion of the transform~\cite{andreasi2005}, so a direct comparison of the 
theoretical response function to the experimental data is possible. This is preferable with respect to  applying the integral 
transform to the experimental data, as the latter usually requires an extrapolation of the data in the high energy tail. Ultimately,
 the advantage of a direct comparison of the theoretical and experimental $R_L$ is that one obtains the $\omega$-dependence of 
the FSI effects as an additional information, as shown in Figure~\ref{fig_FSI}. For the chosen kinematical values, the FSI is 
important also in the region of the quasi-elastic peak.

\begin{center}
\begin{figure}
\centering
\includegraphics[scale=0.19,clip=]{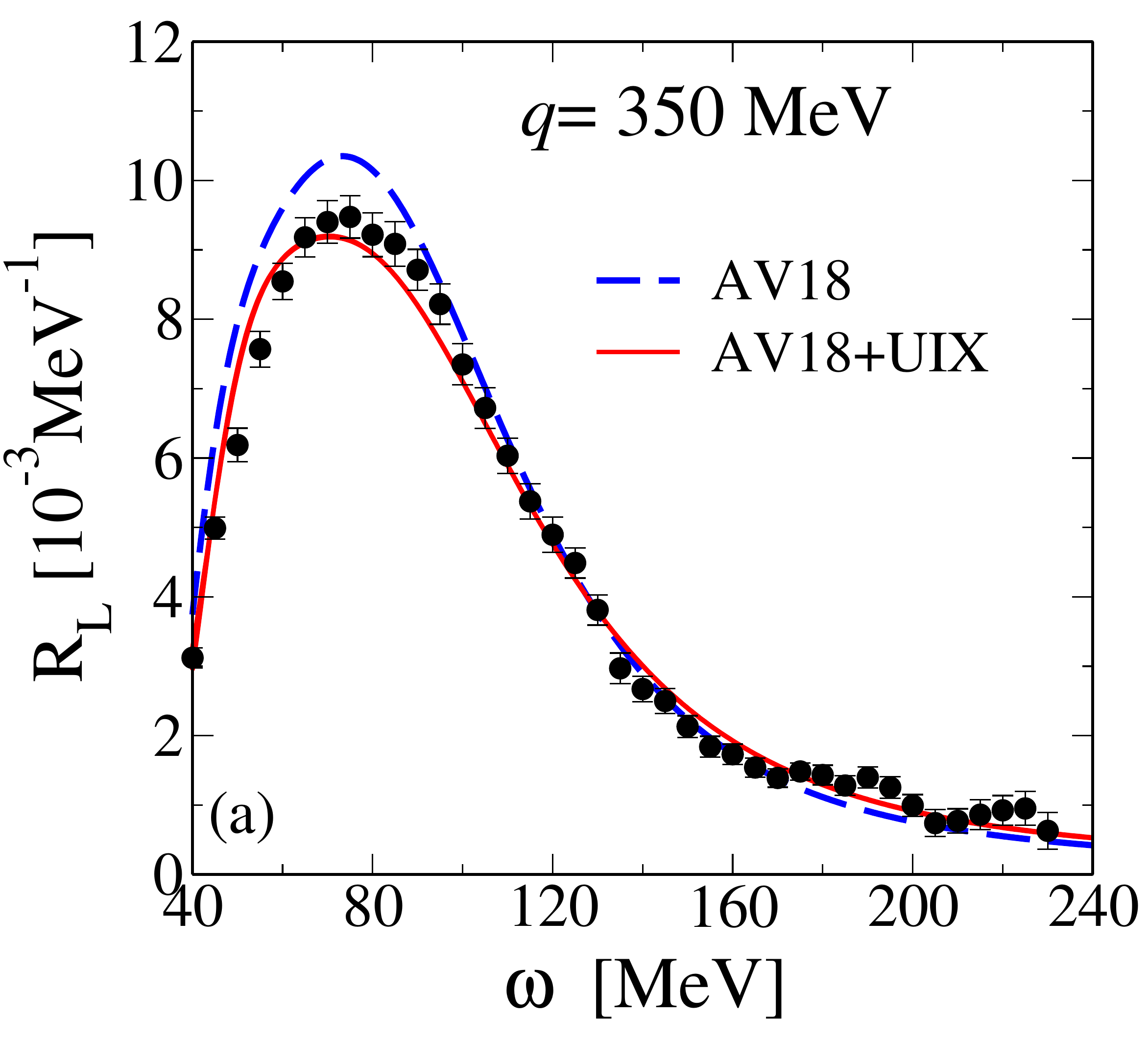}
\includegraphics[scale=0.19,clip=]{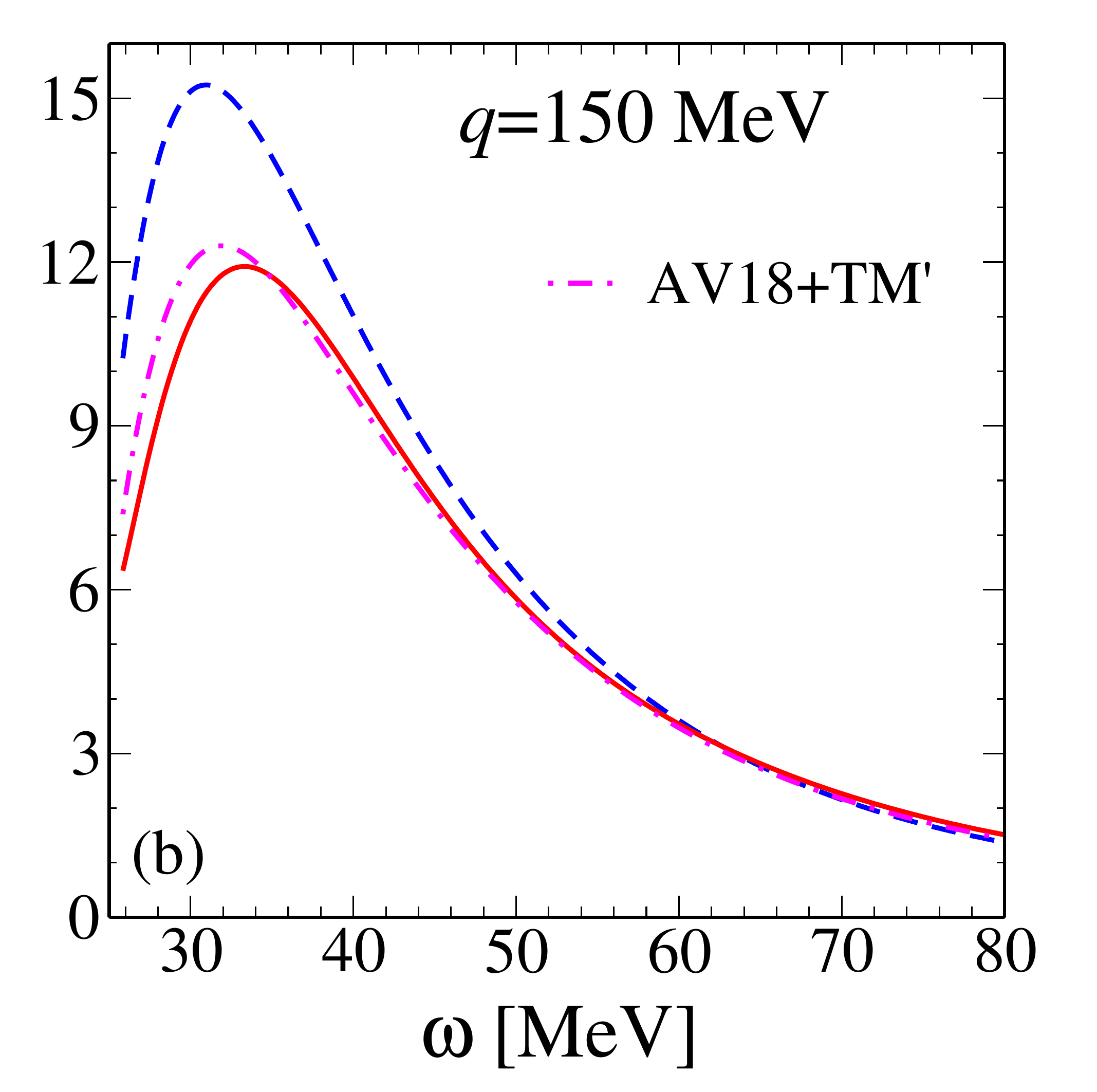}
\includegraphics[scale=0.19,clip=]{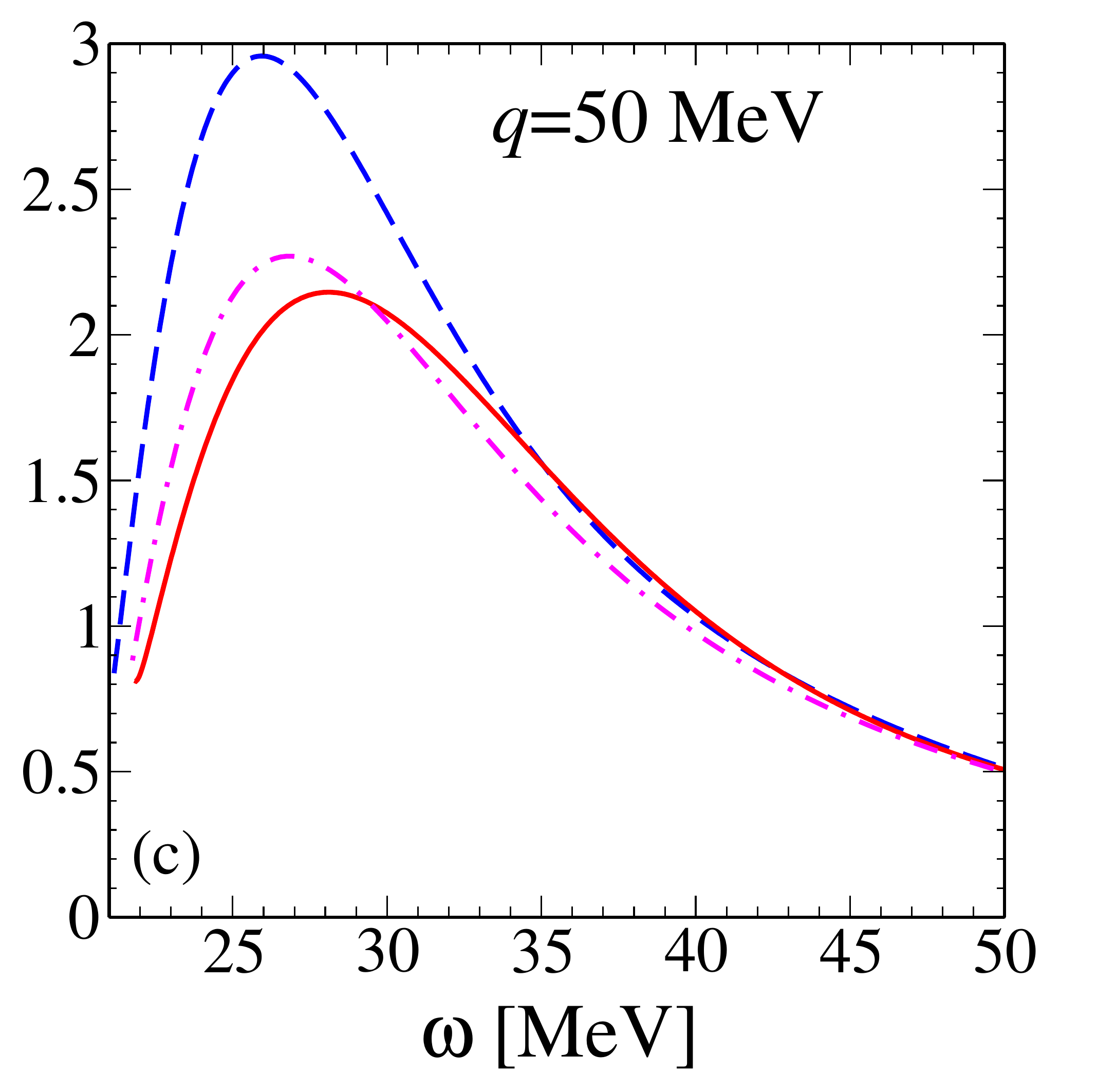}
\caption{(Color online) Effect of 3N forces on the longitudinal response function $R_L$ of $^4$He at different momentum transfers.
Calculations from Refs.~\cite{Bacca:2008tb,Bacca:2009vp} obtained with the AV18 (blue dashed curve), AV18+UIX (red solid curve) and AV18+TM$^\prime$ (magenta dash-dotted curve) potentials. The current operator has been used in IA. Data are from Saclay~\cite{Saclay}.
\label{fig_el_3NF}}
\end{figure}
\end{center}

It is  interesting to study the sensitivity of the longitudinal response functions to different Hamiltonians, with emphasis 
on 3N force effects. In Figure~\ref{fig_el_3NF}, results for $^4$He obtained with the LIT method in 
Ref.~\cite{Bacca:2008tb,Bacca:2009vp} are shown. In those calculations,   conventional potentials have been used 
and the charge operator has been kept in IA.
One observes a rather strong effect of 3N forces:
 the difference between calculations with the AV18 only (blue  dashed curve) and those comprehensive of  the UIX three-body 
force (red solid curve) is increasing at low $q$.  At $q=350$ MeV, as shown in panel $(a)$,  the UIX leads to about a 
$10\%$ reduction of the strength and improves the description of the data.  
At $q=150$ and 50 MeV, presented in panel $(b)$ and $(c)$, where  no experimental data exist, the UIX quenches the 
response function by to up to $50\%$. This fact is confirmed 
when using a  different three-body force like the TM$^\prime$ potential~\cite{TM} (magenta dash-dotted curve).
With different three-body Hamiltonians a $10\%$ variation in $R_L$ is observed. Calculations with the $\chi$EFT approach 
are in progress~\cite{Bacca:2013lya} and it is worth recalling that the TM$^\prime$ force is basically the long range part 
of the $\chi$EFT three-body force at N2LO ($\nu=3$).
Because of the sensitivity of $R_L$ to 3N forces and because of such 10$\%$ variation with different 3N forces, precise 
future experiments can help to better constrain realistic three-body Hamiltonians.
Thus, the longitudinal response 
function of $^4$He is an e.m.~observable, complementary to the hadronic ones, that can potentially allow 
to study and constrain 3N forces.

In case of the transverse response function, $R_T$, where two-body currents contribute already at 
NLO ($\nu=-1$) as suggested by the $\chi$EFT power counting, one can speculate that their strength enhancing effect will  
in part cancel the quenching due to the 3N forces that was observed in $R_L$. However, no calculation studying 
both the effect of switching on and off  3N forces and MEC exists. 
The first calculation of the transverse response function of $^4$He was performed by Carlson and Schiavilla using 
the Laplace transform~\cite{Carlson94}. Conventional three-body Hamiltonians, always comprehensive of 3N forces, 
and MEC have been implemented. A large effect of two-body currents  has been pointed out at momentum transfer $q=300$
 and 400 MeV, leading to an enhancement of the strength by 20$\%$, with respect to the IA. The major contribution was 
shown to come from OPE and proven necessary to describe the data. A comparison to the experiment could, 
though, only be done by Laplace transforming the data, so it was not possible to resolve the $\omega$ dependence of the MEC.
A direct comparison to data was done instead in Ref.~\cite{Bacca_MECHe4} using the LIT method. There, however, a semirealistic 
NN potential was utilized and approximations were introduced in the two-body currents. With this simplified model, two-body 
currents were not found to be very important, probably due to the missing one-pion exchange character both in the NN force 
and corresponding MEC.

\begin{center}
\begin{figure}
\centering
\includegraphics[scale=0.32,clip=]{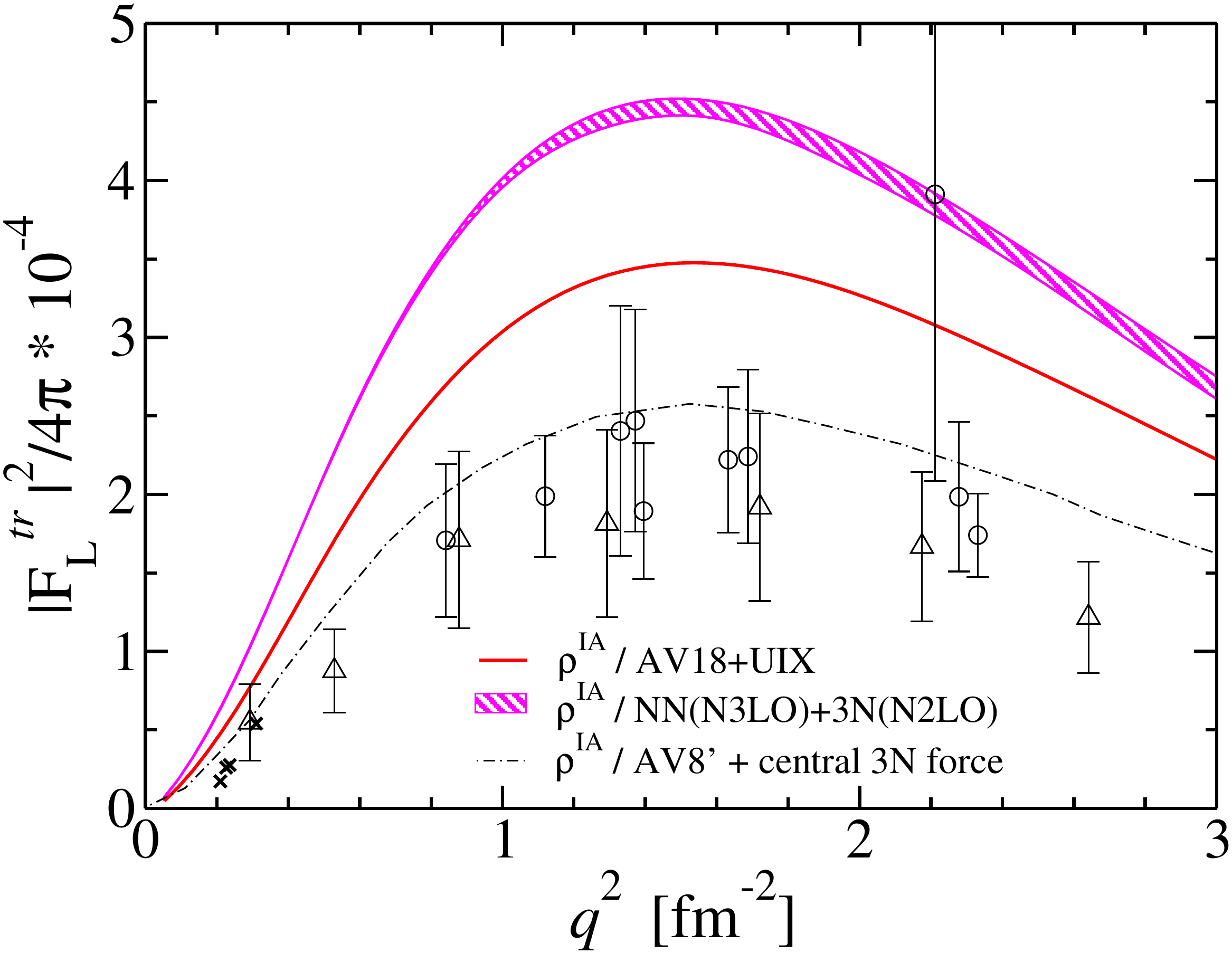}
\caption{(Color online) Transition charge form factor from the ground state to the first $0^{+}$ excited state of $^4$He  with
different  Hamiltonians and in IA for the charge operator.  Theoretical calculations are taken from from Ref.~\cite{Bacca:2012xv} for the AV18+UIX (red solid curve) and the NN(N3LO)+3N(N2LO) chiral forces (magenta band), while the result with the AV8$^\prime$+ central 3N force (black dash-dotted curve) is taken from Ref.~\cite{Hiyama:2004nf}. Experimental data are from K\"obschall {\it et al.}~\cite{Kobschall83} (circles), Frosch~{\it et al.}~\cite{Frosch68} (triangles) and Walcher \cite{Walcher70} (crosses).
\label{fig:finel_He4}}
\end{figure}
\end{center}

As already mentioned,
from inelastic electron scattering experiments other observables can be extracted. For example, 
the inelastic charge form factor, $F^{tr}_{L}(q)$, pertaining to the transition from the ground state to the 
resonant first $0^{+}$  excited state of $^4$He was recently studied with realistic Hamiltonians in Ref.~\cite{Bacca:2012xv}. 
The transition form factor corresponds to the inelastic longitudinal response function $R_{L}(\omega,q)$ 
integrated around the energy of the $0^+$ resonance. Because in this case $J_f=J_0=0$, effectively only the monopole part of the charge operator contributes to this transition.
Several experimental data sets are available~\cite{Frosch68, Walcher70, Kobschall83}, where the background 
of other multipoles and of the monopole high energy tail has been subtracted. 
  The first {\it ab-initio} calculation was performed by Hiyama {\it et al.}~in Ref.~\cite{Hiyama:2004nf} 
with the AV8$^\prime$ two-body potential~\cite{pudliner1997},  a reduction of the AV18 force,  in combination with
a simple central 3N force. The two free parameters of the latter have been calibrated to reproduce the $^3$H and $^4$He 
 binding energies. As shown in Figure~\ref{fig:finel_He4}, a rather good description of the data for $F_L^{tr}(q)$ 
has been achieved (black dash-dotted curve). More recent calculations of $F_{L}^{tr}(q)$ have been  performed with the LIT method, using 
realistic Hamiltonians both from conventional theory as well as from $\chi$EFT. As shown in Figure~\ref{fig:finel_He4}, 
the following Hamiltonians have been used: the AV18+UIX (red solid curve) and the NN(N3LO)+3N(N2LO) (magenta band) chiral 
forces. For the latter, the NN potential was used at N3LO ($\nu=4$)~\cite{Entem03} and the 3N force at N2LO ($\nu=3$).  The band width is obtained by using  two different parameterization of the LECs in the 3N force, following Ref.~\cite{Ga_beta_triton} and Ref.~\cite{Navratil07b}, respectively.
A dramatic dependence of the results on the starting three-body Hamiltonian is observed. Three different Hamiltonians  
 which describe the $^4$He ground-state energy within $1\%$ from experiment show large differences in their predictions 
for $F_{L}^{tr}(q)$.   This is quite surprising and highlights the richness of the dynamical information provided by inelastic observables. 
The failure of the realistic forces to reproduce the available experimental data for $F^{tr}_{L}(q)$ is not believed 
to be cured by two-body charge operators, as they appear only at N3LO ($\nu=0$) in $\chi$EFT and thus are not expected to play a major role.
This example of disagreement between state-of-the-art theoretical calculations and experiment will hopefully stimulate 
further theoretical and experimental activity. From the theoretical point of view, the first excited state of $^4$He 
can be tackled with alternative few-body approaches. Interesting other features, such as its relation to collective modes, 
can also be explored, see, {\it e.g.}, Ref.~\cite{Bacca14}. From the experimental point of view,
the failure of the realistic forces to reproduce the available experimental data for $F_{L}^{tr}(q)$ has motivated 
new proposals to measure the monopole form factor via higher accuracy electron scattering at the S-DALINAC~\cite{Pietralla} and
 via the  $^4$He($^4$He,$^4$He)$^4$He$^*$ reaction at LNS in Catania with the spectrometer
MAGNEX~\cite{catania}.

Regarding exclusive electron scattering reactions, the $^4$He($e,e^\prime p$)$^3$H process was studied,
 {\it e.g.}, by Quaglioni {\it et al.} in
Ref.~\cite{Sofia_electron_exclusive}. Because often ($e,e'p$) experiments are performed to study the structure
of target nucleus~\cite{Dickoff05} and spectroscopic factors are extracted under  the
assumption of direct knock-out of the proton neglecting the FSI, a check on such assumptions for
$A=4$ is very instructive. In Ref.~\cite{Quaglioni05}, by using the LIT method and a simple
semirealistic potential, it was shown that FSI effects are rather large, especially at low
$q$, amounting up to 40$\%$. It was concluded that the extraction of spectroscopic factors is more
viable in the kinematic regions with high momentum transfer and  small missing momentum.
%%%%%%%%%%%%%%%%%
Very recently, a study on exclusive processes including the $^3$He($e,e'p$)$^2$H and $^4$He($e,e'p$)$^3$H
reactions has been carried out in Ref.~\cite{Ford14}, extending the applications of MEC to high energies. It is found that FSI are of utmost importance and necessary 
for a satisfactory description of data.

The  ($e,e'NN$) reaction, in which two nucleons are knocked out and detected, is also considered an important tool to investigate NN correlations in nuclei.
The   $^4$He($e,e'd$)$d$ process was studied theoretically in Ref.~\cite{Andreasi06} using a semirealistic
interaction and an exact and consistent treatment of the FSI. The latter was found to be substantial,
pointing out the importance of having all relevant effects under control if one wants to study NN correlations.
Although several model studies are available in the literature, Unfortunately, fully {\it ab-initio} exclusive calculations, where the Schr\"{o}dinger equation is solved exactly, are not yet available in more complex nuclei, mostly because of
the major stumbling block of dealing with the final states in the continuum. 

\subsubsection{Transition form factors and sum rules in $A>4$ nuclei}

$-\, \, \,\,$ There exist few calculations of transition form factors in $A>4$ nuclei.
Inelastic transition form factors of $^6$Li have been studied in Ref.~\cite{Schiavilla_Li6_PRL}
using VMC wave functions for the ground and excited states obtained from the AV18+UIX potential.
In that work, conventional and consistent MEC have been used and they have been found to significantly
improve the agreement between theory and experiment for the calculated longitudinal and 
transverse transition form factors to a number of low-lying excited states of $^6$Li. 

\begin{center}
\begin{figure}
\centering
\includegraphics[width=14cm]{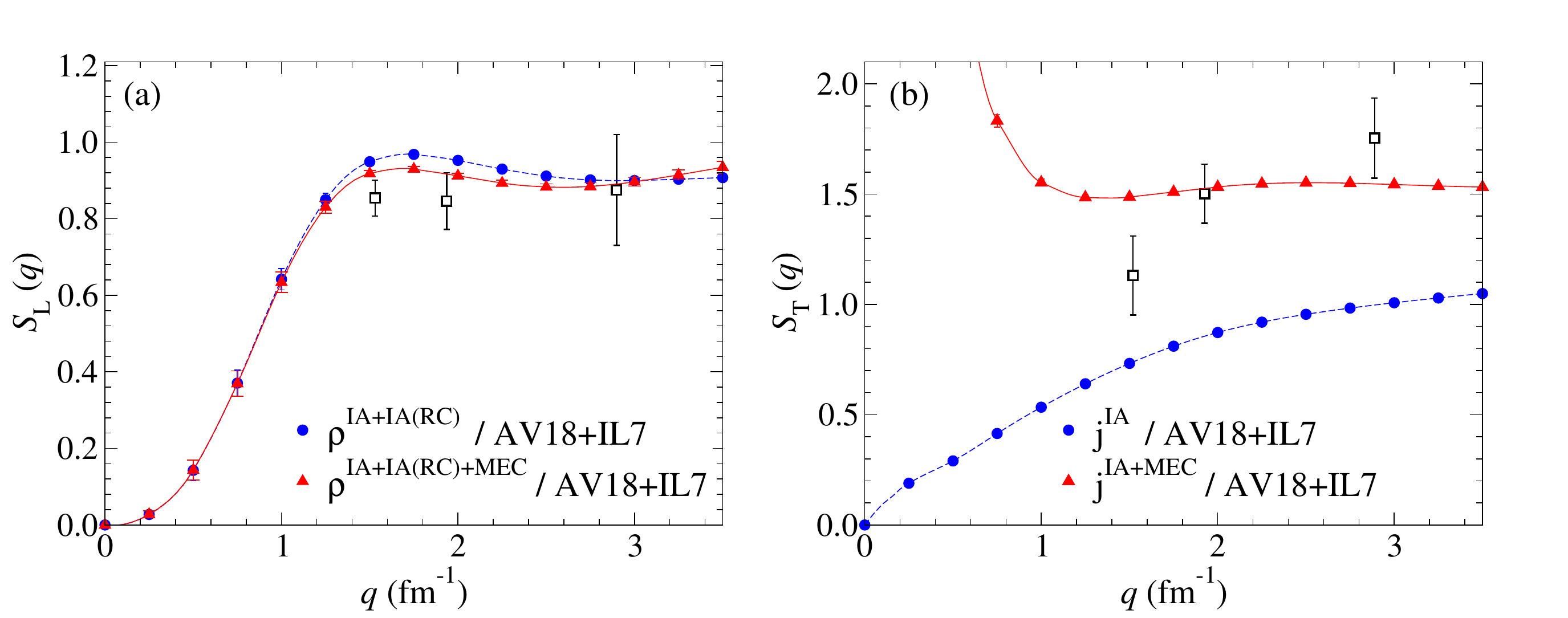}\\
\caption{ (Color online)
Calculated GFMC longitudinal, panel $(a)$, and  transverse, panel $(b)$,  
sum rules for $^{12}$C from Ref.~\cite{Lovato13} 
obtained with the AV18+IL7 Hamiltonian and consistent MEC. 
IA and IA+IA(RC) results (blue dots) and with MEC (red triangles) are compared with 
the experimental  values (empty squares), inclusive of tail corrections
(see text and Ref.~\cite{Lovato13} for more details).}
\label{fig:sum_rule_C12}
\end{figure}
\end{center}

Very recently, GFMC calculations of the charge form factors and sum
rules of e.m.~response functions in $^{12}$C have been carried out by
Lovato {\it et al.} in Ref.~\cite{Lovato13}. 
The study of e.m.~inelastic responses in $^{12}$C is particularly
interesting in view of the recent anomaly observed in the neutrino
quasi-elastic charge-changing scattering data on $^{12}$C measured
at MiniBooNE~\cite{Aguilar08}. In particular, the measured cross
section has been found to be larger than predicted by previous theoretical
calculations. The latter, however, were not carried out within strictly {\it ab-initio}
frameworks. 

The {\it ab-initio} study of Ref.~\cite{Lovato13} has aimed at quantitatively evaluating,
{\it albeit} in processes induced by e.m.~probes, the effect of MEC
on calculated $^{12}$C response functions.  
In particular, Lovato {\it et al.} calculated the $^{12}$C e.m.~sum rules~\cite{Orlandini91}
\be
\label{eq:sumrule}
S_{L/T}=C_{L/T}\int_{\omega_{th}}^{\infty} d\omega \frac{R_{L/T}(q,\omega)}{{G^p_E}^2(Q^2)} \,,
\ee 
where $R_{L/T}$ are the longitudinal and transverse response functions of Eq.~(\ref{lt_responses}), and the $C_{L/T}$ factors are 
defined as
\be
C_L=\frac{1}{Z},~~~~C_T=\frac{2}{Z\mu_p^2+N\mu_n^2}\frac{m^2}{q^2}\,.
\ee
The non-energy weighted sum rules given in Eq.~(\ref{eq:sumrule})
can be expressed as ground-state expectation values by using closure relations, {\it i.e.},
\be
S_{L/T}=C_{L/T} \left[ \langle\Psi_0 | O^{\dagger}_{L/T}({\bf q}) O_{L/T}({\bf q}) | \Psi_0 \rangle - 
| \langle \Psi_0;{\bf q} |  O_{L/T}({\bf q}) | \Psi_0 \rangle  |^2 \right] \,,
\ee
where $O_{L}$ and $O_T$ are the charge and current operators, respectively, and $|\Psi_0;{\bf q}\rangle$ 
denotes the ground state of the nucleus recoiling with total momentum ${\bf q}$.

The GFMC results for $^{12}$C, based on the AV18+IL7 nuclear Hamiltonian and consistent conventional
MEC, are shown in Figure~\ref{fig:sum_rule_C12}. In the figure, blue dots are obtained in IA (and IA+IA(RC) for 
the one-body charge operator), while red triangles include MEC corrections due to $\pi$-like and $\rho$-like
exchanges, to the  $\rho \pi \gamma$ transition current, and to OPE currents involving the excitations 
of $\Delta$ intermediate states. The comparison to the experimental data is 
not straightforward. In fact, in electron scattering, space-like virtual photons are exchanged
for which $|{\bf q}|>\omega$. Therefore, in order to compare the theoretical results with the data
one has to estimate the strengths in the energy region that is not accessed experimentally.
In Figure~\ref{fig:sum_rule_C12}, the experimental points (open squares in the figures) 
have been obtained by integrating  the available data from Ref.~\cite{Jourdan96} over the accessible energy 
interval. The tail contribution, necessary to perform the integral appearing in Eq.~(\ref{eq:sumrule}) 
up to infinity, has been estimated
by assuming that at these energies the $^{12}$C response is proportional to that of the deuteron, 
which can be calculated exactly (see~\cite{Lovato13} for details).
MEC contributions are found to be small in $S_L$, and rather large in $S_T$ where they can increase
the IA results by up to a $50\%$. The study of Ref.~\cite{Lovato13}, and that of Ref.~\cite{Lovato14}
reporting on GFMC calculations of $^{12}$C sum rules of responses induced by neutral weak currents, have found
that calculated MEC effects are large and should not be ignored
in a careful interpretation of available experimental
data.  These findings may have implications for the above mentioned anomaly observed at MiniBooNE, 
and other quasi-elastic scattering data on nuclei.

\section{Photonuclear reactions}
\label{sec:photon}

\begin{center}
\begin{figure}[htb]
\centering
\includegraphics[width=4cm,clip]{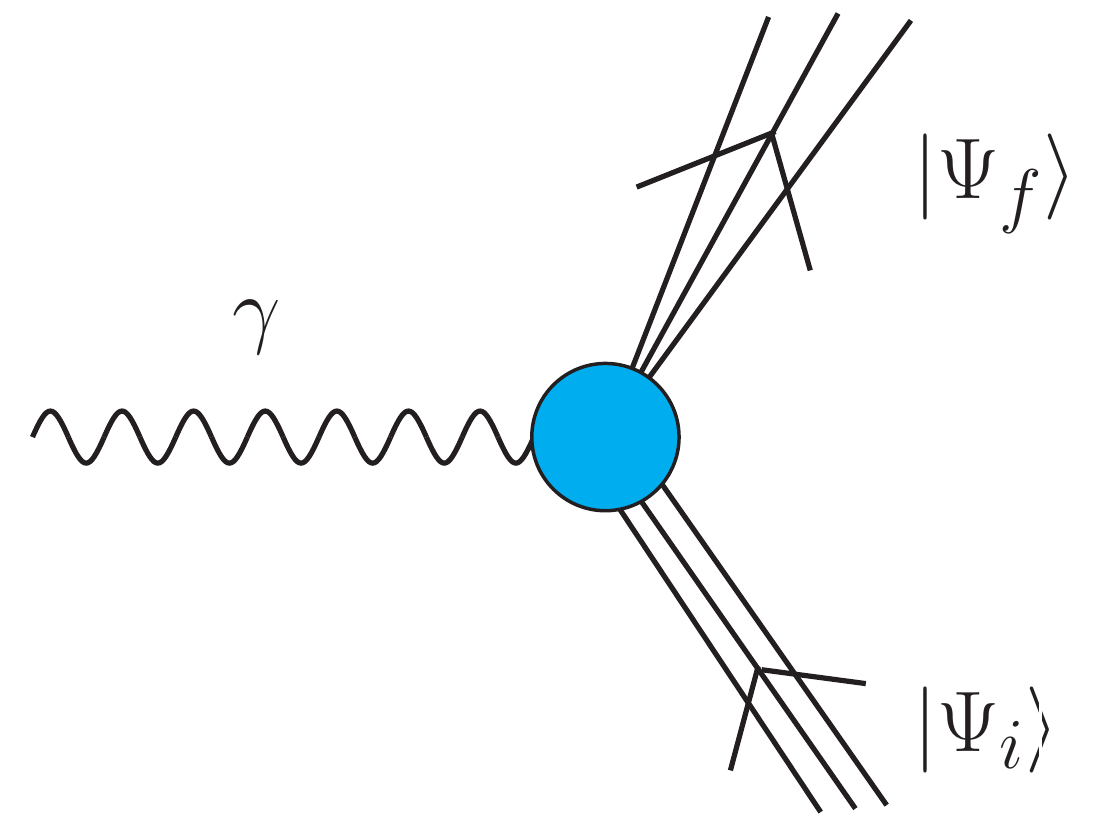}
\caption{(Color online) Feynman diagram of the interaction of a single real photon with the nucleus.}
\label{photodis}     
\end{figure}
\end{center}

In this section, we discuss processes in which real photons interact with the nucleus,
by either being absorbed by it, causing its excitation or break-up, or emitted,
for example, in $\gamma$-decay reactions or as a byproduct of fusion reactions.
To the last class belong radiative capture processes, which, at very low-energies,
are particularly relevant for astrophysical studies. Photo-absorption processes
are schematically represented in Figure~\ref{photodis}, where a real photon $\gamma$
transfers an energy $\omega=|{\bf q}|$ to the nucleus, which undergoes a transition 
from an initial to a final state, denoted by  $|\Psi_i\rangle$ and $|\Psi_f \rangle$, respectively.
Depending on the photon energy, the final state can consist of a  nuclear bound excited state, or an excited state 
in the continuum, where the nucleus breaks up in a number of clusters. At even higher energies, photoproduction of other particles,
such as pions, can be observed, but we will  consider only processes  below such energies.

Below, we devote Section~\ref{sec:photoabs}  to photoabsorption reactions, Section~\ref{sec:radiative_capture} to  
radiative capture reactions, and Section~\ref{sec:e2m1} to e.m.~transitions in low-lying nuclear states.

\subsection{Photoabsorption reactions}
\label{sec:photoabs}

Photoabsorption reactions have been extensively studied in the '70s in 
a number of experiments on stable nuclei. The main feature of the measured cross sections
is a very pronounced peak, referred to as the giant resonance peak, 
that is located at energies $\omega$ of about 10--30~MeV. It took
quite a long time until such observations could be interpreted in terms of microscopic 
{\it ab-initio} theories, and this is due to the difficulty of accounting for correlations
in both the ground and final states. At the above mentioned photon energies, the excited
nucleus can, in fact, break into either different clusters made of lighter nuclei or
even into all its constituents (in the case of very light nuclei). 
Accounting for all these complicated final states
is one of the main difficulties from the theoretical point of view.  

The general form of the nuclear matrix element entering the photodisintegration cross section
is given by
\be
\label{m_e}
N_{\lambda}= \langle \Psi_f |j_{\lambda}({\bf q})| \Psi_i \rangle \,,
\ee
where typically the initial state is the ground state, {\it i.e.}, $\Psi_i=\Psi_0$.
Photodisintegration observables provide a good tool to study two-body e.m.~currents because
they are not dominated by the charge operator 
$\rho({\bf q})$, but rather by the transverse current operator 
$j_{\lambda}({\bf q})= {\bf e}_{\lambda} \cdot {\bf j}({\bf q})$, with $\lambda=\pm 1$, 
where ${\bf e}_{\lambda}$ is a spherical component of the photon polarization vector. 
The latter is orthogonal to  the direction  of propagation ${\bf q}$  of the real photon. 
In the case of unpolarized photons, we have to sum over $\lambda=\pm 1$. In inclusive processes,
{\it i.e.}, $\gamma + A \rightarrow X$, no specific hadronic final state $X$ of the $A$-body
nucleus is measured, thus a sum over all possible final states
has to be performed. The expression for the total cross section then becomes
\be
\label{cs_current}
\sigma_{\gamma}(\omega)=\frac{4\pi^2\alpha}{\omega}~R_T(\omega) \,,
\ee 
where
\be
\label{RT}
R_T(\omega)=\frac{1}{2J_0+1} \!\sum_{\lambda=\pm1} \! \int\!\!\!\!\!\!\!\!\sum_f \left| \langle \Psi_f |j_{\lambda}| \Psi_0 \rangle \right|^2 \delta \left(E_f -E_0-\omega +\frac{\omega^2}{2M_A}\right)
\ee 
is the transverse response function, already introduced in Eq.~(\ref{lt_responses}), but here for $q=\omega$.
The total ground-state angular momentum and energy are denoted by $J_0$ and $E_0$, respectively.
The quantity $\frac{\omega^2}{2M_A}$ in the energy-conserving delta function is the recoil 
energy of the nucleus with mass $M_A$, which can be neglected below pion production.

By calculating the initial and final state wave functions with a given Hamiltonian $H$,
and then sandwiching a one-body or a two-body transverse current operator between them,            
one can study the effect of two-nucleon currents on the calculated cross sections.
Alternatively, one can explore the low-energy Siegert theorem~\cite{Siegert}, which consists of
replacing the total current operator $j_{\lambda}({\bf q})$ by its limit at ${\bf q} \rightarrow 0$. 
By using the continuity equation, one can then connect transverse electric multipoles $EJ$ to the 
longitudinal Coulomb multipoles $CJ$, that are derived from the charge operator $\rho({\bf q})$. 
In this way, the dominant part of the e.m.~two-body currents is implicitly included. 
In the long-wavelength limit, where the first multipole ${J}=1$ prevails and neglecting the $M1$ transition, 
Eq.~(\ref{cs_current}) can be rewritten as
\be
\label{cs_siegert}
\sigma_{\gamma}(\omega)=4\pi^2\alpha~\omega~R^{ E1}(\omega)\,, 
\ee 
with $R^{E1}(\omega)$ being the dipole response function
\be
\label{dipole_resp}
 R^{E1}(\omega)=
\frac{1}{2J_0+1}   \int\!\!\!\!\!\!\!\!\sum_f \left| \langle \Psi_f |D_z| \Psi_0 \rangle \right|^2 \delta(E_f -E_0-\omega)\,.
\ee 
Here,
\be
\label{dipole}
D_z=\sum_i^A \left({z}_i -{Z}_{cm} \right) \left (\frac{1+\tau^3_i}{2} \right)
\ee
is the dipole operator, ${z}_i$ is the $z$-coordinate of the $i$-th nucleon, while
${Z}_{cm}$ denotes the $z-$component of the center of mass of the nucleus. The expression 
in Eq.~(\ref{cs_siegert}) is known as cross section in the unretarded dipole
approximation and is obtained by using the Siegert theorem.  Retardation effects are included when the full Bessel function, entering the general expression of the 
current multipoles, is considered instead of its expression in the long-wave-length limit of $qR \ll 1$, with $R$ being 
the average size of the nucleus
(see, {\it e.g.}~\cite{Arenhoevel91, Carlson98}). 
It is well known that the unretarded dipole cross section
is a very good approximation of the full expression given in Eq.~(\ref{cs_current}) for
energies well below the pion production threshold. 

In the following, we present the most recent {\it ab-initio} calculations of photodisintegration 
cross sections and compare them with the available experimental data. For simplicity, 
we concentrate on unpolarized cross sections and discuss, in dedicated subsections, a variety of light nuclei
ranging from mass number $A=2$ and   $A=16$. Obviously, polarization observables are extremely 
important to understand nuclear dynamics. They have already been investigated for some nuclei and 
we refer to some of these results in our discussion, without, however, going into much details.

\subsubsection{The deuteron}

$-\, \, \,\,$ The photodissociation of the deuteron has been extensively studied 
for a long time both theoretically and experimentally (see, {\it e.g.}, Refs.~\cite{Arenhoevel91, Gilman}).
The deuteron is the simplest nuclear system available in which one can explore the role of two-body
physics coming from two-body current operators. A $\chi$EFT study along these lines 
has been presented in Ref.~\cite{Rozpedzik},
where  OPE currents (diagrams $(b)$ and $(c)$ in Figure~\ref{fig:chi_cnt}) 
and TPE currents (diagrams $(e-h)$ in Figure~\ref{fig:chi_cnt}) have been included.
This calculation is not complete in that contact two-body currents also entering at N3LO 
(see diagrams $(j-o)$ in Figure~\ref{fig:chi_cnt}) have not been included.
Nevertheless, it is interesting to discuss the main findings in comparison to experimental data and to 
calculations  based on the conventional approach.  
\begin{center}
\begin{figure}[htb]
\centering
\includegraphics[width=14cm,clip]{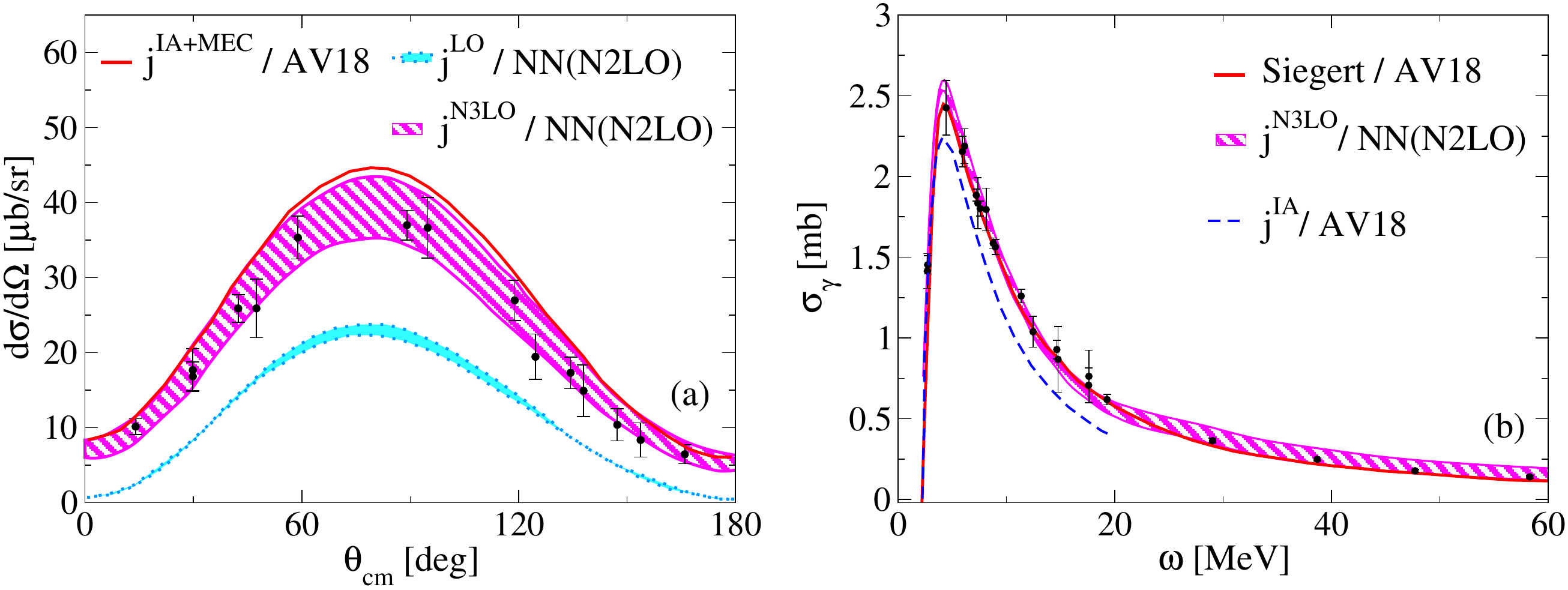}
\caption{(Color online) Panel (a): Deuteron differential unpolarized cross section  at photon 
laboratory energy of $\omega=30$ MeV displayed as a function of the proton 
emission angle.  $\chi$EFT calculations are from Ref.~\cite{Rozpedzik} 
 (with one-body current only in dotted cyan and with one- and two-body currents in the magenta hatched band)
 in comparison to conventional calculations  (red solid curve)
 and to experimental data  by Ying {\it et al.}~from~\cite{Ying}. Panel (b):
 Deuteron total photoabsorption cross section as a function of the photon energy.
Calculations based on Siegert theorem (red solid curve) are from  Ref.~\cite{Bampa}, 
IA calculations (blue dashed curve) are from Ref.~\cite{Marcucci05}, and $\chi$EFT calculations 
with two-body currents (magenta hatched band) are from Ref.~\cite{Rozpedzik}. Experimental
data as given in Ref.~\cite{Arenhoevel91} are represented by the black dots.}
\label{fig_photon_deuteron}  
\end{figure}
\end{center}
In panel $(a)$ of Figure~\ref{fig_photon_deuteron}, the calculated differential cross sections  
from Ref.~\cite{Rozpedzik} are shown. Calculations have been performed by introducing the explicit 
form of the current operator in the relevant matrix elements entering  Eqs.~(\ref{cs_current}) and~(\ref{RT}).
It is apparent that predictions with the one-body current only (cyan dotted band) strongly underestimate 
the data, while, when two-body currents are included (magenta hatched band), an agreement with the data is 
established. As expected, most of the effect of two-body currents comes from the OPE term. 
The band is obtained by using five different parameterizations for the chiral potential at N2LO~\cite{Epelbaum05, Epelbaum05r}, 
and its thickness can be interpreted as an estimate of the theoretical uncertainty.  Such a band is expected to decrease with the
addition of the contact two-body currents and when the chiral order of both potential and currents is increased. 
The $\chi$EFT approach also agrees fairly well with the conventional approach (red solid curve), where the AV18 
potential was used and MEC were included (see Ref.~\cite{Rozpedzik} for details). 

In panel $(b)$ of Figure~\ref{fig_photon_deuteron}, the total photoabsorption cross section is displayed 
as a function of the photon energy. The $\chi$EFT calculation from Ref.~\cite{Rozpedzik} explicitly includes
the effect of two-body currents from the one- and two-pion exchanges and clearly leads to a nice agreement with data. 
The conventional calculation with the AV18 potential (red  solid curve) from Ref.~\cite{Bampa} is instead based 
on the use of the Siegert theorem, and is related to an ${E1}$ response as in Eq.~(\ref{cs_siegert}) and 
({\ref{dipole_resp}}).  Thus,  the main part of the MEC  is included. It is well known from the conventional 
approach that the IA strongly underestimates the total cross section. To highlight this fact, 
in Figure~\ref{fig_photon_deuteron}(b) we report on the conventional calculation by Marcucci {\it et al.} 
from Ref.~\cite{Marcucci05}, where the IA (blue dashed curve) is used for energies below $20$ MeV.
In Ref.~\cite{Marcucci05}, it was also shown that a perfect agreement between conventional calculations
that explicitly include MEC and those obtained with the Siegert theorem is obtained 
below $20$ MeV photon energies. 
Previous studies, reported in Ref.~\cite{Arenhoevel91}, have shown that retardation 
effects and higher order multipoles contribute to an enhancement of the cross section 
of about $10\%$  at $\omega=80$ MeV, while further contributions due to exchange
 currents and relativistic effects cancel each other in the energy range from $5$ to $100$ MeV. 

It  is interesting to note that the $\chi$EFT and the conventional Siegert calculations
are consistent with each other. The main difference in percentage is found in the high energy tail, 
where, in fact, corrections to the unretarded dipole approximation may start to be relevant. 
Nevertheless, the Siegert/AV18 curve falls into the error band of the $\chi$EFT calculation.   

Finally, we would like to mention that polarization observables have been studied in 
the two-body case, both in the conventional approach (see, {\it e.g.},~\cite{Schmitt91,Marcucci05}) and in 
the $\chi$EFT approach (see, {\it e.g.},~\cite{Rozpedzik}), where important effects of exchange currents have been found. 

\subsubsection{The three-body nuclei}

$-\, \, \,\,$ In moving from two- to three-body nuclei, the dynamics becomes richer due to 
the appearance of 3N forces. Because of the present incomplete theoretical underpinning of 3N forces,
a possible way to search for effects of 3N forces is to use high precision NN potentials in three-body
calculations of hadronic observables and look for differences between theoretical predictions
and experimental data. This kind
of studies has been performed for bound-state energies of $^3$H, $^3$He (and $^4$He)~\cite{Nogga2000,Nogga2004}, 
as well as  for nucleon-deuteron collisions, see, {\it e.g.}, Refs.~\cite{Kimiko,d_breakup2}. 
Complementary information can be accessed by studying photodisintegration processes, involving the interaction between 
photons and many-body nuclear currents. Thus, e.m.~reactions with $A=3$ nuclei can be crucial observables to
test, refine and possibly even discriminate among different theoretical approaches for Hamiltonians and currents.

The photodisintegration of three-body nuclei has been investigated by a number of groups using 
different techniques. Calculations using conventional potentials and currents have been performed 
with the  HH method in Refs.~\cite{Viviani:2000, Marcucci05}, with the Faddeev approach
in~Refs.~\cite{Skibinski03,Benchmark_A3_photon,Golack_MEC_A3} and with the LIT method
in~Refs.~\cite{Efros:2000,Benchmark_A3_photon,Bacca_UCOM}. Details on the different
few-body methods can be found in the recent review of Ref.~\cite{orlandini2012}. 
Regarding the effect of conventional two-body currents,
Faddeev  and HH calculations have shown that for the two-body channel 
at low photon energy MEC enhance the differential photodisintegration 
cross section by 50$\%$ in the peak region. 
The first study of three-body photodisintegration processes carried out  within $\chi$EFT
is found in Ref.~\cite{Rozpedzik}, where the two- and three-body photodisintegration 
of $^3$He have been studied with Faddeev-type calculations. In that study,
3N forces were omitted and Coulomb effects were neglected in the scattering states. 
When looking for effects of  one- and two-pion exchange currents
in the differential cross section, the situation resembled very much that one 
of the deuteron presented in Figure~\ref{fig_photon_deuteron}. Two features are worth 
stressing: {\it (i)} the addition of two-body currents strongly enhances the strength
in the differential cross section, and {\it (ii)} the $\chi$EFT approach  
agrees rather well with the conventional calculations with the AV18 potential and corresponding MEC

\begin{center}
\begin{figure}[htb]
\centering
\includegraphics[width=14cm,clip]{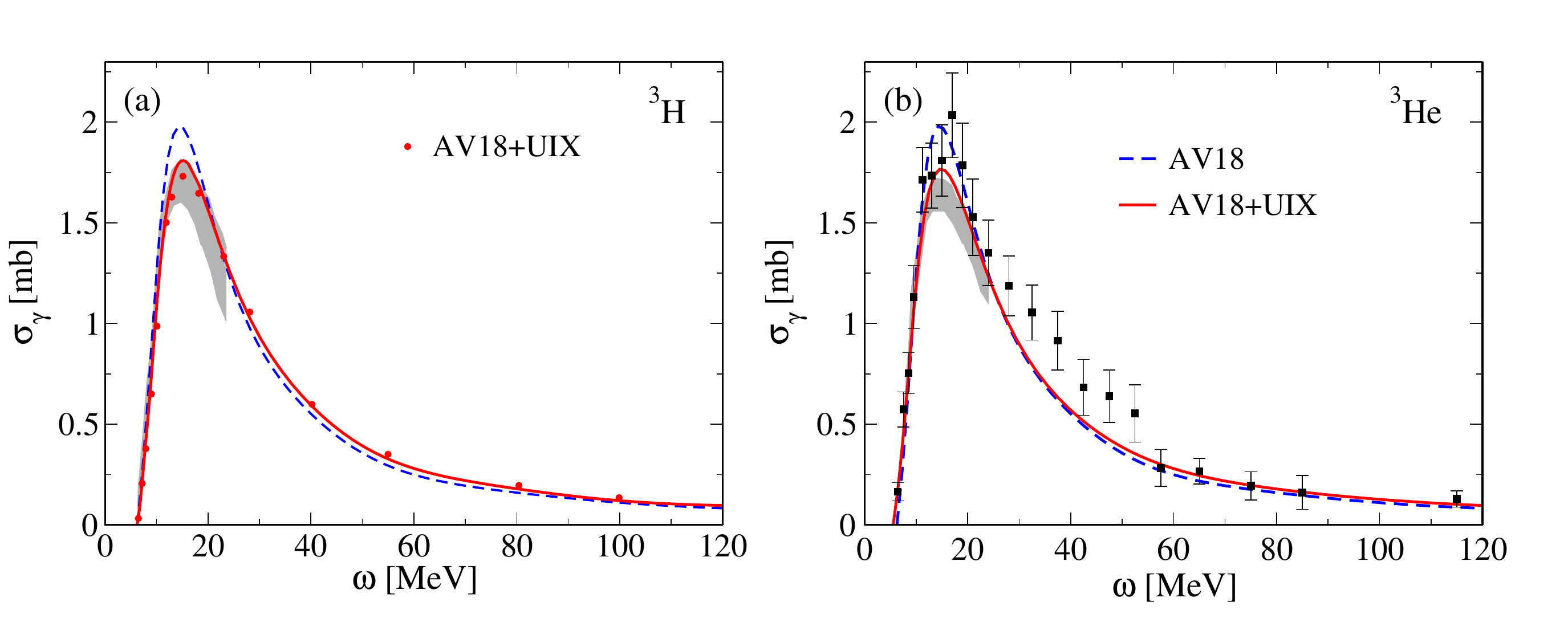}
\caption{(Color online) Total photodisintegration cross sections from Ref.~\cite{Benchmark_A3_photon} 
for $^3$H, panel $(a)$, and $^3$He, panel $(b)$, as a function of the photon energy in comparison to the 
experimental data by Faul {\it et al.} from Ref.~\cite{Faul} (gray band), and by Fetisov {\it et al.} 
from Ref.~\cite{Fetisov} (squares). Calculations obtained with the LIT method (blue dashed curve with the AV18 
and red solid curve with the AV18+UIX potential) compared to the results obtained in the Faddeev approach 
(red dots with the AV18+UIX potential).}
\label{fig_photon_A3}     
\end{figure}
\end{center}

We now turn our attention to study the sensitivity of the calculated cross sections to 
different Hamiltonians used to generate the nuclear wave functions. Specifically, 
we are interested to see what happens when 3N forces are switched on and off.
In Ref.~\cite{Golack_MEC_A3}, it has been shown that cross sections calculated  
with explicit two-body currents or with the Siegert theorem 
nicely agree with each other
at low-energies.  Thus, in a Siegert calculation of the photodisintegration cross section the only 
input that can be changed is the nuclear Hamiltonian. 
In Figure~\ref{fig_photon_A3}, we show results from Ref.~\cite{Benchmark_A3_photon} 
obtained with the LIT method using the AV18 (blue dashed curve) and AV18+UIX (red solid curve)
potentials for $^3$H $(a)$ and $^3$He $(b)$, in comparison with available experimental data. 
Similar results were obtained with slightly different three-body Hamiltonians 
in Ref.~\cite{Efros:2000}. We  emphasize that, due to the Siegert theorem,
these results also include 3N currents effects in the electric dipole response, 
that otherwise should have been included explicitly as it has been done in Ref.~\cite{Marcucci05}.
In Figure~\ref{fig_photon_A3}$(a)$,  we also show the calculation for $^3$H
performed within the Faddeev approach~\cite{Benchmark_A3_photon} (red dots) 
using the same interaction, {\it i.e.}, the AV18+UIX, and dipole transition operator. 
One can appreciate that the agreement between the two different methods is very good.
The small difference between the two approaches can be interpreted as the numerical
error of a theoretical calculation with this interaction. The latter  is  smaller 
than the theoretical deviation obtained when using a Hamiltonian without 3N forces
and smaller than the experimental error bar indicated by the gray band.

In Refs.~\cite{Efros:2000, Benchmark_A3_photon}, it has been found that the 
addition of the 3N forces reduces the strength of the calculated cross sections
by roughly $10\%$ at the peak region, leading to a better agreement with experimental data,
as can be especially appreciated in the $^3$H case illustrated in Figure~\ref{fig_photon_A3}$(a)$. 
This behavior observed in presence of the 3N forces is related to the additional binding 
they provide in the three-body system, which, in a na\"ive picture, reduces the probability
to disintegrate the nucleus. In fact, for the triton, the binding energy is $B.E.=7.60 (8.47)$ MeV
with the AV18(AV18+UIX) potential. In Ref.~\cite{Efros:2000},
the isospin $T=3/2$ channel (which is exclusively a three-body break-up channel) 
was found to be more sensitive to 3N forces than the $T=1/2$, their main effect
being an enhancement of the tail of the cross section, amounting to 15$\%$ at 70 MeV.

Other  non-local two-body Hamiltonians have been used to calculate the total cross 
section in the unretarded dipole approximation. Among these are the the potential obtained within the unitary correlation operator method (UCOM)~\cite{UCOMpotential}
used in Ref.~\cite{Bacca_UCOM}, and the $J$-matrix inverse scattering  potential (JISP)~\cite{JISPpotential} used in Ref.~\cite{JISP_HH}.
It is interesting to see how these potentials compare to other interactions and to experimental data, 
as they may induce strong exchange currents due to their non-locality.
In case of the UCOM potential, the results lie between the AV18 and AV18+UIX curve,
while for the JISP potential, the cross section has a higher tail with respect to the AV18+UIX curve,
indicating that strong exchange currents are induced with this potential.
Due to the large experimental error bar, it is not quite possible to discard such a
potential model based solely on the three-body data.

The contribution of two- and three-body break-up channels to the total cross 
section has also been studied with the Faddeev approach.
The three-body break-up cross section becomes larger than the two-body break-up one at 
14 MeV and at higher energies it is the dominant channel. Similarly to the inclusive case, 
in the exclusive two-body channel, {\it e.g.}, $\gamma+^3$H$\rightarrow d+ n$, the addition of
3N forces leads to a lower peak in the cross section and to a better agreement with experiment. 
However, the data are quite scattered, making it hard to draw final conclusions on different potentials.
The photodisintegration of $^3$He into three-nucleons was extensively studied in Ref.~\cite{Skibinski03} 
with conventional Hamiltonians and currents. Specifically, effects due to 3N forces have been sought for by 
comparing results with and without 3N forces in various kinematic regions. Very large effects of 3N 
forces amounting up to a 85$\%$ were found, a finding that could help planning for future experiments.

We recall that studies on polarization observables on $^3$He have been performed, {\it e.g.},
in Refs.~\cite{Skibinski_pol,Viviani:2000,Marcucci05}, where
a sensitivity to both 3N forces and to the detail of the many-body contributions 
to the nuclear current operator has been observed.  In Ref.~\cite{Rozpedzik},
polarization observables have been studied within the $\chi$EFT approach and 
very strong effects of exchange currents have been found in the photon analyzing power, 
leading even to a completely different shape of the angle-dependent curves when going
from the simple IA  picture to the one that  includes two-body e.m.~currents. 
Furthermore, recent studies on neutral pion production off three-body nuclei 
carried out in $\chi$EFT have also pointed out to large effects of two-body currents~\cite{Lenkewitz}.

\subsubsection{The $^4$He nucleus}

$-\, \, \,\,$ Particular attention has been recently devoted to the study of the $^4$He photodisintegration reaction.
This process is particularly interesting, from both a theoretical and an experimental point 
of view,
due to a number of reasons.
First, with a binding energy per nucleon of about 8 MeV/A, $^4$He can be considered the 
link between the very light nuclei,$^2$H, $^3$H, and $^3$He, and heavier systems.
Second, because NN  and 3N forces are typically calibrated on two- and three-body nuclei,
$^4$He is the most natural testing ground for microscopic nuclear forces. Third, due to 
its larger density with respect to the lighter nuclei, intermediate and short range physics effects, 
such as those generated by 3N forces, are expected to be more important. In fact,  
conventional 3N forces contribute  17$\%$ of additional binding energy, 
leading to 24.27 (28.42) MeV with the AV18 (AV18+UIX) potential.
This correction due to 3N forces
is much larger than the 10$\%$ effect found in $^3$H.  
Finally, because of gauge invariance, nuclear forces also manifest themselves
as exchange currents which are very important in photonuclear reactions,
making the photodisintegration of $^4$He particularly interesting to study further effects of 3N forces.

The theoretical computation of such process is complicated by the difficulty of calculating 
the four-body continuum states. For the ground state of $^4$He, several few-body methods 
can be applied to precisely calculate its properties using realistic forces (as it has been shown 
in the famous benchmark study reported in Ref.~\cite{kamada2001}). The same is not true for break-up observables
where many channels in the final state are open. However, one can avoid the complications of
the explicit calculations of continuum final states by using a bound-state formulation of 
the continuum problem. This is what is achieved for, example, with the LIT method, and with the
complex scaling method, see, {\it e.g.}, the review articles of Ref.~\cite{orlandini2012} and \cite{Carbonell}.

From the experimental point of view, most of the efforts have been devoted to the 
measurements of exclusive channels, like $^4$He$(\gamma,n)^3$He and  $^4$He$(\gamma,p)^3$H.
For the  $^4$He$(\gamma,p)^3$H  reaction, the available data are scattered in the 
energy range between 20 and 35 MeV. The current experimental status is nicely summarized 
in Ref.~\cite{Raut:2012zz}, where data from both  photon beams and as well as those
deduced from the time reversed radiative-capture reaction using the principle of detailed balance are used. 
For the $^4$He$(\gamma,n)^3$He reaction, the experimental situation is unsettled 
(see Ref.~\cite{Tornow:2012zz} for a summary on this topic). 
\begin{center}
\begin{figure}[htb]
\centering
\includegraphics[width=14cm,clip]{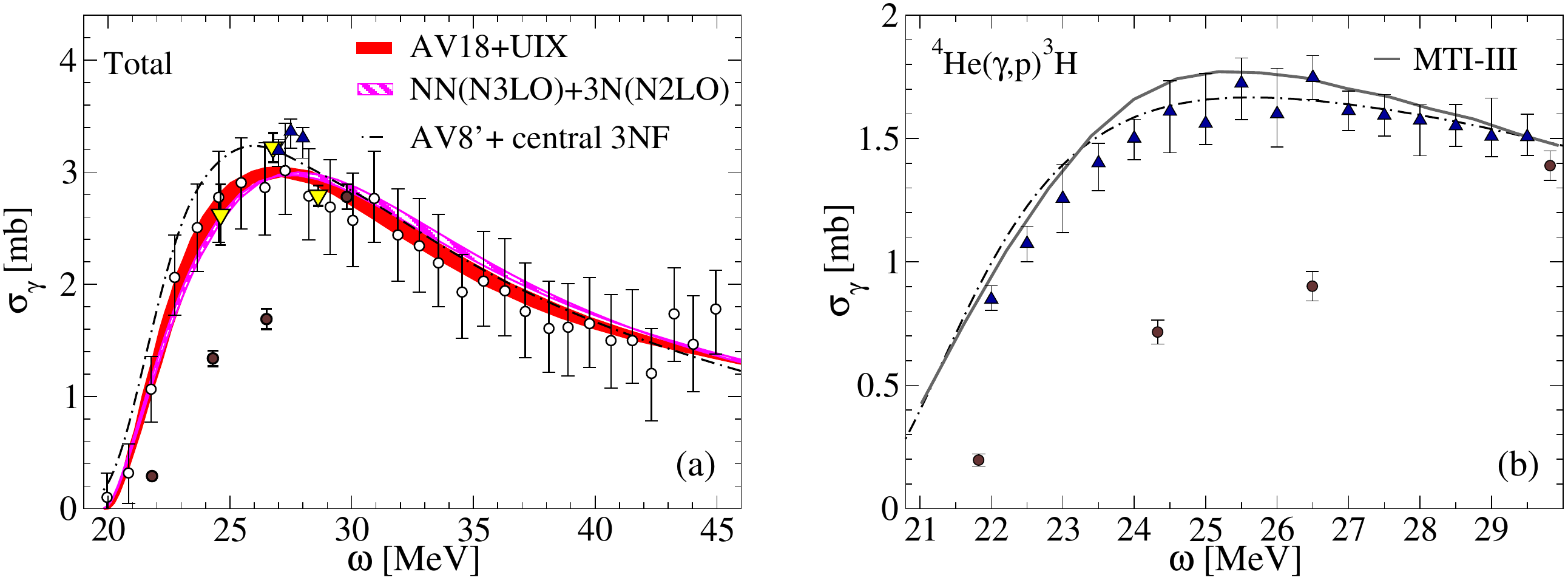}
\caption{(Color online) $^4$He  photoabsorption cross section based on the Siegert theorem  
for the total process, panel $(a)$,
and the exclusive channel $^4$He$(\gamma,p)^3$H, panel $(b)$. Theory
is compared to  experimental data:  
empty circles by Nakayama {\it et al.} from Ref.~\cite{Nakayama}, yellow triangles-down by 
Nilsson {\it et al.} from Ref.~\cite{Nilsson}, blue triangles-up by Raut {\it et al.} from 
Ref.~\cite{Raut:2012zz,Tornow:2012zz} and brown circles by Shima {\it et al.} 
from Ref.~\cite{Shima:2005ix} (see text for more details).}
\label{fig_photon_he4}      
\end{figure}
\end{center}
In Figure~\ref{fig_photon_he4}, we present {\it ab-initio} theoretical results in 
comparison with the most recent experimental data (from 2005 or later). On the 
left, we show the total cross section and on the right we present the $^4$He$(\gamma,p)^3$H exclusive 
reaction. 
For the total cross section, the first realistic calculations  have been performed in 
Ref.~\cite{gazit2006} with the AV18+UIX potential (red solid band), using the LIT method in 
conjunction with the EIHH expansion.
The $\chi$EFT approach has been utilized, instead, in Ref.~\cite{Qua07} (magenta hatched band), 
where the NN force was taken at N3LO ($\nu=4$) and the 3N force at N2LO ($\nu=3$) with a fit of the LECs from 
Ref.~\cite{Navratil07b}.
There, the LIT method has been used in combination with the NCSM ~\cite{stetcu2007}. 
For these realistic three-body Hamiltonians the results are shown with a band, 
corresponding to the theoretical uncertainty of the used computational method. 
An other recent calculation has been performed by 
Horiuchi {\it et al.}~\cite{Horiuchi:2012sn} (dash dotted curve) using the complex
scaling method  with the two-body AV8$^\prime$ potential~\cite{pudliner1997} 
(a simplified version of the AV18 NN potential) supplemented by a simple central 
3N force from Ref.~\cite{Hiyama:2004nf}. 
These three theoretical calculations are shown in Figure~\ref{fig_photon_he4}(a). 
They are all based on the use of the Siegert theorem, {\it albeit} with different Hamiltonians, 
which all reproduce the experimental binding energy of $^4$He within one percent.
Because they form a kind of theoretical  band, with a variation of the order of 
$10\%$ at the peak, we can be quite confident in saying that there is a solid understanding 
of nuclear dynamics in this regime.
The large differences at threshold are due to the inaccuracy of the complex scaling 
method (see Ref.~\cite{Horiuchi:2012sn}) and should not to be considered as an effect of 
the potential. 

When we compare  against recent experimental data,
the most striking disagreement between theory and experiments, amounting up to a 
factor of 2, is with the data from Shima {\it et al.}~from Ref.~\cite{Shima:2005ix}. 
They have been obtained with a quasi-monoenergetic photon beam and a time projection chamber,
where a simultaneous measurement of both the $^4$He$(\gamma,n)^3$He and  $^4$He$(\gamma,p)^3$H 
reactions has been performed. The data in the energy range below 30 MeV, where the 
large disagreement exist, have been remeasured by the same group, confirming their
findings~\cite{Shima:last}. The latter are, however, in disagreement with previous 
data from Nilsson {\it et al.}~on the neutron channel~\cite{Nilsson} 
(for the total cross section we show them below the three-body disintegration 
threshold, where $\sigma_{\gamma}\simeq 2 \sigma(\gamma,n)$) and with data from 
Nakayama {\it et al.}~\cite{Nakayama}, measured via the $^4$He($^7$Li,$^7$Be) reaction. 
To clarify the situation, an experiment has been carried out at the High Intensity 
Gamma-Ray source~\cite{Raut:2012zz, Tornow:2012zz}. Both the exclusive two-body channels have been
measured (in Figure~\ref{fig_photon_he4}$(a)$ we sum the two channels and show just the points
in the energy range below the three-body break-up).
This last experiment has confirmed that the peak of the cross section is 
at around 25-26 MeV, in nice agreement with the theory and with the data from Nilsson {\it et al.}
and Nakayama {\it et al.}, and in strong disagreement with the data from Shima {\it et al}.
Discarding this specific set of data, it can be said that there is nice agreement between 
theory and experiment.

For the exclusive $^4$He$(\gamma,p)^3$H, in Figure~\ref{fig_photon_he4}$(b)$ we show the calculations 
with the MTI-III potential from Ref.~\cite{Quaglioni05} (grey solid curve), performed with the LIT method, 
and those with the AV8$^\prime$ with a central 3N force from Ref.~\cite{Horiuchi:2012sn} 
(dash-dotted curve), performed with the microscopic R-matrix method.  
We compare them to the exclusive 
measurements from Tornow {\it et al.}~\cite{Tornow:2012zz} and  from  
Shima {\it et al.}~\cite{Shima:2005ix}. Also in this case, the theory supports a higher 
cross section in disagreement with the data from Shima {\it et al}. One can also see 
that the calculation with the more realistic AV8$^\prime$  potential and the central 
3N force is actually in better agreement with the data from  Tornow {\it et al}. Efforts  
are being directed towards a calculation of this cross section based on the AV18+UIX
with the LIT method~\cite{Nevo}.  

Finally, we would like to comment on the role of 3N forces.
In Ref.~\cite{gazit2006}, they were found to lead to a 6$\%$ decrease of the 
cross section peak, but also to a large enhancement, up to 35$\%$ , of the strength at 
energies above 50 MeV. Such  3N force dependence cannot be simply interpreted 
as a binding effect. The only inclusive experimental data that cover a 
wide range in energy are from Arkatov {\it et al.}~\cite{Arkatov}.  When the UIX 
potential is added to the AV18 two-body force, the agreement with data 
is actually improved in the energy range from $\omega=50$ to 80 MeV. Beyond that,
and up to pion threshold, the AV18 potential does provide the best description of 
the data. Finally, we  remind that other non-local two-body potentials, like the JISP, 
actually do lead to a cross section which is higher and not in agreement with the experimental data, 
as pointed out in Ref.~\cite{JISP_HH}. This is presumably due to large exchange currents 
induced by the non-locality in the potential.

\subsubsection{The $A=6$ and $7$ nuclei}

$-\, \, \,\,$ We now consider slightly heavier nuclei. Traditionally, nuclei with a number of nucleons 
between 4 and about 12 are considered as a bridge between the few- and the many-body systems. 
Furthermore, with increasing mass numbers, nuclear spectra become more complex and 
new interesting phenomena arise. For example, within the isobar nuclei  with mass 
number $A=6$, we have a two neutron-halo nucleus, {\it i.e.},  $^6$He~\cite{Tan96}.
This happens also to be the lightest halo nucleus in the nuclear chart, and, as such, has 
received quite a bit of attention both from the theoretical and experimental point of view 
(see, {\it e.g.}, Ref.~\cite{Brodeur}).  $^6$He is a very short lived nucleus (half life of 
$t_{1/2}$ = 807 ms) that can be compared to the stable $^6$Li nucleus. An interesting 
question to ask is: does the interaction of a real photon with these two isobar 
analog nuclei lead to  different structures in the photodisintegration cross section? 
Obviously, to answer this question one has to face two challenges, one 
theoretical and one experimental:  {\it (i)} the exact calculation of six-body final 
state wave functions with various open channels is currently out of reach, and {\it (ii)} 
experiments with real photons cannot be done on unstable nuclei, so alternative
techniques need to be looked for.
\begin{center}
\begin{figure}[htb]
\centering
\includegraphics[width=9cm,clip]{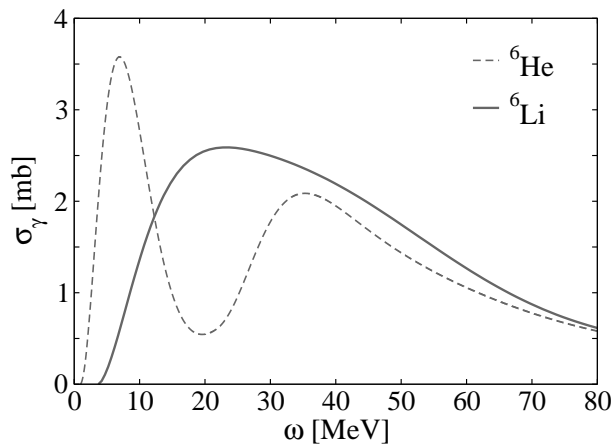}
\caption{Theoretical photoabsorption cross section of $^6$He and $^6$Li from 
Ref.~\cite{BaB04} calculated with the AV4$^\prime$ potential using the Siegert theorem.}
\label{fig-6b}    
\end{figure}
\end{center}  
The theoretical problem has been circumvented by using the LIT method. 
The application of this method in conjunction with the EIHH expansion has 
allowed to study the photodisintegration of the six-body nuclei in Refs.~\cite{bacca2002,BaB04}.
Simple semi-realistic interactions have been used, that just fit the $S$-wave or 
(partially)  the $P$-wave phase-shifts in the NN collisions.
Even though they do miss some of the complexity of realistic potentials, 
such as the tensor and spin-orbit force,  they  decently reproduce  (within $10\%$)
the photodisintegration of lighter nuclei, as discussed earlier.
The studies on the six-body nuclei have shown that
the halo structure of the rare $^6$He  isotope leads to considerable differences 
from the stable $^6$Li nucleus in photodisintegration. As shown in Figure~\ref{fig-6b}, 
for $^6$He the dipole cross section exhibits two
well separated peaks, while a single resonant shape is observed for $^6$Li.
The first peak corresponds to the break-up of the neutron halo, while the 
second peak corresponds to the break-up of the $\alpha$-particle. 
These results do not depend on the employed NN interaction:
three different semi-realistic interactions have been used, and they have been found to
lead to very similar structures. The curves in Figure~\ref{fig-6b} have been obtained 
with the AV4$^\prime$ potential~\cite{AV4prime}, which is a truncated version of
the AV18 potential, that leads to a reasonable description (10-15$\%$ from experiment) 
of the binding energy of these nuclei and includes some $P$-wave components,
which are found to be important in $p$-shell nuclei~\cite{BaB04}.

\begin{center}
\begin{figure}[htb]
\centering
\includegraphics[width=14cm,clip]{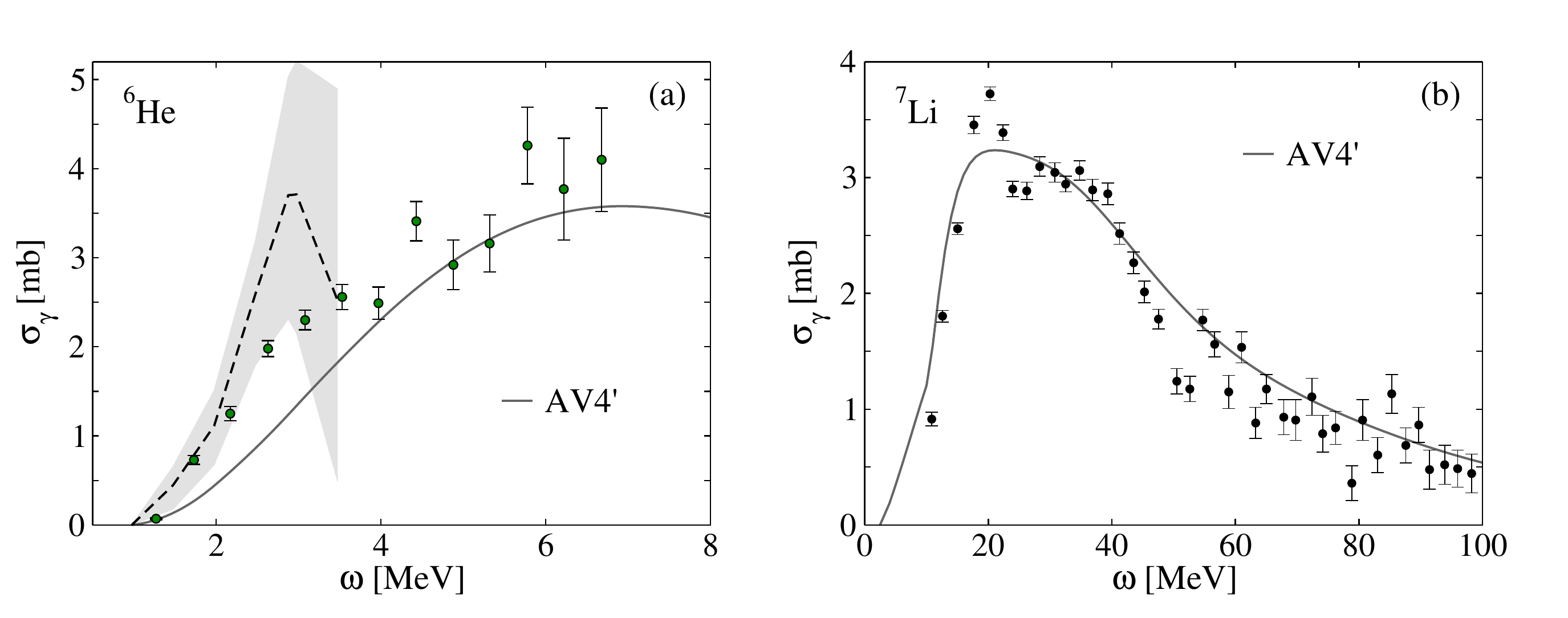}
\caption{(Color online) $^6$He and $^7$Li total photoabsorption cross sections calculated with the
AV4$^\prime$ potential from Refs.~\cite{BaB04} in comparison to experimental data:
for $^6$He, green circles denote data by Aumann {\it et al.}~\cite{Aumann}, while the gray band denotes data
by Wang {\it et al.}~\cite{Wang}; for $^7$Li, dark circles represent data by Ahrens {\it et al.}~\cite{ahrens1975}.}
\label{fig-he6-li7}   
\end{figure}
\end{center}

Photonuclear experiments on unstable system cannot be done. However, the low-energy photodisintegration 
cross section, which is related to the dipole response function via the Siegert theorem as in Eq.~(\ref{cs_siegert}),
can be inferred from Coulomb excitation experiments, see, {\it e.g.}, the review \cite{Gade2008}. In this case, a beam of rare isotopes
is used as projectile  in collisions with
heavier targets. In specific circumstances,  the Coulomb interaction between the projectile 
and the target either dominates or can be separated from the nuclear part. From the Coulomb interaction
one can extract the ${E1}$ strength and then reconstruct the photoabsorption cross section. 
For $^6$He, this kind of experiments has been done at GSI by Aumann {\it et al.}~\cite{Aumann}
with high projectile energy (240 MeV/nucleon) and at NCSL by  Wang {\it et al.}~\cite{Wang} with lower projectile 
energy (25.2 MeV/nucleon).  In Figure~\ref{fig-he6-li7}(a), we show these experimental 
results in comparison to the theoretical calculations from Ref.~\cite{BaB04}.
As one can see, the theory decently describes the data for $\omega>5$ MeV, but underpredicts the strength very close to threshold. The spin-orbit force, which is missing in the AV4$'$ potential, 
could possibly enhance the strength in this energy region.

For $^6$Li, which is a stable isotope, several experiments on photonuclear reactions exist, 
however none of them corresponds to an inclusive measurement. Because summing all the different
channels coming from different experiments is delicate and because the experimental situation 
for the exclusive processes is not settled, we prefer, instead, to show the comparison of 
theory with experiment for $^7$Li.  Calculations on $A=7$ nuclei have been performed in Refs.~\cite{BaA04} using
the LIT with an EIHH expansion 
and the AV4$'$ potential. Inclusive data for 
$^7$Li have been measured by Ahrens {\it et al}.~\cite{ahrens1975} and are shown  
in Figure~\ref{fig-he6-li7}(b). One can observe that, despite the simple potential used, 
the theoretical curve nicely describes the gross feature of the experimental data, {\it i.e.},
a steep rise, a broad maximum and a slow fall-off. $^7$Li, as  $^6$Li, is a stable nucleus and 
show only one large bump, very differently from the two separated peaks found in $^6$He by theory. 
These theoretical calculations have stimulated new experimental efforts devoted to measure, {\it e.g.}, the 
photoabsorption of $^{6,7}$Li  at the High Intensity Gamma-Ray Source~\cite{Pywell} and at MAX-lab in Lund~\cite{Maddalena,Duncan}.
The predicted presence of the second peak in the $^6$He photodisintegration cross section 
is still waiting for an experimental observation.

\subsubsection {Towards medium-mass nuclei}

$-\, \, \,\,$ First principle calculations of photonuclear reactions, where one starts from realistic NN forces
that reproduce (some) NN phase-shifts, have been available only for
the light systems, in the recent past. For the medium-mass nuclei, instead, several studies exist in the 
literature, where the photodisintegration is investigated using Skyrme-functionals 
(see, {\it e.g.}, Refs.~\cite{Erler2011, Nakatsukasa2012}), which are typically calibrated 
on finite nuclei. Alternative (macroscopic)  approaches, like, {\it e.g.}, Halo EFT
have also been recently investigated, see Ref.~\cite{Hammer2011}.
 The reason for the lack of {\it ab-initio} calculations in
the medium-mass regime is that one has to account for the very strong correlations
induced by realistic interactions in both the ground state and in the final states in 
the continuum.   The difficulty of theoretically describing
the continuum states can be circumvented with the LIT method which reduces the 
problem to a bound-state-like equation.
 As shown above, the LIT method has been used together with HH expansions and 
with the NCSM. Both these methods do not lend themselves to a straightforward application
to the medium-mass nuclei. In a recent paper~\cite{Bacca:2013dma}, it has been shown that 
the LIT method can be used in conjunction to CC theory, a many-body approach 
well suited for the medium-mass and heavy nuclei~\cite{CC_review}.
The combination of the LIT with CC theory~\cite{long_paper}
has allowed to extend the previous mass limits of the theory 
and address the inclusive photodisintegration in $^{16}$O.
The method has been first benchmarked with the exact EIHH on $^4$He using the same nuclear Hamiltonian. 
Despite the approximations introduced in the CC calculations, 
a very nice agreement has been found between the two-methods for $^4$He.

\begin{center}
\begin{figure}[htb]
\centering
\includegraphics[width=9cm,clip]{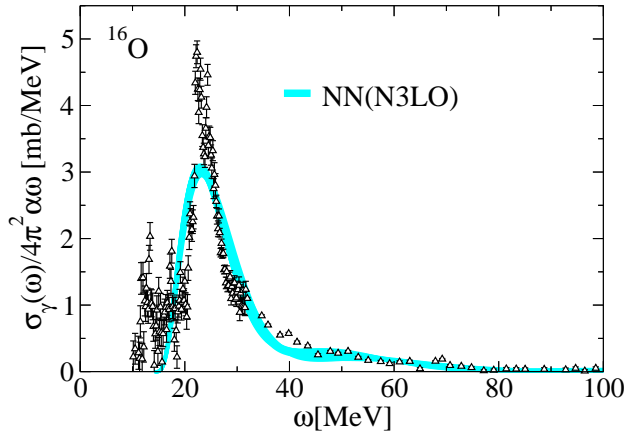}
\caption{(Color online) Comparison of the  $^{16}$O dipole
  response function calculated in coupled-cluster theory from Ref.~\cite{Bacca:2013dma} against experimental data  by
  Ahrens {\it et al.} from Ref.~\cite{ahrens1975}.
}
\label{fig-O16}
\end{figure}
\end{center}

In Figure~\ref{fig-O16}, we present the theoretical results for the dipole response function of $^{16}$O
in comparison to the available  data from Ahrens {\it et al.}~\cite{ahrens1975}. The experimental
cross section is transformed into a dipole response function using Eq.~(\ref{cs_siegert}).
For the starting Hamiltonian, the $\chi$EFT approach has been taken, where the NN interaction at N3LO ($\nu=4$) from Entem and Machleidt~\cite{Entem03} 
(inclusive of Coulomb force) has been used,  but 3N forces have been omitted. 
The band of the cyan curve 
is obtained by inverting the LIT with a slightly different regularization procedure, 
see, {\it e.g.}, Refs.~\cite{Bacca:2013dma, andreasi2005} for more details.
It is apparent that the experimental location of the peak is correctly reproduced by 
the calculation. Also the total experimental dipole strength is reproduced, while the width 
of the theoretical resonance is broader than the experimental one.
The investigation of the impact of the neglected 3N forces and  higher order correlations 
on the photonuclear cross section of medium-mass nuclei is currently underway.
This first case study paves the way for many future investigations of continuum
responses in medium-mass nuclei for both stable and unstable isotopes, which will
be accessible using the LIT method in conjunction with CC theory.

\subsection{Radiative capture reactions}~\\
\label{sec:radiative_capture}

Radiative capture processes are the time inverse of nuclear photoabsorption reactions
at few MeVs (or even keVs) above threshold, where only one channel is energetically open.
The two processes are connected to each other via the detailed balance principle,
see {\it e.g.}, Ref.~\cite{ThompsonNunes}. Thus, conclusions similar to those obtained 
from studies on photodisintegration processes
are drawn from studies on radiative capture reactions.  
Besides the important role played in nuclear structure studies, results from {\it ab-initio} calculations
of low-energy radiative captures involving light nuclei significantly affect important areas of
astrophysics (see, {\it e.g.}, the review articles of Refs.~\cite{Adelberger98,Adelberger11}). 

For reactions involving charged objects, one typically introduces the astrophysical
$S$-factor, which is related to the radiative capture cross section, $\sigma_c(E)$, via
\begin{equation}
S(E)=\sigma_c(E)E \exp{[2\pi \eta(E)]}\, ,
\end{equation} 
where $E$ is the center of mass energy of the reactants and $\eta(E)$ is the Sommerfield parameter, 
which takes into account the probability of penetrating through the Coulomb barrier. $S(E)$ is a much
more slowly varying function of $E$ compared to $\sigma_c(E)$. The latter falls exponentially due
to the Coulomb repulsion. The $S$-factors at low (virtually zero) energies 
is particularly interesting for astrophysical studies, as at these energies nucleosynthesis
reactions take place in the stars. For example, $S$-factors of radiative capture reactions
are used as inputs parameters of ({\it i})  theoretical calculations aimed at determining the primordial
abundances of light elements~\cite{Smith92}, and  ({\it ii}) the standard solar model, relevant
for studies on stellar structure and evolution~\cite{Adelberger98,Adelberger11}. An accurate knowledge
of the pertinent cross sections is then a crucial prerequisite for the above mentioned
astrophysical studies. The experimental evaluation of such cross sections is often problematic,
leading to scarce and/or inaccurate data. Due to the level of reliability reached by
 {\it ab-initio} calculations, cross sections derived from them are used as inputs,
instead. This strategy has been recently implemented in, {\it e.g.}, Refs.~\cite{Nollett11,Nollett14},
where the calculated $^2$H and $^7$Li yields of big-bang nucleosynthesis
rely on the theoretical estimates of the $p(n,\gamma)d$ and $d(p,\gamma)^3$He cross sections,
obtained within pionless EFT~\cite{Rupak99}, and the conventional
approach~\cite{Viviani:2000,Marcucci05}, respectively.  
  
After the review by Carlson and Schiavilla~\cite{Carlson98}, a number of {\it ab-initio}
calculations of e.m.~reactions in light nuclei relevant to astrophysics has appeared in the
literature. Some of these calculations have been already reviewed
by Marcucci {\it et al.} in Ref.~\cite{Marcucci04}. Below, we report on a few of
the most recent highlights, with emphasis on advances brought in by the use of nuclear $\chi$EFTs
and by the introduction of novel computational methods.

\subsubsection{Systems with $A\le4$}

$-\, \, \,\,$ Low-energy proton and neutron captures on  few-nucleon systems are 
particularly interesting  from the standpoint of nuclear structure studies. Indeed, MEC effects were
at first quantified as a correction to the calculated thermal neutron radiative capture
on proton cross section, and were found to provide the 10\% contribution needed to resolve the
discrepancy with the experimental datum~\cite{Riska72}. 

Thermal neutron radiative captures are induced by the $M1$ transition operator
that connects the initial two-cluster state in relative S-wave and the final bound state.
The experimental values of the cross sections for thermal neutron captures on $^1$H, $d$, and
$^3$He in mb are: ($332.6\pm 0.7$)~\cite{Mughabghab81}, ($0.508 \pm 0.015$)~\cite{Jurney82},
and ($0.055\pm 0.003$)~\cite{Wolfs89,Wervelman91}, respectively. The drop in the values in going from 
$A=2$ to $A=3,4$ is due to the so called `pseudo-orthogonality' between the initial and
final wave functions in the reactions involving $A=3,4$ nuclei.
In fact, the $^3$H and $^4$He wave functions,  $\Psi_3$ and $\Psi_4$ respectively,
are approximately eigenfunctions of the LO (or IA) one-body $M1$ operator ${\bm \mu}(\rm IA)$ (see Eq.~(\ref{eq:M1_IA})),
namely $\mu({\rm IA})_z  \Psi_3 \simeq \mu_p \Psi_3$ and $\mu({\rm IA})_z  \Psi_4 \simeq 0$,
where $\mu_p$=2.793 n.m.~is the proton magnetic moment---the experimental
value of the $^3$H magnetic moment is 2.979 n.m, while $^4$He has no
magnetic moment. If small components in wave functions, generated by tensor
components in the nuclear potentials, are neglected, then the matrix elements 
$\langle\Psi_3\!\mid\!\mu{(\rm IA)}_z\!\mid\!\Psi_{1+2}\rangle$
and $\langle\Psi_4\!\mid\!\mu({\rm IA})_z\!\mid\!\Psi_{1+3}\rangle$ vanish
due to orthogonality between the initial and final states.
In the case of the deuteron, instead,  the $M1$ operator can connect the large deuteron S-wave component
to the $T$=1 $^1$S$_0$ $np$ scattering state.

Due to this suppression at the IA  level, radiative
capture cross sections are particularly sensitive both to small components
in the wave functions (and, therefore, indirectly to the nuclear Hamiltonians
utilized to generate them), and to many-body components in the e.m.~current operators. 
Early calculations of the $nd$ and $n\,^3$He radiative capture cross sections
in IA predicted only $\sim 50\%$~\cite{Friar90} and $\sim 10\%$~\cite{Carlson90},
respectively, of the corresponding experimental values. Studies that account for corrections
due to conventional MEC reported a calculated value of the $nd$ radiative capture
within 15\% of the experimental value~\cite{Viviani96}, while 
the calculated value of the $n\,^3$He cross section  was found to be 86 $\mu$b~\cite{Schiavilla92},
to be compared to the experimental value of 55$\pm$3 $\mu$b.

\begin{center}
\begin{figure}
\centering
\includegraphics[width=4in]{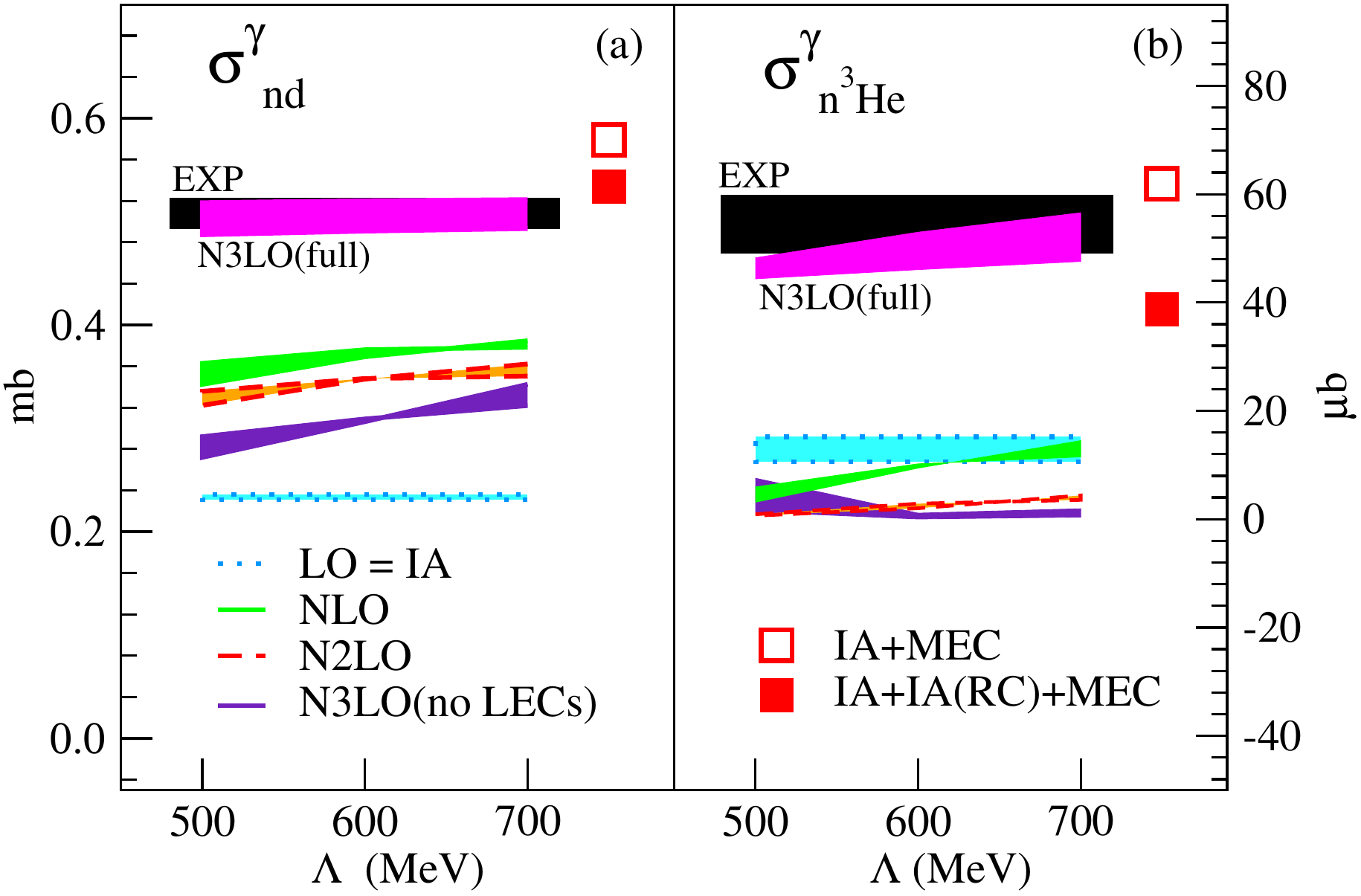}\\
\caption{(Color online) Results for $\sigma_{nd}$ $(a)$, $\sigma_{n^3{\rm He}}$ $(b)$ from Ref.~\cite{Girlanda10}, obtained by including cumulatively the LO, NLO, N2LO, N3LO(no LECs), and
N3LO(full) contributions to the e.m.~current operator from $\chi$EFT. Also shown are predictions
obtained in the conventional approach (squares labeled
IA+MEC and IA+IA(RC)+MEC, which includes relativistic corrections to the IA operator).  
The experimental data (black) are from
Ref.~\cite{Jurney82} for $nd$ and Ref.~\cite{Wolfs89} for $n\, ^3$He. The band height represents the experimental error bar.}
\label{fig:ng-cap}
\end{figure}
\end{center}

In  recent years, methods for solving the
$A=3$ and $4$ Schr\"odinger equations have been refined~\cite{Deltuva07,Lazauskas09a,Viviani10},
and highly accurate nuclear wave functions for these systems are now available.  
Theoretical calculations of the $nd$ and $n\,^3$He radiative capture cross sections~\cite{Girlanda10}
based on nuclear wave functions obtained from the HH techniques~\cite{Kievsky08} 
are shown in Figure~\ref{fig:ng-cap}. The bands represent results from hybrid and chiral 
calculations that use the $\chi$EFT currents developed in Refs.~\cite{Pastore08,Pastore09}.
The thickness of the bands represents the spread in the calculated values
corresponding to the two  considered nuclear Hamiltonians, namely the AV18+UIX %~\cite{AV18,Pudliner95}
and the NN(N3LO)+3N(N2LO), where the two-- and three--body chiral interactions
are form Refs.~\cite{Entem03} and~\cite{Ga_beta_triton}, respectively. The sensitivity
to the regularization cutoff $\Lambda=500$--$700$ MeV is shown on the $x$-axis.
The procedure implemented to fix the LECs entering the e.m.~currents
is different from that one utilized for the calculations of Refs.~\cite{Piarulli12,Pastore12}
that have been discussed in Secs.~\ref{sec:gs-mag-mom} and~\ref{sec:a3-ffs}, respectively.
In particular, here, the  $\Delta$-resonant saturation  argument has been exploited to 
infer the ratio between the isovector LECs entering the tree-level current at N3LO ($\nu=1$)---see 
panel $(k)$ of Figure~\ref{fig:chi_cnt}. Therefore, one is left with four e.m.~LECs. Namely, an isoscalar
and an isovector LEC associated with the tree-level current at N3LO, see panel $(k)$ of Figure~\ref{fig:chi_cnt},
plus an isoscalar and an isovector LEC associated with a contact-like current at N3LO,
see panel $(j)$ of Figure~\ref{fig:chi_cnt}. These four LECs have been fixed so as to reproduce
the deuteron magnetic moment, the isoscalar and isovector combinations of the trinucleon
magnetic moments, and the $np\rightarrow d\gamma$ cross section at thermal neutron energy~\cite{Girlanda10}.
In addition, LECs of minimal nature entering the N3LO contact currents have been taken from a
NLO ($\nu=2$) $\chi$EFT potential~\cite{Pastore09} rather than from the N3LO ($\nu=4$)~\cite{Entem03} $\chi$EFT potential. 

In Figure~\ref{fig:ng-cap}, results obtained with the complete N3LO e.m.~operator are shown by the magenta (light)
band labeled N3LO(full), and are in very satisfactory agreement
with data (black band).  Their sensitivity to the cutoff
is negligible ($\sim 10$\%) for the $nd$ ($n\, ^3$He) capture.
As discussed above, these processes are strongly suppressed at IA or LO ($\nu=-2$):
the calculated $\sigma_{nd}^\gamma$(LO) and $\sigma_{n\, ^3{\rm He}}^\gamma$(LO)
are less than half and a factor of five smaller than the measured values, respectively.
The IA(RC) correction at N2LO ($\nu=0$), corresponding to the relativistic correction to the LO 
e.m.~current illustrated in panel $(d)$ of Figure~\ref{fig:chi_cnt}, along with the correction labeled
N3LO(no-LECs), corresponding to the one-loop and minimal contact currents at N3LO,
is found to have opposite sign with respect to the LO contribution. 
The N3LO(full) contributions, corresponding to the tree-level and non-minimal
contact currents at N3LO---see panels $(j)$ and $(k)$ of Figure~\ref{fig:chi_cnt}--are large and crucially
important for bringing theory into agreement with experiment. 
Results obtained within the conventional model are also shown in Figure~\ref{fig:ng-cap}.
These calculations use the MEC of Ref.~\cite{Marcucci05} that have been constructed
from the AV18+UIX nuclear Hamiltonian, along with the $A=3,4$ improved
nuclear wave functions~\cite{Viviani10}. In particular, empty red squares
represent calculations obtained with the non-relativistic IA operator 
plus conventional MEC corrections. The full red squares add to the empty ones
the IA(RC) contribution  (or, the N2LO current contribution illustrated in panel 
$(d)$ of Figure~\ref{fig:chi_cnt}).
These relativistic corrections have been neglected in previous conventional
calculations~\cite{Friar90,Carlson90,Schiavilla92,Viviani96,Marcucci05},
however, they are found to be significant and, at least in the
case of the $nd$ capture, they bring the prediction of the conventional 
approach within 4\% of the experimental data, as opposed to the 15\% estimate from
previous studies~\cite{Viviani96}. Despite this satisfactory result,
the description of the $n^3$He remains problematic. We note that
hybrid studies based on the $\chi$EFT currents by Park {\it et al.}~\cite{Park96}
have reported values for the $nd$ and $n^3$He capture cross sections 
about 6\% and 15\% smaller than measured, with a cutoff sensitivity
of about 15\%~\cite{Song07,Song09,Lazauskas09}.    

Proton capture reactions on neutron or few-body nuclei have been calculated in the past with 
conventional potentials and currents by Viviani {\it et al.}~\cite{Viviani:2000}, 
by Golak {\it et al.}~\cite{Golack_MEC_A3}, and by Marcucci {\it et al.}~\cite{Marcucci05}. 
At energies of the order of a few MeV and up, where this process is dominated by an $E1$ transition,
conclusions similar to those obtained from studies on photodisintegration reactions are drawn.
In particular, the one-body current alone does not suffice to describe the data, however,
when MEC contributions are included, explicitly or implicitly via Siegert theorem, an improved description of the 
data is obtained. 
At the low energies relevant for astrophysics, the capture process is instead dominated by 
the $M1$ multipole. Because there is no Siegert theorem for magnetic transitions, MEC have to be 
included explicitly. This has been accomplished in Ref.~\cite{Marcucci05}, where MEC contributions
have been found to be large and necessary for a satisfactory description of 
the LUNA data for the $^2$H$(p,\gamma)^3$He reaction. 

The $d(d, \gamma)^4$He reaction has been recently
investigated with {\it ab-initio} methods by Arai {\it et al.}~\cite{Arai11}. 
This radiative capture can impact studies on the abundances of primordial elements. 
This process occurs predominantly via an $E2$ transition at low-energy. 
In Ref.~\cite{Arai11}, calculations with a realistic potential and with 
simple semirealistic central forces, have been compared, and it has been found that 
the $S$-factor for the $d(d, \gamma)^4$He reaction is very sensitive to the tensor
forces.

\subsubsection{Systems with $A>4$}

$-\, \, \,\,$ Many of the relevant nuclear reactions in primordial nucleosynthesis and 
in solar neutrino production involve light nuclei with mass number $A>4$~\cite{Adelberger98, Adelberger11}.
The first steps towards {\it ab-initio} calculations of a few key reactions have been 
taken using Quantum Monte Carlo methods. For example, the  $\alpha$ capture 
reactions $^2$H$(\alpha, \gamma)^6$Li, $^3$H$(\alpha, \gamma)^7$Li, and  $^3$He$(\alpha, \gamma)^7$Be 
have been evaluated in Refs.~\cite{Nollett01-1,Nollett01-2} using VMC wave functions for the 
initial clusters and the final nucleus. However, a phenomenological relative wave function between 
the two initial clusters was used. 
Similarly, a first attempt to calculate the $^7$Be$(p, \gamma)^8$B reaction was proposed 
by Navratil {\it et al.}~\cite{Navratil06} using the interior overlap functions obtained from NCSM 
 to constrain the tail of a phenomenological Wood-Saxton potential.

A full {\it ab-initio} calculation requires the use of the same Hamiltonian 
for both the initial continuum state and the bound final state. This has been only 
recently achieved for nuclei with $A>5$. Below, we  highlight two examples.

\begin{center}
\begin{figure}[htb]
\centering
\includegraphics[width=9cm,clip]{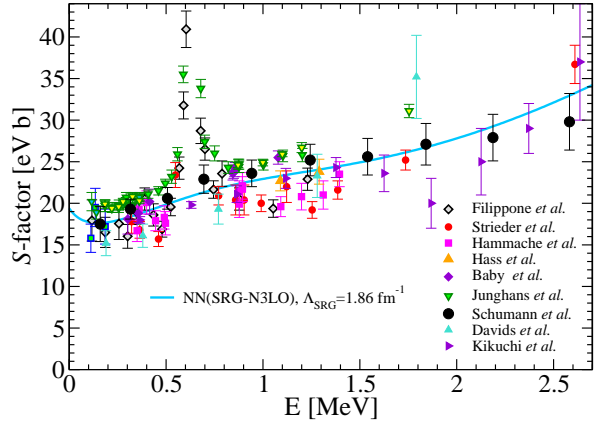}
\caption{(Color online) $^7$Be$(p, \gamma)^8$B $S$-factor: calculation by Navratil {\it et al.}~\cite{Navratil:2011sa}
in comparison to the available experimental data shown as in Ref.~\cite{Navratil:2011sa}.}
\label{fig:Be7p} 
\end{figure}
\end{center}

The $^7$Be$(p, \gamma)^8$B $S$-factor  is an important input in modeling the 
neutrino flux coming from our Sun~\cite{Adelberger11}.  An {\it ab-initio} calculation of this process
has been performed by Navratil {\it et al.}~\cite{Navratil:2011sa} using the NCSM with the resonating
group method~\cite{quaglioni2008, Quaglioni09}, which allows to describe continuum states. The used starting
interaction is a $\chi$EFT two-body potential at N3LO ($\nu=4$)~\cite{Entem03}, which has then been softened 
by a Similarity Renormalization Group (SRG) transformation~\cite{Bogner2006}. The latter is characterized 
by an evolution parameter,
 $\Lambda_{\rm SRG}$,
which has been chosen by tuning the calculated $^8$B separation energy close to 
the experimental value. In fact, the low-energy behavior of the $S$-factor is very sensitive 
to the threshold energy.

This process has been investigated using different experimental techniques, including direct 
measurements with proton beams on $^7$Be targets and indirect Coulomb excitations with $^8$B beams
 hitting on a heavy target and breaking into proton and $^7$Be. Data are shown in Figure~\ref{fig:Be7p} 
for the astrophysical $S$-factor. Even though they  are a bit scattered, they clearly display a resonant 
feature due to the $1^+$ excitation, which however is very narrow and does not affect much the $S$-factor
 at zero energy, where the process is dominated by a dipole transition.
The theoretical curve shown in Figure~\ref{fig:Be7p} does not include any $M1$ transitions,
 thus it does not show the resonant feature. However, it nicely reproduces the shape of the $E1$ contribution to the $S$-factor.  
Because the dipole operator is protected by the Siegert theorem, the leading part of the MEC is implicitly included.
SRG renormalizations of the dipole operator and overall 3N forces have been conjectured to be small, 
and omitted for the time being~\cite{Navratil:2011sa}.
\begin{center}
\begin{figure}[htb]
\centering
\includegraphics[width=9cm,clip]{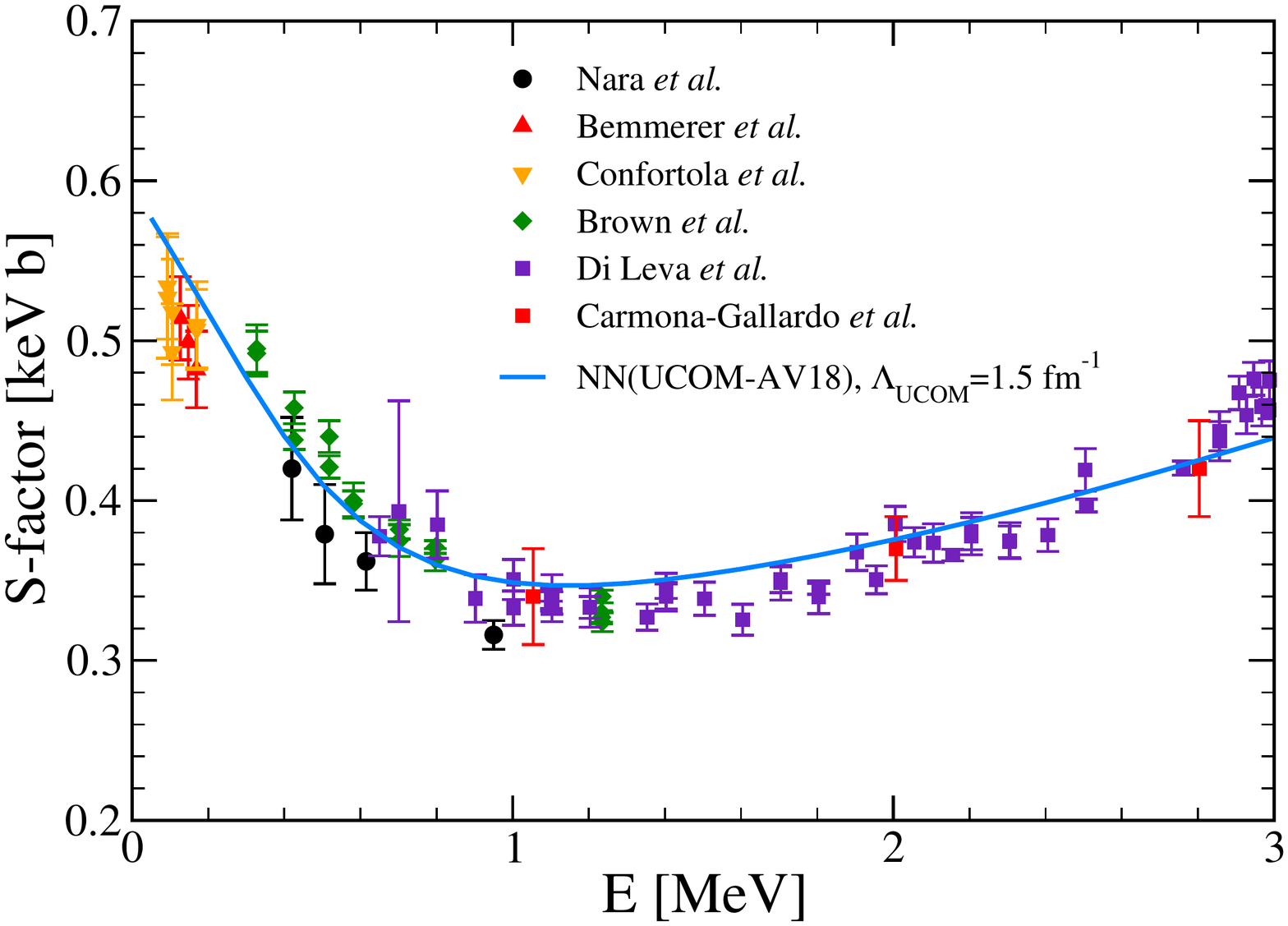}
\caption{(Color online) $^3$He$(\alpha, \gamma)^7$Be $S$-factor: calculation by Neff~\cite{Neff2011}
in comparison to the available recent experimental data from the Weizmann Institute by  Nara {\it et al.}~\cite{Nara04}, from the LUNA collaboration by Bemmerer {\it et al.}~\cite{Bemmerer06} and Confortola {\it et al.}~\cite{Confortola07},  from Seattle by Brown {\it et al.}~\cite{Brown07}, from Bochum with the recoil separator ERNA  by Di Leva {\it et al.}~\cite{DiLeva09} and from Madrid by Carmona-Gallardo  {\it et al.}~\cite{Carmona12}.}
\label{fig:He7alpha}       
\end{figure}
\end{center}

The $^3$He$(\alpha, \gamma)^7$Be radiative capture takes place in the solar Hydrogen burning 
reaction chains. It is important in determining the high energy solar neutrino flux and 
in understanding the abundance of primordial $^7$Li. It has been measured by several groups, 
but the difficulty of reaching the very low solar energies makes it hard to extract the needed 
$S$-factor at zero energy. 
This reaction has been recently calculated by Neff~\cite{Neff2011} within 
the Fermionic Molecular Dynamics (FMD)~\cite{Neff2008}, which allows to consistently 
obtain bound and scattering states from a soft effective interaction. 
%Another advantage of the FMD basis is its flexibility to describe both polarized clusters and shell-model-like configurations.
The starting potential used in the calculations is the AV18 %~\cite{AV18} 
two-body force, which has been then softened by using the unitary correlation operator 
method~\cite{UCOMpotential, Roth2010}. Here, the evolution parameter, $\Lambda_{\rm UCOM}$, is tuned so as to reproduce the
 $^3$He+$^4$He threshold in $^7$Be.
The  calculation carried out with a dipole transition operator is shown 
in Figure~\ref{fig:He7alpha} in comparison to recent measurements. As one can see,
 both the energy dependence  and the absolute normalization are in good agreement with experiment.
The calculation provides a zero-energy $S$-factor of 0.592 keVb. Also in this case 3N forces and
 renormalizations to the dipole operator have been omitted.

The two examples above represent a break-through in the development of calculational techniques
that can tackle radiative capture reactions in larger nuclei. However, further studies are needed 
to assess the theoretical sensitivity of these processes to intermediate and short range physics, entering, {\it e.g.}, 
in the form of 3N forces or MEC. In the future, we can expect more investigations along these lines
also from the newly developed method to calculate radiative captures from lattice EFT introduced in Ref.~\cite{Rupak2013}, and firstly applied to the $p(n, \gamma)d$ reaction.

Finally, we would like to mention that alternative low-energy effective field theories, such as Halo EFTs, 
can be applied to the study of radiative capture reactions. This has been recently accomplished, {\it e.g.}, 
in Refs.~\cite{Rupak11,Zhang14-1,Zhang14-2} for $A=8$ systems.

\subsection{Electromagnetic transitions in low-lying nuclear states}
\label{sec:e2m1}

In this last section we discuss a number of calculated reduced e.m.~transition probabilities
for $E2$ and $M1$ operators in low-lying excited states. The measurement of these
observables can involve different nuclear reaction mechanisms. Excited states can be
populated, for example, via pure hadronic reactions, or by e.m.~induced reactions.
To the former class belongs, for example, the $d+^6$Li$\rightarrow p+^7$Li$^*$ reaction,
which can be exploited to populate the $^7$Li excited states. To the second class belong, for example, 
radiative capture processes, such as the $\alpha\alpha$ radiative capture
reaction (or {\it bremsstrahlung}) that is used to populate low-lying excited 
states of $^8$Be~\cite{Datar13}. 
Nuclear states' lifetimes and associated reduced transition
probabilities are then inferred from the observed photons emitted in the decay process.
Transition probabilities for stable targets, {\it e.g.}, $^7$Li and $^9$Be, 
can also be (indirectly) accessed through ($e$,$e^\prime$) scattering experiments.
Therefore, technically speaking, some of the transitions we discuss do not strictly belong
to this part of the review devoted to processes involving real photons.
 
There are a number of {\it ab-initio} calculations of e.m.~transitions carried out in IA,
among which the most notable and recent are, for example, transitions
occurring in low-lying states of $^{6-7-8}$Li~\cite{Cockrell12},
$^{10}$Be~\cite{McCutchan09,Liu11}, $^{10}$C and  $^{10}$B~\cite{McCutchan12},
$^{12}$C~\cite{Epelbaum12_hoyle_BE2, Maris14}, and  $^{16}$O~\cite{Epelbaum13paa}.
Because our interest lies in going beyond the IA, which is crucial for magnetic transitions,
below we focus the discussion on recent GFMC calculations reported in Ref.~\cite{Pastore12}.
In that work, two-body components in the e.m.~current operator
have been explicitly accounted for in the  $M1$ transitions induced by the 
operator defined in Eq.~(\ref{eq:mag-mom}).
In particular, $\chi$EFT operators developed in Refs.~\cite{Pastore08,Pastore09,Pastore11,Piarulli12}
have been used in hybrid calculations based on nuclear wave functions obtained from the
AV18+IL7 
nuclear Hamiltonian. Because the $M1$ and the $E2$ operators can connect the same states, it is interesting
to compare magnetic dipole and electric quadrupole transitions.  
The  $E2$ transitions can be easily obtained by using the Siegert theorem, {\it i.e.},
writing the quadrupole operator of Eq.~(\ref{eq:q2-mom}) as
\begin{equation} 
 Q({\rm IA}) = \sum_i e_{N,i}\, r_i^2 \, Y_2(\hat{\bf r}_i) \ ,
\end{equation}
where $Y_2$ is the spherical harmonic of rank 2. 
The decay widths in units of MeV are obtained from the reduced transition probabilities  defined in
Eqs.~(\ref{eq:be2}) and~(\ref{eq:bm1}), via~\cite{Ring80}
\begin{eqnarray}
\label{eq:gammaE}
 \Gamma({E2})& = &0.241 \left (\frac{\Delta E}{\hbar c} \right)^5 B(E2) \ , \\
\label{eq:gammaM}
 \Gamma({M1}) &= &0.890 \left (\frac{\Delta E}{\hbar c} \right)^3 B(M1) \ ,
\end{eqnarray}
where $\hbar c$ is in units of MeV~fm, and $\Delta E$ is the experimental energy
difference between the final and the initial state (in units of MeV), as obtained
from the values reported in Refs.~\cite{Tilley02,Tilley04}.

\begin{center}
\begin{figure}
\centering
\includegraphics[height=.4\textheight,keepaspectratio=true]{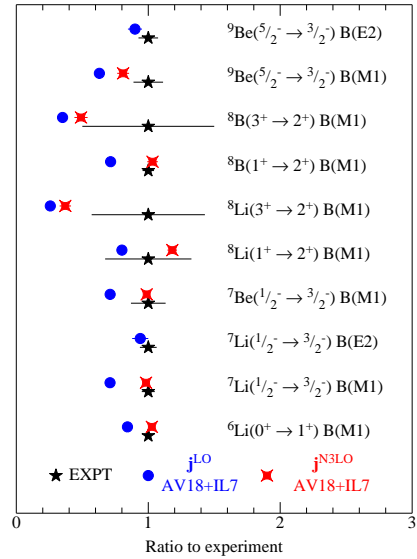}\\
\caption{(Color online) Transition widths from Ref.~\cite{Pastore12} normalized to the experimental values~\cite{Tilley02,Tilley04}
           for $A=7$--$8$ nuclei. The notation is as in Figure~\ref{fig:mm}.}
\label{fig:e2m1}
\end{figure}
\end{center}

Results from these hybrid calculations in $A\leq9$ nuclei are shown in Figure~\ref{fig:e2m1}.
In this figure, we show the ratios to the experimental values of the widths~\cite{Tilley02,Tilley04},
which are represented by the black stars, along with their associated experimental error bars.
Predictions in IA are represented by blue dots\footnotemark[1]\footnotetext{Note that the $E2$ transitions implicitly include the effect of two-body currents via the Siegert theorem, where the charge density is used in IA (which explains the use of the blue color).}, while those obtained with the full e.m.~current operator
are represented by red diamonds. Predictions for the $A=6,7$ nuclei as well as that for the 
($J_i^{\pi_i}=1^+\!\rightarrow\! J_f^{\pi_f}=2^+$) transition in $^8$B
are in very good agreement with the experimental data,
and corrections from two-body e.m.~current operators are essential to reach
the agreement with them. For the remaining $M1$ transitions
in $A=8$ and $9$ nuclei, the comparison becomes more difficult, as the experimental
errors are too large to allow for conclusive statements on the agreement between
theory and experiment.  We also note that theoretical 
description of the two $E2$ transitions reported in Figure~\ref{fig:e2m1} is in good agreement with the experiments.
This is presumably due to the fact that two-body effects in the $E2$ multipole
are implicitly accounted for via Siegert theorem.  
The hybrid GFMC calculational scheme described above has been very recently
applied to evaluate about a dozen $M1$ transitions occurring in low-lying states
of $^8$Be~\cite{Pastore14}, and corrections from two-body e.m.~currents
have been found to provide $\sim 20-30$\% of the total calculated transition
matrix elements.  

In Ref.~\cite{Pastore12}, a number of predictions (not shown in the figure) for $E2$ transitions
in low-lying excited states along with a prediction for the 
($J_i^{\pi_i}=\frac{1}{2}^-\!\rightarrow\!J_f^{\pi_f}=\frac{3}{2}^-$) $M1$
transition in $^9$Li, obtained with the full $\chi$EFT operator at N3LO ($\nu=1$), have
been provided. The possibility to experimentally verify such predictions on  $^9$Li
is being discussed at TRIUMF~\cite{TRIUMF_proposal}.
Experimental efforts aimed at both increasing the precision of the
existing measured transitions, and at investigating currently unknown transitions
would play an important role in assessing the accuracy of the available theoretical nuclear
models. Work along these lines has been recently presented in Ref.~\cite{Datar13},
where the authors have measured the ($J_i^{\pi_i}=4^+\!\rightarrow\! J_f^{\pi_f}=2^+$) 
$E2$ transition in $^8$Be via $\alpha\alpha$ radiative capture
reaction, and have reduced the experimental error by more than a factor
of three with respect to previous existing measurements. Along with the improved
experimental datum, Ref.~\cite{Datar13} reported a theoretical GFMC calculation
of the $B(E2)$ associated with that transition. However, the comparison between
the experiment and theory was complicated, in this case, by the resonant nature
of the $J^{\pi}=2^+$ and $4^+$ states, which tend to break apart into two $\alpha$ particles.
The effect of the continuum in the resonance has been investigated within a cluster approach
(see, {\it e.g.}, Ref.~\cite{Jeremy}),
and extensions of these studies to the {\it ab-initio} framework are being presently investigated.

\section{Summary and outlook}
\label{sec:conclusions}

In this review, we provided a summary on the present status of
{\it ab-initio} calculations of e.m.~observables in light nuclei. 
We presented calculations in which nuclei are described in terms of non-relativistic
nucleons interacting via many-body potentials and e.m.~probes interact with the nucleons
via many-body e.m.~currents. We discussed reactions
occurring at energies below the pion threshold  and 
focused on studies released after the 1998 review article by Carlson and
Schiavilla~\cite{Carlson98}. The years following the publication of that review
have witnessed a tremendous progress of computational techniques and resources, 
as well as the development of novel many-body methods,
which have made it possible to extend {\it ab-initio} studies to nuclei with $A\sim 16$.
In conjunction with these technological developments, conventional nuclear Hamiltonians
and e.m.~currents have been improved  to reach a high level of sophistication. Meanwhile,
nuclear $\chi$EFTs have evolved into an intense and prolific field of research and
a description of both nuclear potentials and interactions of nuclei with external 
e.m.~probes is now accessible also from the $\chi$EFT perspective.  

The study of e.m.~reactions presents numerous advantages from the theoretical and
experimental points of view. The interaction of light nuclei with e.m.~probes is 
perturbative, and thus it is well described in terms of a single-photon exchange. 
In addition, nuclear structure effects are present only
in the nuclear targets, as opposed to hadronic reactions in which one has 
to worry about structure effects entering both hadronic probes and hadronic targets.
Furthermore, cross sections for e.m.~reactions, while being usually smaller than those 
associated with hadronic reactions,  are  comparatively bigger than those associated 
to weak probes, such as neutrinos. These features make e.m.~reactions ideal tools
to study nuclear dynamics.  

Obviously, a theoretical description of this kind of reactions is a demanding
task owing to the presence of both strong and e.m.~interactions.  
% 
%  the presence in this kind of reactions of both strong and e.m.~interactions 
% is a demanding test for the theoretical description.
The {\it ab-initio}  description of e.m.~observables is, in general, very satisfactory.
Whenever possible, we showed comparisons between calculations performed by various
groups using different techniques and/or different dynamical schemes.  What can be
inferred from such comparisons is that $(i)$ the theoretical accuracy is extremely 
well under control for very light nuclei (see, {\it e.g.}, the longitudinal response 
functions in $^3$He--Figure~\ref{fig:InelasticA3}), and $(ii)$ the theoretical prediction 
is very robust since the conventional and $\chi$EFT approaches 
agree, in most cases, with each other and with experimental data (see, {\it e.g.}, 
the low-momentum elastic form factors of deuteron, $^3$He/$^3$H and $^4$He--Figures~\ref{fig:gc-d}, 
\ref{fig:gm-d}, \ref{fig:ff-charge-a3}, \ref{fig:ff-mag-a3}, and~\ref{fig-el_he4}, 
as well as the photodisintegration cross section of $^4$He--Figure~\ref{fig_photon_he4}). 

There are, however, observables for which a solid understanding of the 
dynamics has not been achieved yet (see, {\it e.g.}, the inelastic monopole transition 
form factor of $^4$He--Figure~\ref{fig:finel_He4}). There, calculations with different 
realistic Hamiltonians disagree with each other and with experiments. Such challenging 
observables are very interesting, in that they provide an alternative tool to better understand 
and constrain the present knowledge of the nuclear dynamics.

In general, comparisons of theoretical calculations with available experimental data indicate that,  
if one aims at a well-founded and accurate description of the experimental data, 
$(iii)$  many-body components in the e.m.~currents and $(iv)$ many-body potentials in 
the nuclear Hamiltonians need to be accounted for.

Two-body e.m.~currents, especially those of one-pion range, are found to be significant in a number of e.m.~observables. 
For example, while they have been found to provide a $15\%$ correction to the calculated nuclear
magnetic moments of $A=3$ nuclei, their contribution has been found to be as large as $40\%$ of the total
calculated magnetic moment of $^9$C (Figure~\ref{fig:mm}), a sizable correction which cannot be neglected. 
Similarly, two-body e.m.~currents have been recently found to provide corrections at the $\sim 20-30\%$ level
in a number of calculated $M1$ transitions between low-lying excited states of $^8$Be~\cite{Pastore14}.  

Three-body forces are also found to play a crucial role  (see, for example, the longitudinal 
response functions of $^3$He and $^4$He--Figures~\ref{fig:InelasticA3} and~\ref{fig_el_3NF}). 
This  highlights that the study of e.m.~reactions complements that of hadronic reactions in the
quest to better understand the role of 3N forces. 

Furthermore, the development of new  many-body approaches, able to overcome stumbling 
blocks such as the {\it ab-initio} description of reactions
involving more than four nucleons, has just started to reveal its potential. Recent 
application of such methods to, for example, the photodisintegration of $^{16}$O (Figure~\ref{fig-O16}) 
and  radiative capture reactions (Figure~\ref{fig:Be7p} and~\ref{fig:He7alpha}),
are just the first examples of interesting problems that can now be tackled and more 
is expected to come in the future. In fact, the inclusion of the important two-body 
currents and 3N forces is presently being actively pursued by the theoretical community.
When these further steps are achieved, the present uncertainty due to 
the use of truncated dynamical input (for $A>4$) nuclei will be largely reduced. In the mean time, 
new experimental activity is being planned, as we referred to whenever possible, 
and several laboratories in the world are aiming at reducing the error bars of 
previous measurements and/or measuring new quantities predicted by the theory, with 
the common aim to further understand nuclear dynamics.

\ack{ 
This work was supported by the Natural Sciences
and Engineering Research Council (NSERC) and the National Research
Council of Canada (S.B.), and by the National Science Foundation,
grant No. PHY-1068305  and  the U.S.~Department of Energy, Office of Nuclear Physics, 
contract No.~DE-FG02-09ER41621 (S.P.). 

We would like to thank W.~Leidemann, G.~Orlandini, R.~Schiavilla, 
M.~Schindler and R.B.~Wiringa for a critical reading of the manuscript 
and for useful discussions at various stages of this review. We are grateful to
K.~Nollett,  A.~Kievsky, M.~Piarulli, S.C.~Pieper  and M.~Viviani for useful discussions.
We are thankful to A.~Lovato, T.~Neff, D.R.~Phillips and S.~Quaglioni for providing us with 
their original data and with helpful elucidations on their calculations. We thank Y.~Tanaka
for helping with the graphics, and B.~Davids and E.~Ricard-McCutchan for useful discussions
on experimental techniques. Finally, we owe sincere thanks to all our colleagues and friends 
from the various institutions, both experimentalists and theorists, with whom  
we collaborate on the quest of understanding the nuclear dynamics with e.m.~probes.
}

\section*{Acronyms and abbreviations}

\subsection*{Miscellaneous}
\begin{longtable}{@{}p{2cm}@{}p{\dimexpr\textwidth-1cm\relax}@{}}
\nomenclature{3N}{Three-nucleon}%
\nomenclature{4N}{Four-nucleon}%
\nomenclature{$\chi$EFT}{Chiral effective field theory}%
\nomenclature{CT}{Contact term}%
\nomenclature{EFT}{Effective field theory}%
\nomenclature{e.m.}{Electromagnetic}%
\nomenclature{IA}{Impulse approximation}%
\nomenclature{LEC}{Low Energy Constant}%
\nomenclature{MEC}{Meson exchange currents}%
\nomenclature{NN}{Nucleon-nucleon}%
\nomenclature{NR}{Non-relativistic}%
\nomenclature{OPE}{One-pion exchange}%
\nomenclature{QCD}{Quantum chromodynamics}%
\nomenclature{RC}{Relativistic correction}%
\nomenclature{SNPA}{Standard nuclear physics approach}%
\nomenclature{SRG}{Similarity renormalization group}%~\cite{Bogner2006}}%
\nomenclature{TPE}{Two-pion exchange}%
\nomenclature{UCOM}{Unitary correlation operator method}%~\cite{UCOMpotential}}%
\end{longtable}

\subsection*{Nuclear potentials}
\begin{longtable}{@{}p{2cm}@{}p{\dimexpr\textwidth-1cm\relax}@{}}
\nomenclature{AV18}{Argonne-$v_{18}$ NN potential~\cite{AV18}}%
\nomenclature{AV8$^\prime$}{Argonne-$v_{8}^\prime$ NN potential~\cite{pudliner1997}}%
\nomenclature{IL2}{Illinois-2 3N potential~\cite{IL7}}%
\nomenclature{IL7}{Illinois-7 3N potential~\cite{IL}}%
\nomenclature{JISP}{$J$-matrix inverse scattering potential~\cite{JISPpotential}}%
\nomenclature{TM}{Tucson-Melbourne 3N potential~\cite{TM}}%
\nomenclature{UIX}{Urbana IX 3N potential~\cite{Pudliner95}}%
\end{longtable}

\subsection*{Computational methods}
\begin{longtable}{@{}p{2cm}@{}p{\dimexpr\textwidth-1cm\relax}@{}}
\nomenclature{CC}{Coupled-cluster~\cite{CC_review}}%
\nomenclature{EIHH}{Effective interaction hyperspherical harmonics~\cite{EIHH,barnea2001}}%
\nomenclature{FMD}{Fermionic Molecular Dynamics~\cite{Neff2008}}
\nomenclature{GFMC}{Green's function Monte Carlo~\cite{pieper2001}}%
\nomenclature{HH}{Hypershperical harmonics~\cite{Kievsky08}}%
\nomenclature{LIT}{Lorentz integral transform~\cite{efros1994,Efl07}}%
\nomenclature{NCSM}{No-core shell model~\cite{navratil2009,Barrett13}}
\nomenclature{VMC}{Variational Monte Carlo~\cite{pieper2001}}%
\end{longtable}

%\appendix
%\section{First part}
%Here I am trying out a citation~\cite{ncsm}, 
%\cite{BaA04}
%\section{Second part}

% \nomenclature{UTC}{Coordinated Universal Time} 
% \nomenclature{ADT}{Atlantic Daylight Time} 
% \nomenclature{EST}{Eastern Standard Time}
% \printnomenclature

%Bibliography stile does not seem to be working properly. Could not fix it.
\bibliography{bibliography}
%\bibliographystyle{unsrt}
%\bibliographystyle{science}

%\printglossaries

\end{document}